\documentclass[numberedappendix]{emulateapj}
\pdfoutput=1
\usepackage{apjfonts}
\usepackage{natbib}
\usepackage{graphics}
\usepackage{multirow}
\usepackage{amsbsy}
\usepackage{mathrsfs}
\usepackage[table]{xcolor}
\def\bicep{{\sc Bicep}}

\def\wmap{{\sc Wmap}}

\def\healpix{{\tt Healpix}}

\def\deg{^{\circ}}

\def\muK{~\mu{\rm K}}

\newcommand{\ukrts}{ $\mu\mathrm{K}_{\mathrm{\mbox{\tiny\sc cmb}}}\sqrt{\mathrm{s}}$ }

\def\aap{Astr. \& Astroph.\ }

\def\apjl{Astrophys.\ J.\ Lett.\ }
\def\apjs{Astrophys.\ J.\ Suppl.\ }

\shorttitle{\bicep\ Galactic Observations}
\shortauthors{Bierman et al.}

\begin{document}

\title{A Millimeter-Wave Galactic Plane Survey with the BICEP Polarimeter}

\author{
E. M. Bierman\altaffilmark{1},
T. Matsumura\altaffilmark{2},
C. D. Dowell\altaffilmark{2,3},
B. G. Keating\altaffilmark{1},
P. Ade\altaffilmark{4},
D. Barkats\altaffilmark{5},
D. Barron\altaffilmark{1},
J. O. Battle\altaffilmark{3}, \\
J. J. Bock\altaffilmark{2,3},
H. C. Chiang\altaffilmark{2, 6},
T. L. Culverhouse\altaffilmark{2},
L. Duband\altaffilmark{7},
E. F. Hivon\altaffilmark{8},
W. L. Holzapfel\altaffilmark{9},
V. V. Hristov\altaffilmark{2}, \\
J. P. Kaufman\altaffilmark{1},
J. M. Kovac\altaffilmark{2, 10},
C. L. Kuo\altaffilmark{11}, 
A. E. Lange\altaffilmark{2},
E. M. Leitch\altaffilmark{3},
P. V. Mason\altaffilmark{2}, 
N. J. Miller\altaffilmark{1},
H. T Nguyen\altaffilmark{3},\\
C. Pryke\altaffilmark{13}, 
S. Richter\altaffilmark{2},
G. M. Rocha\altaffilmark{2},
C. Sheehy\altaffilmark{13},
Y. D. Takahashi\altaffilmark{9},
K. W. Yoon \altaffilmark{14}
}

\altaffiltext{1}{University of California - San Diego, USA}
\altaffiltext{2}{California Institute of Technology, USA}
\altaffiltext{3}{Jet Propulsion Laboratory, USA}
\altaffiltext{4}{University of Wales, UK}
\altaffiltext{5}{Joint ALMA Office-NRAO, Chile}
\altaffiltext{6}{Princeton University, USA}
\altaffiltext{7}{Commissariat \`{a} l'\'{E}nergie Atomique, France}
\altaffiltext{8}{Institut d'Astrophysique de Paris, France}
\altaffiltext{9}{University of California - Berkeley, USA}
\altaffiltext{10}{Harvard University, USA}
\altaffiltext{11}{Stanford University, CA, USA}
\altaffiltext{12}{Universite Paris XI, France}
\altaffiltext{13}{University of Chicago, USA}
\altaffiltext{14}{National Institute of Standards and Technology, USA}

\email{ebierman@physics.ucsd.edu}

%%%%%%%%%%%%%%%%%%%%%%%%%%%%%%%%%%%%%%%%%%%%%%%%%%%%%%%%%%%%%%%%%%%%%%%%%%%%%%%%%%%%%%%%%%%%%%%%%%%%%%%%%%%%%%%%%%%%%%%%%%%%%%%%%%%%%%%%%%%%%%%%%%%%%%%%%%%%%%%%%%%%%%%%%%%%%%%%%%%%%%%%%%%%%%%%%%%%%%%%%%%%%%%%%%%%%%%%%%%%%%%%%%%%%%

\begin{abstract}
In addition to its potential to probe the Inflationary cosmological paradigm, millimeter-wave polarimetry is a powerful tool for studying the Milky Way galaxy's composition and magnetic field structure.  Towards this end, presented here are Stokes $I$, $Q$, and $U$ maps of the Galactic plane from the millimeter-wave polarimeter BICEP covering the Galactic longitude range $260\deg < \ell < 340\deg$ in three atmospheric transmission windows centered on 100, 150, and 220 GHz.  The maps sample an optical depth  $1 \lesssim A_V \lesssim 30$, and are consistent with previous characterizations of the Galactic millimeter-wave frequency spectrum and the large-scale magnetic field structure permeating the interstellar medium.  Polarized emission is detected over the entire region within two degrees of the Galactic plane and indicates that the large-scale magnetic field is oriented parallel to the plane of the Galaxy.  An observed trend of decreasing polarization fraction with increasing total intensity rules out the simplest model of a constant Galactic magnetic field throughout the Galaxy.  Including \wmap\ data in the analysis, the degree-scale frequency spectrum of Galactic polarization fraction is plotted between 23 and 220 GHz for the first time.  A generally increasing trend of polarization fraction with electromagnetic frequency is found, which varies from 0.5\%-1.5\% at frequencies below 50 GHz to 2.5\%-3.5\% above 90 GHz.  The \bicep\ and \wmap\ data are fit to a two-component (synchrotron and dust) model showing that the higher frequency \bicep\ data are necessary to tightly constrain the amplitude and spectral index of Galactic dust.  Furthermore, the dust amplitude predicted by this two-component fit is consistent with model predictions of dust emission in the \bicep\ bands.  The polarization angles in all three bands are generally perpendicular to those measured by starlight polarimetry and show changes in the structure of the Galactic magnetic field on the scale of $~60\deg$.  Map noise and systematic effects are characterized and the resulting maps and derived parameters are corrected for spectral mismatch leakage and time-series filtering effects.  The effort to extend the capabilities of \bicep\ by installing 220 GHz polarization band hardware into the existing \bicep\ focal plane is described along with the results of the data analysis from the new band.  

\end{abstract}

\keywords{Surveys --- Submillimeter ---
          cosmic microwave background polarization ---
          cosmology: observations --- diffuse radiation --- 
          Galaxy: structure }

%%%%%%%%%%%%%%%%%%%%%%%%%%%%%%%%%%%%%%%%%%%%%%%%%%%%%%%%%%%%%%%%%%%%%%%%%%%%%%%%%%%%%%%%%%%%%%%%%%%%%%%%%%%%%%%%%%%%%%%%%%%%%%%%%%%%%%%%%%%%%%%%%%%%%%%%%%%%%%%%%%%%%%%%%%%%%%%%%%%%%%%%%%%%%%%%%%%%%%%%%%%%%%%%%%%%%%%%%%%%%%%%%%%%%%%%%%%%%%%%%%%%

\section{Introduction}\label{sec:intro}
Emission from the Milky Way Galaxy at millimeter wavelengths is both a rich source of astrophysical information and a potential contaminant for cosmic microwave background (CMB) observations.  The density of gas and dust in the interstellar medium (ISM) varies from very low ($<1$ particle$/$cm$^3$) in diffuse regions to very high ($>10^6$ particles$/$cm$^3$) in molecular clouds and complexes (collections of star-forming cloud cores at approximately the same distance and age).  Measurements in the millimeter band have the potential to probe a wide range of ISM densities while near-IR bands~\citep{Martin1990} have enough resolution to probe medium to high density ISM regions.  In general, large-scale diffuse emission from the Galaxy at frequencies below 90 GHz is dominated by the synchrotron mechanism in ionized gas, and to a lesser extent free-free along with thermal emission from rotational and vibrational modes of dust.  Emission at frequencies above 90 GHz is dominated by vibrational modes of Galactic dust~\citep{Whittet1992}.  The exact composition and emission spectrum of the Galaxy varies across the sky and can be determined with multi-wavelength observations in the infrared and millimeter measuring continuum emission, spectral lines, and polarization.

Polarized radiation probes various aspects of the ISM and Galactic magnetic field.  Optical starlight polarization~\citep{Davis1951} is due to the absorption of background light by dust grains aligned perpendicular to the Galactic magnetic field.  Polarimetric observations of stars have revealed that the polarization and hence the Galactic magnetic field in the diffuse component are oriented preferentially parallel to the Galactic plane~\citep{Mathewson1970}.  Dust grains emit radiation in the infrared and millimeter bands, polarized along their long dimensions, making thermal dust emission polarized orthogonal to the Galactic magnetic field~\citep{Lazarian2007}.  While not dominant at any millimeter-wave band, free-free emission from electron-ion scattering can contribute to the measured intensity near 90 GHz; however, this emission is not polarized.  The desire to characterize the ISM and the Galactic magnetic field provides the motivation for millimeter-wave continuum polarimetry of the Galaxy~\citep{Hildebrand1988,Hildebrand1999,Benoit2004,Chuss2005,Matthews2009,Culverhouse2010}.  

The main goal of the Background Imager of Cosmic Extragalactic Polarization (\bicep, ~\cite{Keating2003a,Chiang2010}) and its successors\footnote{list of future space, balloon, and ground CMB experiments: http://cmbpol.uchicago.edu/workshops/technology2008/depot/meyer-stephan.pdf} is to search for the unique CMB B-mode polarization pattern due to primordial gravitational waves, which has an amplitude determined by the energy scale of inflation~\citep{Seljack1997letter,KamKos1997}.  Polarized Galactic emission is an astronomical foreground~\citep{Bock2008} that may need to be confronted to make this measurement, and therefore, the properties of the emission also motivate the investigation~\citep{Jones2003,Ponthieu2005,Tucci2005,Eriksen2006,Hansen2006,Amblard2007,Larson2007,Eriksen2008,Leach2008,Miville2008,Dodelson2009,Dunkley2009}.  Only a few experiments have explored large-scale Galactic polarization properties at frequencies between 90 GHz and 350 GHz, such as Archeops (353 GHz,~\cite{Benoit2004}), \wmap\ (94 GHz,~\cite{Kogut2007}), and QUaD (100 and 150 GHz,~\cite{Culverhouse2010}).  To further study emission from sources other than the CMB, \bicep\ was upgraded from a two-band experiment (100 and 150 GHz) to a three-band polarimeter with the addition of 220 GHz capability for the second and third seasons.  This paper discusses unique aspects of the \bicep\ Galactic observations, including the Galactic maps with the additional 220 GHz channels (Section~\ref{sec:maps}), explains the data analysis methodology(Section~\ref{sec:scanstrategy}-\ref{sec:mapmaking}), and discusses the analysis of the polarized Galactic signal(Section~\ref{sec:flatvsslope}-\ref{sec:propvsfreq}).

%%%%%%%%%%%%%%%%%%%%%%%%%%%%%%%%%%%%%%%%%%%%%%%%%%%%%%%%%%%%%%%%%%%%%%%%%%%%%%%%%%%%%%%%%%%%%%%%%%%%%%%%%%%%%%%%%%%%%%%%%%%%%%%%%%%%%%%%%%%%%%%%%%%%%%%%%%%%%%%%%%%%%%%%%%%%%%%%%%%%%%%%%%%%%%%%%%%%%%%%%%%%%%%%%%%%%%%%%%%%%%%%%%%%%%%%%%%%%%%%%%%%

\section{BICEP Instrument and Mapmaking} \label{sec:instrument}

\subsection{Brief Instrument Description}\label{sec:instrumentsub}
For a complete description of the \bicep\ telescope see \cite{Yoon2006} and \cite{Yuki2010}.  \bicep\ is an on-axis refracting telescope with a 250 mm aperture and can scan in azimuth and elevation as well as rotate around the optical axis (boresight) of the telescope, which is less than 0.01$\deg$ away from the center feed.  \bicep's small aperture allows the entire optical system to be cooled to cryogenic temperatures in a vacuum cryostat, sealed with a millimeter-wave transparent foam window.  While located at the South Pole, the telescope mount is enclosed at room temperature within the observatory, protected by a fabric bellows structure.  To control the response of the beam sidelobes and minimize ground contamination, the telescope has an inner co-moving absorptive forebaffle and a fixed reflective outer ground screen, similar to that used in POLAR~\citep{Keating2003b}.  The readout electronics are sealed in an RF-tight cage and consist of detector AC biasing circuits, analog preamplifiers, lock-in amplifier cards, and cold JFETs in the cryostat.

The focal plane is comprised of 49 pairs of polarization sensitive bolometers (PSBs)~\citep{Jones2002}, two orthogonal detectors per feed, whose responses are summed or differenced to measure total intensity and polarization, respectively.  For the first observing season in 2006, the focal plane had twenty-five feeds tuned for the 100 GHz atmospheric transmission window and twenty-four tuned for the 150 GHz atmospheric window.  For the second and third observing seasons, the focal plane consisted of twenty-five 100 GHz feeds, twenty-two 150 GHz feeds, and two new feeds tuned for the 220 GHz atmospheric transmission window (discussed in Appendix~\ref{sec:s220}).  Three corrugated feedhorn sections, cooled to $\le 4$ K, couple radiation from the two high-density polyethylene lenses to each PSB while also providing a sharp low-frequency cutoff for the band.  The high-frequency cutoff is defined by a set of metal mesh filters~\citep{Ade2006} attached to the feedhorn stack.  The filters and PSBs are cooled to 250~mK by a $^4$He-$^3$He-$^3$He-sorption refrigerator system~\citep{Duband1990}.  Teflon filters block out-of-band infrared radiation to minimize optical loading.

\subsection{Scan Strategy}\label{sec:scanstrategy}

\begin{figure} %figure 1
\resizebox{\columnwidth}{!}{\includegraphics{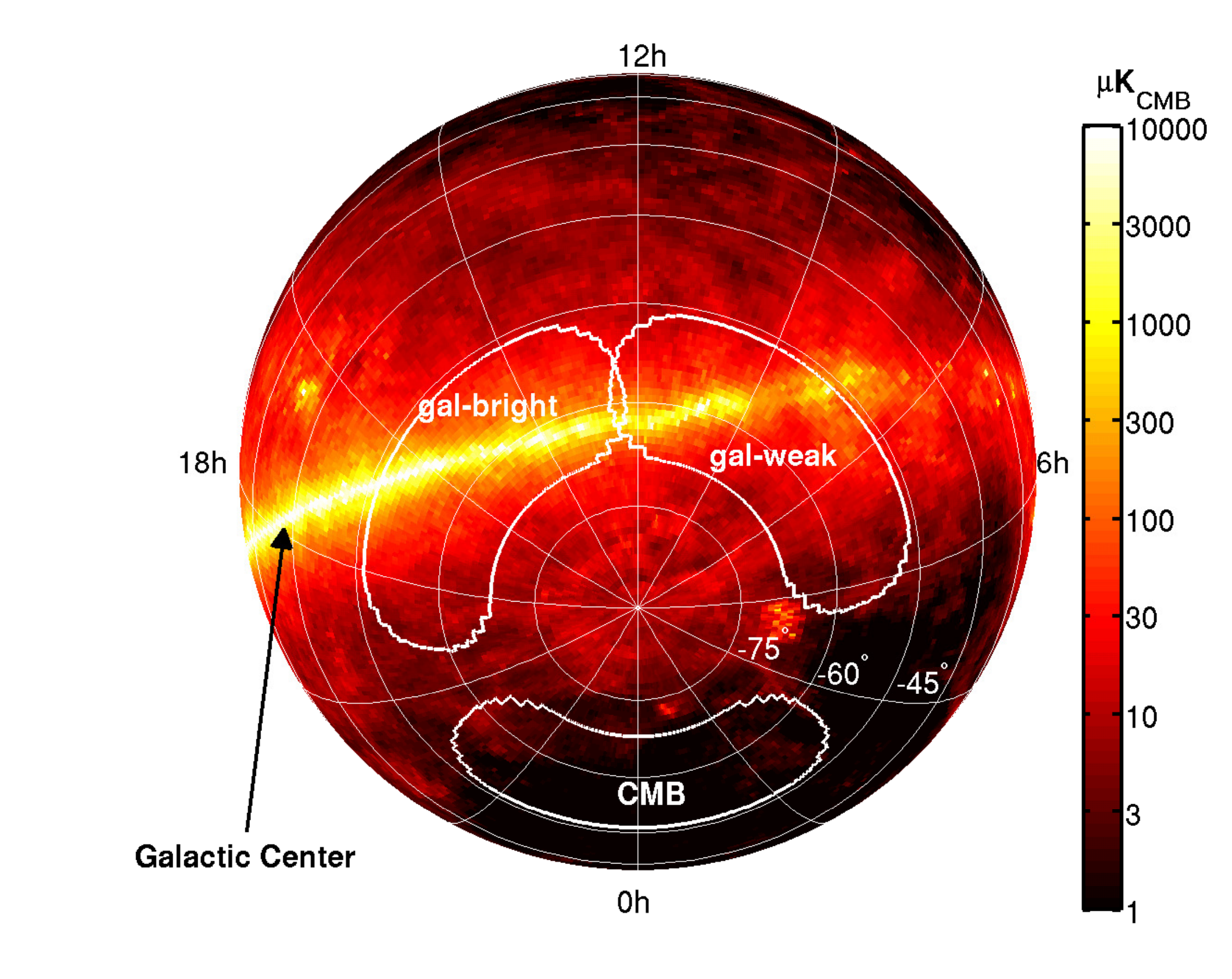}}
\caption{150 GHz FDS Model~8 dust emission prediction ~\citep{Fink1999}, shown in equatorial coordinates.  \bicep's primary CMB observing field is called ``the southern Galactic hole'', a region of low dust emission used for optimal B-mode detection.  The two Galactic fields are used for studying astronomical foregrounds and Galactic physics.  The Gal-weak region spans the Galactic plane from Galactic longitude $260\deg < \ell < 300\deg$, while the Gal-bright region spans the plane from $300\deg < \ell < 340\deg$.}
\label{fig:coverage_fds}
\end{figure}

Figure~\ref{fig:coverage_fds} shows the three main \bicep\ observing regions overlaid on the Galactic dust model of ~\cite{Fink1999} (hereafter FDS) evaluated at 150 GHz. Approximately 10,000 hours were spent observing a region predicted to have minimal astronomical foreground contamination in an attempt to detect the B-mode signature of inflationary gravitational waves (``CMB'' region), 945 hours were dedicated to observing the Galactic plane in a region near the center of the Galaxy (``Gal-bright'' region, $300\deg < \ell < 340\deg$) and 1484 hours were spent observing the Galactic plane in a region farther from the Galactic center (``Gal-weak'' region, $260\deg < \ell < 300\deg$).  

\bicep\ observes all regions in a similar manner.  The smallest observing unit is a `half-scan'; a uni-directional azimuthal telescope movement at constant elevation that lasts 27 seconds.  To avoid potential thermal disturbances, 3.5 seconds is cut at the beginning and end of each half-scan.  The choice of scan speed is bounded at low frequency by the atmosphere and detector stability, and bounded at high frequency by the bolometer time constant.  Within those constraints, the telescope scan speed is chosen to be 2.8$\deg/\rm{sec}$ to minimize microphonics and thermal drifts.  A `scan-set' lasts approximately one hour and consists of 50 half-scans each in the positive and negative azimuthal directions at a given elevation.  A calibration period at the beginning and end of each scan-set consists of a small elevation movement, called an `el-nod', which serves as the primary relative gain calibration within a feed and across the focal plane.  Atmospheric loading is proportional to the line-of-sight air-mass, which is well modeled as $\csc(\theta_{el})$ plus an offset, where $\theta_{el}$ is the elevation angle.  El-nods produce a similar detector response (approximately 100 mK peak to peak) across all detectors for an elevation change of one degree.  To normalize the response over time, the average response for all the detectors in each band is calculated and divided out.  As opposed to tracking the celestial observing center, \bicep\ centers each scan-set about a fixed azimuth angle causing sky sources to move relative to the scan center,  while stationary ground and scan thermal/optical contamination remains fixed.  This has the added benefit of grouping both the scan and ground contaminations into one contamination (``scan-fixed contamination'').~~\cite{Chiang2010} takes this process a step further and removes a template of the scan-fixed contamination from each scan-set; however, this additional step was not necessary in this paper because of the much larger polarization signal relative to the noise.

After each scan-set, the telescope is stepped in elevation by 0.25$\deg$ and moved in azimuth to locate the next scan-set about the center of the observing region.  Each set of scan-sets (called a ``phase'') consists of seven (lasting six hours) or ten (lasting nine hours) steps at one of four orientations about the boresight (0$\deg$, 180$\deg$, 135$\deg$, 315$\deg$) centered in the elevation range $55\deg$ to $60\deg$.  Observations of the Gal-weak field were carried out mostly in six hour phases in Austral winter.  Observations of the Gal-bright field consisted primarily of nine hour phases during Austral summer 2008, although there were a few other six hour phases executed at various times during the three years of observing.   

Timestream statistics (such as the variance, skew and kurtosis of a half-scan) and el-nod calibration values are used to determine nominal observing conditions.  These statistics allowed the data to be cut on various time scales such as per-phase, per-scan-set and per-scan bases.  Gal-bright observations use 763 out of 945 total possible hours and Gal-weak observations use 1463 out of 1484 total possible hours based on el-nod cuts, while scan statistics cut approximately 5\% of the remaining data.

\subsection{Spectroscopic Characterization}\label{sec:specchar}

This section highlights the spectral characteristics of the \bicep\ telescope, and the effects on the resulting maps.  For a brief discussion of other \bicep\ characterizations, including 220 GHz detector properties, see Appendix~\ref{sec:signalcharacterization}; for a more thorough discussion of 100 and 150 GHz detectors see \cite{Yuki2010}.  The key instrumental properties of \bicep\ are summarized in Table~\ref{tab:bicepsummary} as described in Appendix~\ref{sec:signalcharacterization} and \cite{Yuki2010}.

For all PSBs used in \bicep, at all three frequency bands, detected radiation from the sky produces a bolometer signal,
\begin{eqnarray} \label{eq:model}
d(t)&=&K_t * {\Big\{}n(t) + ~g(t) \int d\nu A_{e}(\nu) F(\nu) \int d\Omega\ P(\Omega) \nonumber \\
& & \ \ \ \ \ \ \ \ \ \ \ \ \ \ \ \ \ \Big( I + \gamma (Q~\cos(2 \psi) + U~\sin(2 \psi) \Big) {\Big\}}, 
\end{eqnarray}
where $I$, $Q$, and $U$ are the total intensity and two linear polarization Stokes parameters on the sky respectively. \bicep\ is incapable of measuring the fourth Stokes parameter V, which accounts for circular polarization.  However, for the CMB and Galaxy emission, $V$ is expected to be negligible compared with the two linear Stokes parameters.  $A_e$, the effective antenna area, is assumed to be proportional to $\lambda^2$.  To recover the underlying sky signal from the detector voltage timestreams $d(t)$, each of the following parameters is calibrated as described in~\cite{Yuki2010}: $\psi$, the detector polarization angle; $\epsilon$, the cross-polarization response; $\gamma=\frac{1 - \epsilon}{1 + \epsilon}$, the resulting polarization efficiency; $P(\Omega)$, the antenna response as a function of angular position $\Omega$; $F(\nu)$, the end-to-end detector spectral response including filters, feedhorns, lens, etc.; $g(t)$, the detector responsivity; $n(t)$, the noise; $K_t$, the time-domain bolometer transfer function and filtering due to the electronics.

\begin{table}
   \caption{Telescope Characteristics}
   {~~~~~Characteristics of \bicep\ from its three observing seasons.  While not an exhaustive list of all possible categories and errors, the listed parameters show the properties of the polarimeter relevant for this paper.  Section~\ref{sec:specchar} discusses the spectroscopic characterizations in more detail, while a brief discussion of the other \bicep\ characterizations are in Appendix~\ref{sec:signalcharacterization} and Section~\ref{sec:generalmapnoise} with further discussion in \cite{Yuki2010}.  Characteristics for 100 and 150 GHz are consistent with~\cite{Yuki2010} except for spectral characterizations.\\}

   %\centering 
   \begin{tabular*}{3.4in}{ p{1.9in} p{0.35in}  p{0.35in}  p{0.35in} }
     \hline
     \hline
     Instrument Property (Band Average)        & 100     & 150     & 220    \\
     \hline
     Number of Feeds (2006, 2007-2008)    & 25, 25      & 24, 22      & 0, 2       \\

     Polarization Orientation Uncertainty & $<0.7\deg$  & $<0.7\deg$  & $<0.7\deg$ \\
     Pair-Relative Polarization Orientation Uncertainty          & $0.1\deg$   & $0.1\deg$   & $0.1\deg$  \\

     Polarization Efficiency, $\gamma$    & 0.92        & 0.93        & 0.85       \\

     Optical efficiency (OE)              & 20.8\%      & 19.8\%      & 15.8\%     \\

     Gaussian Beam Width (FWHM)           & 0.93$\deg$  & 0.60$\deg$  & 0.42$\deg$ \\

     Differential Pointing / Beam Size    & 1.0 \%      & 1.8 \%      & 2.6 \%     \\

     Ghost Beam Power                     & 0.41\%      & 0.50\%      & 1.3\%      \\
     Ghost Beam Power, Pair-Difference    & 0.02\%      & 0.04\%      & 0.04\%     \\

     Spectral Band Centers, flat source (GHz)$^{a}$  & 95.5   & 149.8   & 208.2    \\
     Spectral Band Centers, Galaxy (GHz)            & 96.3    & 152.4   & 212.2    \\
     Spectral Gain Mismatch$^{b}$         & 0.17\%      & 0.19\%      & 0.72\%      \\

     Relative Gain Uncertainty$^{b}$      & $0.8\%$     & $1.3\%$     & $<10\%$    \\ 
     Absolute Gain Uncertainty           & $2\%$       & $2\%$       & $15\%$     \\

     NET (\ukrts)                                      & 530         & 450         & 1040       \\
     NEQ per feed$^{c}$ (\ukrts)          & 410         & 340         &  880       \\

     \hline
   \end{tabular*}   \label{tab:bicepsummary}
\footnotetext
{\scriptsize These values are slightly different for 100 and 150 GHz from~\cite{Yuki2010} due to measurement uncertainties.}

\footnotetext
{\scriptsize These values are computed before correcting for spectral gain mismatch from FTS measurements.}

\footnotetext
{\scriptsize These are the noise values used throughout this paper to calculate the white noise levels of the maps.}

 \end{table}

\cite{Yuki2010} explored the leakage of total intensity to polarization for each feed, estimated using individual PSB pair-sum and pair-difference maps from the CMB region.  The relative gain uncertainty is similar to spectral gain mismatch but can include other effects such as thermal response mismatch.  The maps were cross-correlated, showing the relative gain uncertainty was less than 0.8\% and 1.3\% for 100 and 150 GHz feeds respectively.  It was found that the relative gain mismatch for the 220 GHz pixels was 10\% and visual inspection of the maps showed the two 220 GHz detectors gave inconsistent polarization results.  The cause of this inconsistency was found to be mostly attributable to spectral gain mismatch.  

\begin{figure}  %figure 2
\resizebox{\columnwidth}{!}{\includegraphics{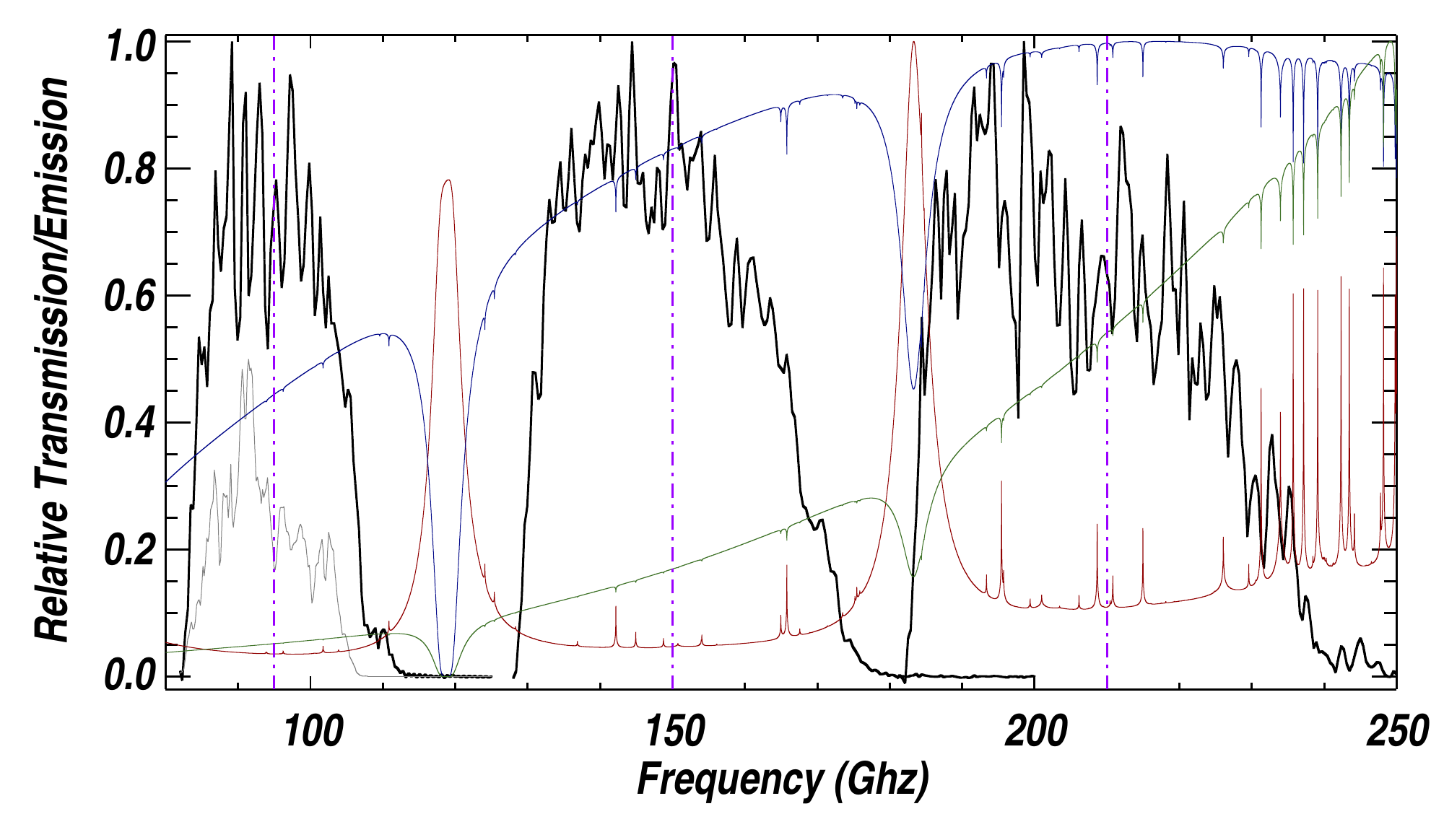}}
\caption{\bicep's three electromagnetic frequency spectra, $F(\nu)$, normalized to unity (black) with the band centers (shown as vertical purple dashed lines).  Also plotted is the average spectrum for \wmap's 94 GHz band normalized to 0.5 (gray) and the spectral radiance, $S(\nu)$, for a model of atmospheric emission at the South Pole (red; ~\cite{Grossman1989}), the spectrum (blue) of the CMB anisotropy (temperature derivative of the Planck function evaluated at 2.725 K~\citep{Fixsen2002}) and a typical Galactic emission spectrum (green) in the \bicep\ observing region, as seen through the atmosphere.}
\label{fig:bicep_spectra}
\end{figure}

A careful campaign to characterize all of \bicep's feeds was undertaken in January 2008 at the South Pole using a high resolution (up to 250 MHz resolution) polarized Fourier transform spectrometer (FTS).  The optical alignment and power falling on each pair of bolometers in a feed was calibrated before a set of eight independent spectral measurements were taken.  Figure~\ref{fig:bicep_spectra} shows $F(\nu)$, the average spectrum for each band with the FTS's source spectrum divided out, assuming the FTS's source filled the pixel's beam.

If not calibrated properly, mismatched spectra can cause spurious polarization in detector differences and gain errors among different feeds.  For example, the \wmap\ satellite ~\citep{Jarosik2007} notes that passband mismatch is a problem; solved by fitting out for a spurious map component.  The value of the mismatch is quoted as 1\% on average with a maximum of 3.5\% for the 23 GHz band~\citep{Page2007}, in agreement with pre-flight spectral measurements.  \bicep\ does not have sufficient polarization angle coverage for each feed to fit out the spurious component; however, it is possible to use the measured spectrum to mitigate this effect.

\bicep's spectral response mismatch leaks intensity into polarization and is caused by the combination of two effects.  First, several feeds have mismatched spectra, most notably the two 220 GHz feeds.  Second, each PSB is relatively calibrated using the change in atmospheric loading with elevation; however, atmospheric emission has a different spectrum than Galactic emission or the CMB.  The spectral mismatch leakage is calculated using the measured instrumental spectral response, a model of the atmospheric emission at the South Pole~\citep{Grossman1989}, and a model of the typical Galactic spectrum in the 220 GHz observing region (Section ~\ref{sec:ivsfreq}).

The spectral gain mismatch ($\xi$) is :
\begin{eqnarray}  \label{eq:spectralmismatch1}
  \xi=\frac{G_A - G_B}{G_A + G_B}
\end{eqnarray}
where the responsivities, $G_{A,B}$, represent the two PSBs in a feed and are given by
\begin{eqnarray} \label{eq:spectralmismatch2}
G_{A,B} = \frac{\Gamma_{atmosphere}}{\Gamma_{Galaxy}} \\
\Gamma_{source} = \int F(\nu) S(\nu) \lambda ^ 2 d\nu
\end{eqnarray}
where $S(\nu)$ is the spectral radiance $\Big( \frac{W}{sr~m^2~Hz} \Big)$ of the source emission spectrum of the atmosphere or typical Galactic source, and $\lambda ^2$ accounts for the throughput of the optics, which are assumed to be single-moded.  

\begin{figure}  %figure 3
\resizebox{\columnwidth}{!}{\includegraphics{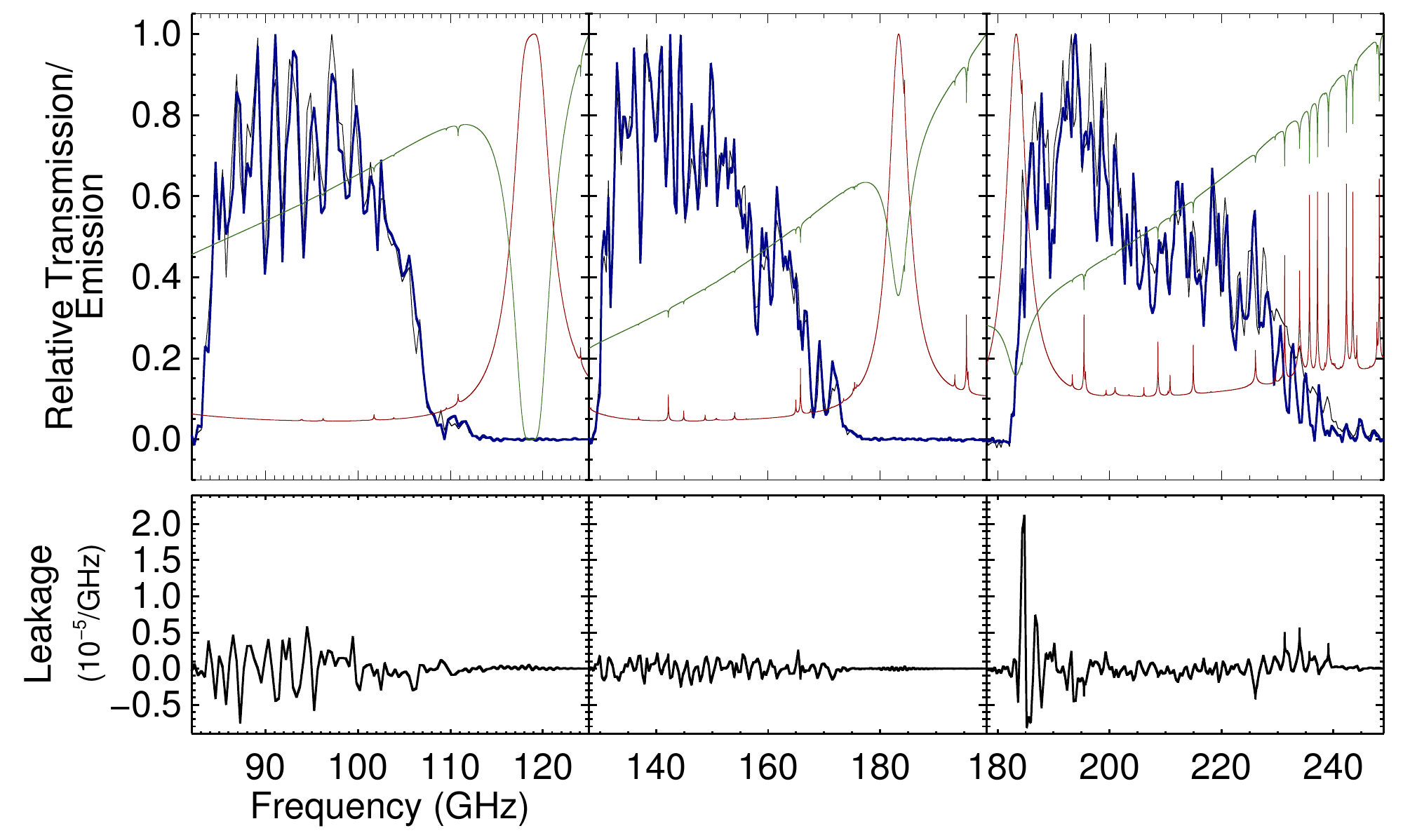}} %plot_bicep_spectra_3.pro
\caption{Spectral gain mismatch for one feed at each of \bicep's three bands.  The top plots show the measured spectrum for the two PSBs in a feed, ``A'' (Black) and ``B'' (Blue), along with a model of atmospheric transmission at the South Pole during median Austral winter conditions (red; ~\cite{Grossman1989}) and a typical Galactic source spectrum (green) in the 220 GHz analysis region, as seen through the atmosphere.  The bottom three plots show the leakage fraction per frequency, $d\xi$, if calibrated off the atmospheric emission but observing the Galactic source spectrum as shown.  The integral of the leakage fraction over the bounds shown gives the total spectral leakage measured for that feed, $\xi$.  The leakage at 220 GHz comes from the combination of a small difference in lower band edge and a large amount of power from the atmospheric water line at 185 GHz.}
\label{fig:bicep_spectral_mismatch}
\end{figure}

Shown in Figure~\ref{fig:bicep_spectral_mismatch} are examples of the spectral gain mismatch between two PSBs in a feed for each band and the spectral gain mismatch per frequency given by:
\begin{eqnarray} \label{eq:spectralmismatchperfreq}
\frac{d\xi}{d\nu}= \Big( \frac{ F_A(\nu)}{\Gamma_{A,Galaxy}} - \frac{ F_B(\nu)}{\Gamma_{B,Galaxy}} \Big) S(\nu) \lambda ^ 2 / (G_A + G_B)
\end{eqnarray}
The average magnitudes of the spectral gain mismatch are 0.17\%, 0.19\%, 0.72\% for 100, 150, and 220 GHz, respectively, using a typical Galactic source spectrum and median precipitable water vapor conditions at the South Pole during Austral winter.  Changing the observed source or atmospheric conditions can have an appreciable change in these numbers.  For example, one of the 150 GHz feeds has a 0.12\% leakage during typical Austral summer conditions but can change in value by 0.5\% depending on the atmospheric model.  In this paper, typical values are used to correct the maps for this effect and simulations are carried out to probe the effects of the uncertainty in this calculated parameter.

The absolute gain calibration for \bicep\ maps at each frequency, co-added over all detectors, is determined by comparing total intensity cross power spectra from the \bicep\ CMB region to the spectra found by \wmap\ in the same region (see~\cite{Chiang2010} and \cite{Yuki2010} for more details).  Spectral gain mismatch between feeds could potentially introduce systematic effects into the analysis via feed calibration differences.  Comparing the two 220 GHz-feed intensity maps, a difference of 30\% is found.  While this discrepancy was initially suspected to be due to spectral gain mismatch between different feeds, an investigation into the origin of this discrepancy did not find that this was the cause.  While calibration per feed is important, this effect does not leak intensity power to polarization, and the quantities studied in this paper are mostly relative quantities insensitive to this systematic.  Therefore a calibration error per feed will not significantly affect the results of this paper.  

While the three bands are called ``100 GHz'', ``150 GHz'', and ``220 GHz'', these are not the actual band centers.  The band center for a given source is calculated as:
\begin{eqnarray} \label{eq:bandcenter}
\nu_0=\frac{\int^{\nu_H}_{\nu_L} \! \nu \ F(\nu) S(\nu) {\lambda^2} \ d\nu}{\int^{\nu_H}_{\nu_L} \! F(\nu)  S(\nu){\lambda^2} \ d\nu}, 
\end{eqnarray}
where $F(\nu)$ is the measured average spectrum for each band, $S(\nu)$ is the source emission spectrum, and $\lambda^2$ accounts for the throughput of the receiver.  
For a flat spectral source, this gives 95.5, 149.8, and 208.2 GHz for 100, 150, and 220 GHz bands respectively. 
The dominant source of uncertainty is due to the optical setup and whether the FTS source is beam filling or not.  
This can change these values by 0.5, 0.8, and 1.0 GHz for 100, 150 and 220 GHz respectively. 
Changing the integration limits can also change these values by approximately 0.1 GHz for each band.
If the band center is determined using a CMB source as seen through the atmosphere this gives band centers 96.2, 150.6, and 208.8 for 100, 150 and 220 GHz.
If the band center is determined using a typical Galactic source with dust and synchrotron emission, as seen through the atmosphere, this gives band centers 96.3, 152.4, and 212.2 for 100, 150 and 220 GHz.  Another way to calculate the average band center is to compute the band center for each detector separately and then average over all the detectors in a band.  Doing this for the Galactic spectrum gives average band centers of 96.3, 152.3, 212.0 GHz with uncertainties in the average of 0.1, 0.3, and 1.0 GHz for 100, 150 and 220 GHz respectively.  The standard deviation of the distributions over the detectors are 0.4, 1.4, and 2.1 GHz for 100, 150 and 220 GHz respectively, showing how different the spectra of a given detector in a band can be.
Out-of-band high frequency response is less than -25 dB, which was characterized by using high-pass thick-grille filters and a chopped source. 

\subsection{Time Domain Data Processing}\label{sec:filtering}
   Following the el-nod calibration and correction for spectral gain mismatch, the sum and the difference of a PSB pair are calculated.  Differencing the calibrated orthogonally linearly polarized detectors removes most of the unpolarized atmospheric response and unpolarized, scan-fixed, contamination.  Atmospheric $1/f$ noise dominates the $rms$ of the pair-sum timestreams for a typical half-scan.  These fluctuations can be less than 1 mK in good weather or reach approximately $300$ mK in bad weather, with a $1/f$ knee as high as a few Hertz.  The data also contain a sub-dominant scan-fixed contamination, which does not integrate down as uncorrelated noise.  To remove both $1/f$ atmospheric and scan-fixed contamination, the data from each half-scan are high-pass filtered by removing a second-order polynomial fit to each half-scan, at the cost of removing some Galactic signal as well.  The pair-sum and pair-difference timestreams are treated separately but with the same filtering.

Polynomial removal causes an obvious distortion of the signal when the scans include the Galactic plane.  To reduce this effect, the Galaxy is masked during the determination of the polynomial fit (``polynomial mask'').  The preferred filtering scheme uses a second-order polynomial while masking out samples $\left|b\right|<4\deg$.  This scheme achieves a compromise between noise filtration and signal preservation.  This process is not perfect and some residual filtering effects remain in the maps, as discussed in Section~\ref{sec:maperrors}.  A dedicated study of different filtering techniques was undertaken but none improved the maps significantly without causing worse filtering effects or adding additional noise. For example, maps with only a DC offset removed have less filtering applied to the data but also have large scale non-physical features that makes quantitative analysis unreliable.

\begin{figure} % Figure 4 galpaperplots_filterplot
\resizebox{\columnwidth}{!}{\includegraphics{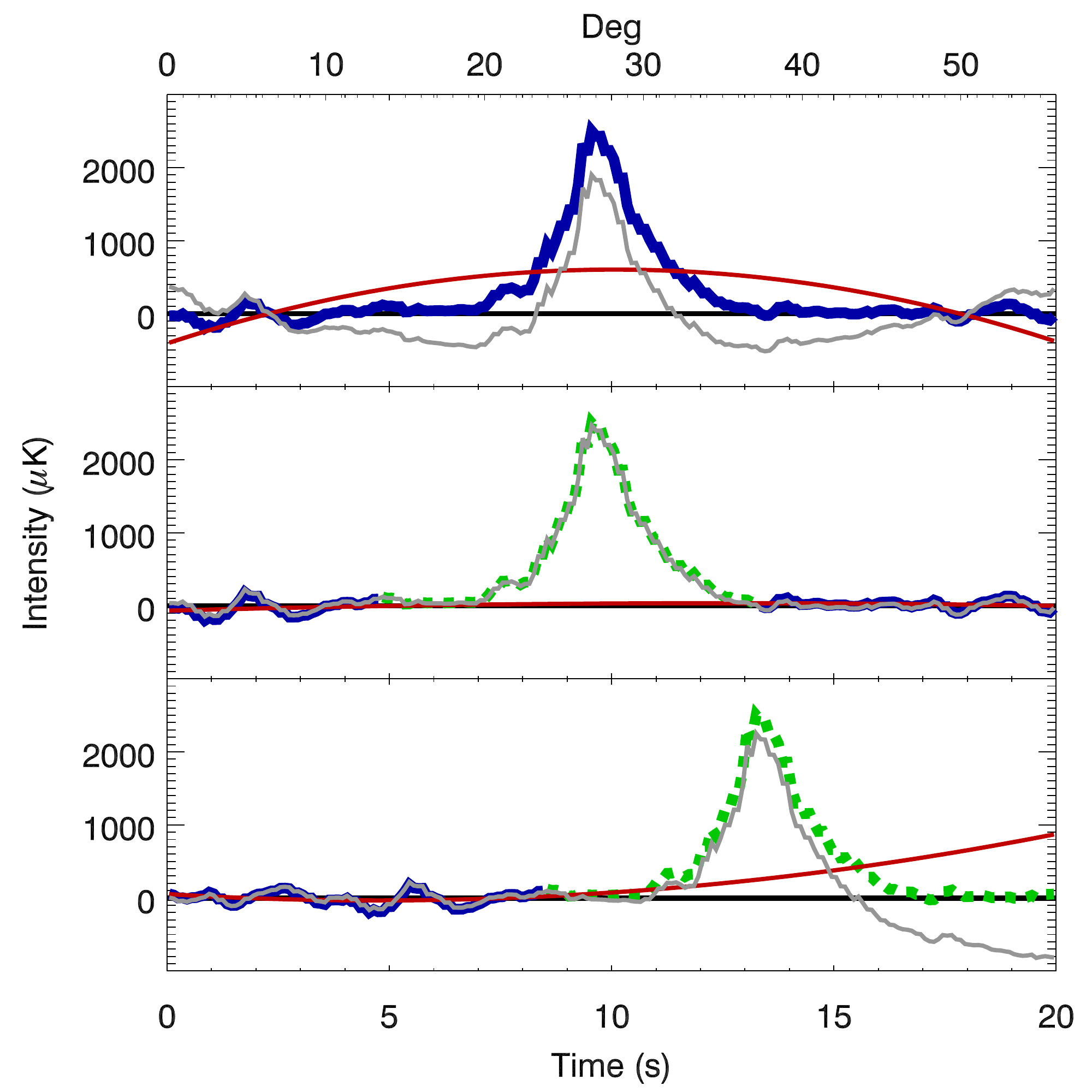}}
\caption{A detector timestream simulation showing three different polynomial filtering methods.  The top plot shows a typical timestream resulting from a scan across the Galaxy (blue) with a fitted second-order polynomial (red) subtracted off, causing a significant distortion of the Galaxy (gray).  The middle plot shows the same scan, except the Galaxy (green) has been excluded from the fit.  The polynomial has been interpolated across the Galaxy leading to minimal filtering effects.  The bottom plot shows a typical scan that ends on the Galaxy, requiring the polynomial to be extrapolated onto the Galaxy.  The extrapolated polynomial causes severe distortion of the maps, requiring these scan portions to be excluded from the analysis.}
\label{fig:gal_mask}
\end{figure}

\begin{figure}  %figure 5
  \begin{center}
    \resizebox{\columnwidth}{!}{\includegraphics{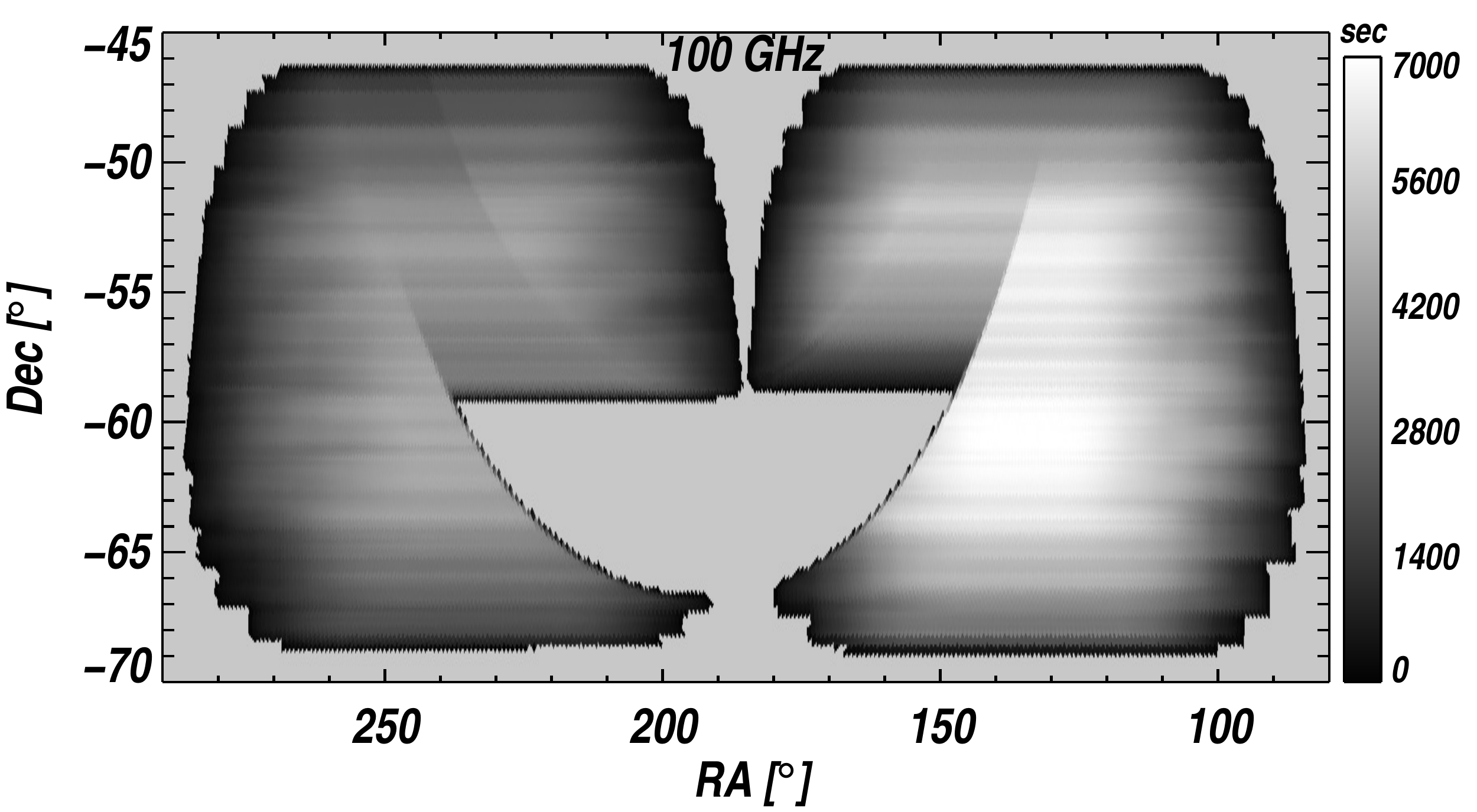}}
    \resizebox{\columnwidth}{!}{\includegraphics{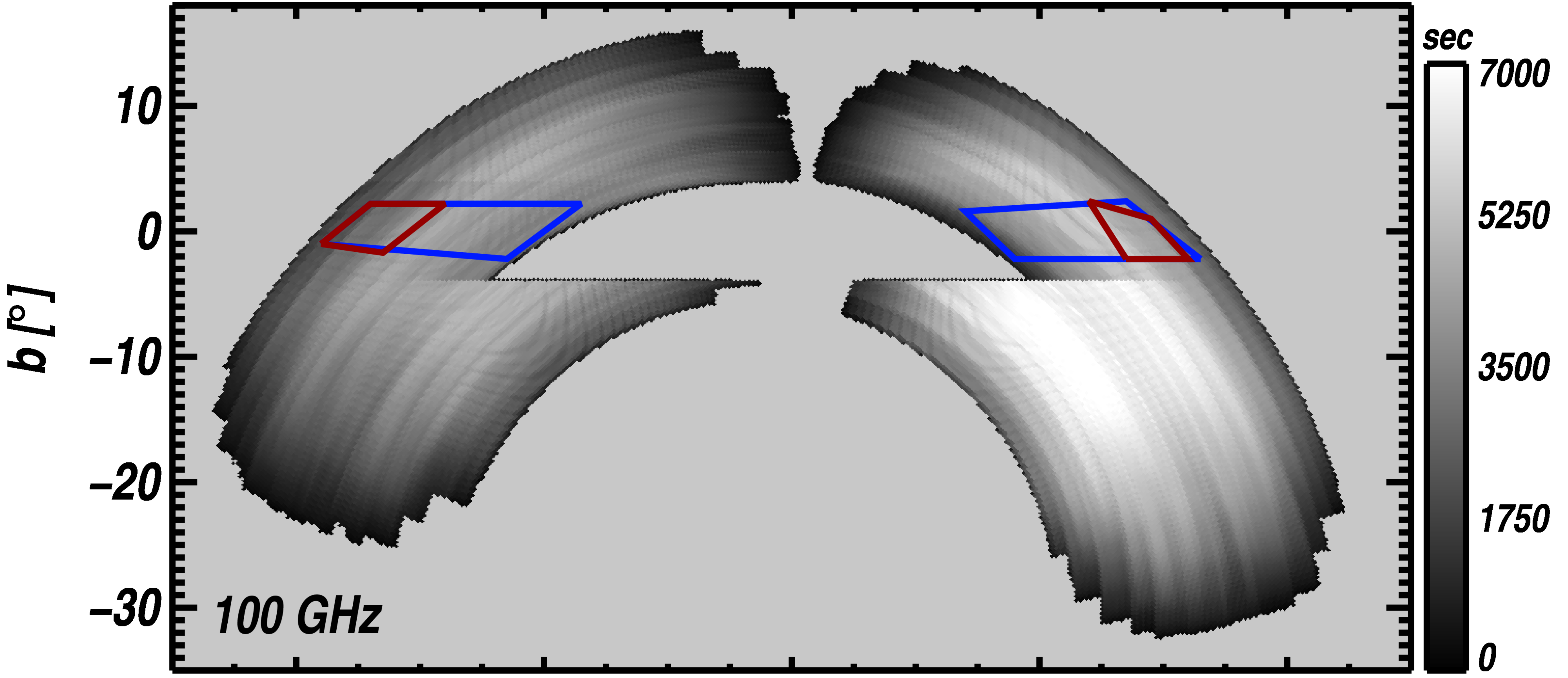}}
    \resizebox{\columnwidth}{!}{\includegraphics{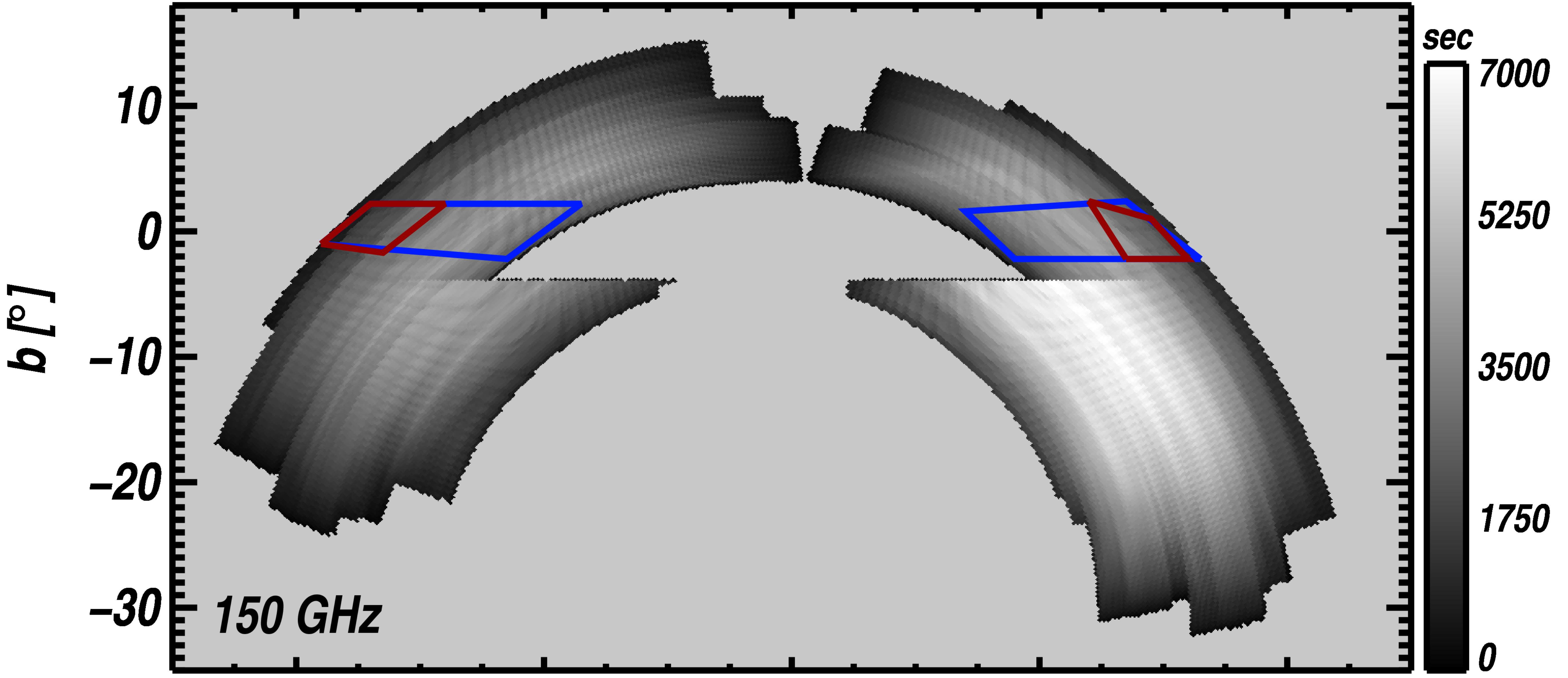}} 
    \resizebox{\columnwidth}{!}{\includegraphics{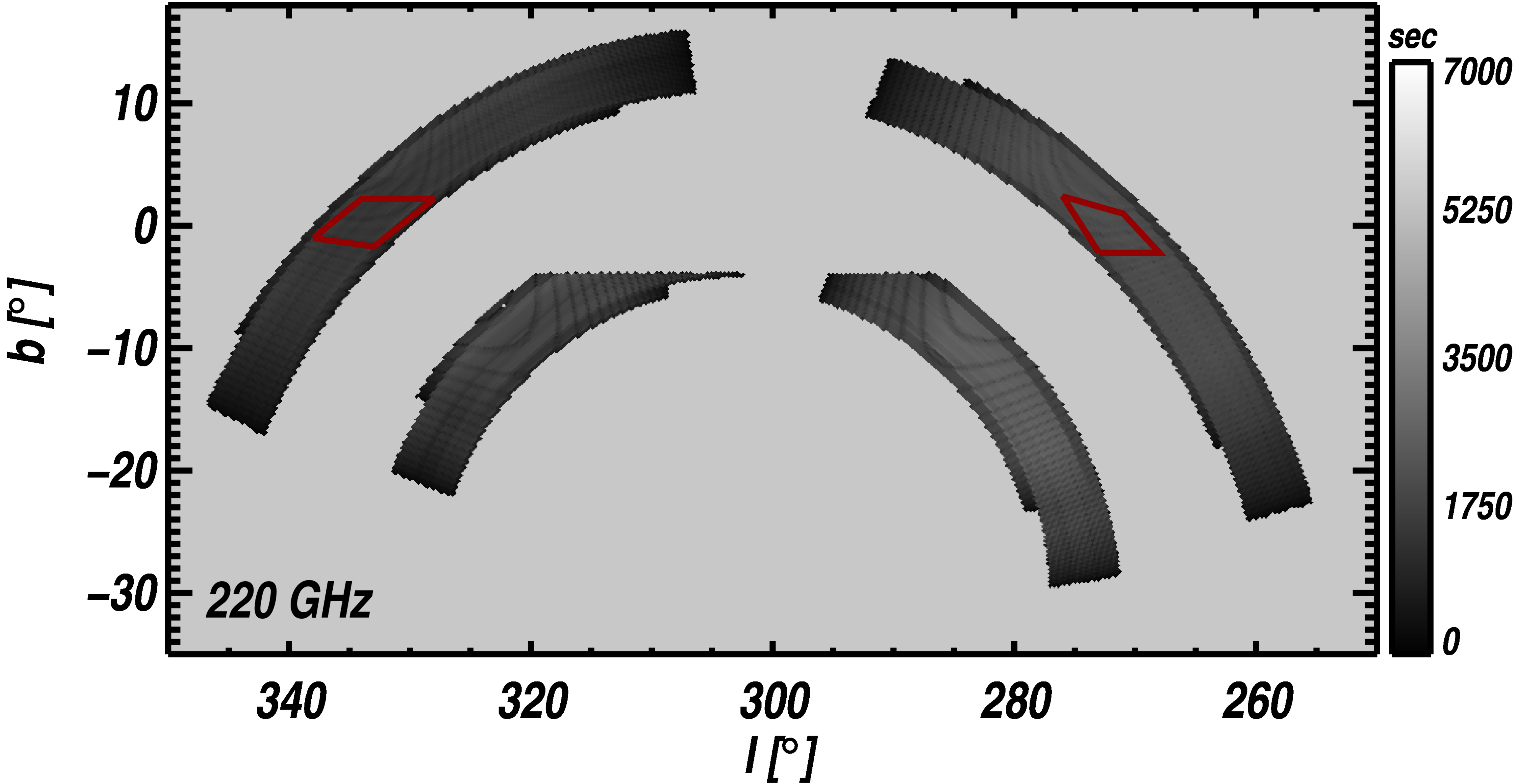}}

  \end{center} 
  \caption{Integration time per 0.25$\deg$ \healpix\ map pixel for 100 GHz (top: Celestial coordinates, second from top: Galactic coordinates), 150 GHz (second from bottom; Galactic Coordinates), and 220 GHz (bottom; Galactic Coordinates) derived from all three seasons co-added over all four boresight angles of \bicep\ observations.  The wedge-shaped cuts in the map without integration time are from the omission of parts of scans with extrapolated polynomial fits (Section~\ref{sec:filtering}). The 100 GHz celestial coordinate map is an example of the native observation reference frame from the South Pole.  The red outlined area (called the ``220 GHz analysis region'') is an area of the sky where analysis at all three bands can be carried out.  This region is a subset of the blue outlined area on the 100 and 150 GHz maps (called the ``100/150 GHz analysis region'') where the final analysis can only be carried out at 100 and 150 GHz (See Section~\ref{sec:results}).}
\label{fig:Integration_time_B}
\end{figure}

There is an added complication in this scheme for scans that end within the masked region (Figure~\ref{fig:gal_mask}).  A polynomial constrained by measurements on both sides of the Galactic plane closely approximates the low-frequency drifts within the interpolated region.  However, scans not constrained on both sides of the plane require the polynomial to be extrapolated beyond the fitted region.  Extrapolated polynomials tend to diverge because there are no data constraining the fit.  Therefore, the filtering scheme is modified so that scan portions that end within the masked region are excluded, giving rise to maps with a missing wedge (Figure~\ref{fig:Integration_time_B}).  Additionally, a brief measurement on both sides of the plane was not enough to constrain the polynomials sufficiently.  Therefore, scans were required to have at least 10 samples on either side of the Galactic plane to be used in the mapmaking process.

 \subsection{Mapmaking}\label{sec:mapmaking}
Once the data-processing steps are completed, the Stokes parameters $I$, $Q$, and $U$ are derived using standard techniques~\citep{Jones2007}.  Equation~\ref{eq:model} can be simplified to
 \begin{eqnarray} \label{eq:signal_model_g}
 d_{A,B}^{'}&=&g_{A,B} \Big ( I~+~\gamma_{A,B}~ \Big( Q~\cos(2~\psi_{A,B})~+~U~\sin(2~\psi_{A,B}) \Big) \Big),
 \end{eqnarray}
assuming the beam functions are the same for a given pair of PSBs and the responsivities,  $g_{A,B}$, still include the spectral gain mismatch after el-nod calibration.  The Stokes parameters $I$, $Q$, and $U$ now represent quantities integrated over $\Omega$ and $\nu$, and $\{A,B\}$ refers to the two orthogonal bolometers within a given feed.

Before calculating the sum and difference data of the two PSBs in a feed, the spectral gain mismatch factors calculated in Equation~\ref{eq:spectralmismatch1} are corrected:
 \begin{eqnarray} \label{eq:signal_model}
 d_A~=d_A^{'} \times (1 + \xi) \\
 d_b~=d_b^{'} \times (1 - \xi), \nonumber
 \end{eqnarray}
where $\xi$ is the spectral gain mismatch and the $d_{A,B}^{'}$ denotes the uncorrected timestream data.  The pair-sum ($d_{+}$) and pair-difference ($d_{-}$) timestreams can then be determined:
 \begin{eqnarray}  \label{eq:signal_dif_sum}
 d_+=\frac{d_A \ + \ d_B}{2} &=& \ I + \alpha_{+} Q + \beta_{+} U \approx  I \\
 d_-= \frac{d_A \ - \ d_B}{2} \ &=& \ \alpha_{-} Q + \beta_{-} U \approx Q~\cos(2 \ \psi_A)~+~U~\sin(2 \ \psi_A), \nonumber 
 \end{eqnarray}
which give rise to the Stokes parameters $I$, $Q$, and $U$ where $\alpha_{\pm}$ and $\beta_{\pm}$ account for polarization angle:
 \begin{eqnarray} \label{eq:alpha_beta}
 \alpha_{\pm} \ &\equiv& \ \frac{\gamma_A \ \cos(2 \ \psi_A) \ \pm \gamma_B \ \cos(2 \ \psi_B)}{2} \\
 \beta_{\pm}  \ &\equiv& \ \frac{\gamma_A \ \sin(2 \ \psi_A) \ \pm \gamma_B \ \sin(2 \ \psi_B)}{2}. \nonumber
 \end{eqnarray}

The A and B bolometers are assumed to be nearly perpendicular, so the pair-sum gives the total intensity to a very high precision.  If $\gamma$ is also assumed to be equal to one, then $\alpha_{\pm}$ and $\beta{\pm}$ simplify to \{0,~$\cos(2 \psi_A)$\} and \{0,~$\sin(2 \psi_A)$\}.  However, $\gamma$ is closer to 90\%, and can differ by several percent between a given pair of bolometers within a feed, so no simplification is made during mapmaking using the pair-difference signal.  After the spectral gain mismatch is corrected and after the sum and difference are taken, the resulting pair-sum and pair-difference timestreams are polynomial filtered as discussed in Section~\ref{sec:filtering}.

 The $I$, $Q$, and $U$ maps are given by:
 \begin{eqnarray} \label{eq:linear_fitting}
   m_{(I,Q/U)} = (A_T N^{-1} A)^{-1} A_T N^{-1} d_{(+,-)},
 \end{eqnarray}
which is a noise weighted linear least squares regression where $m$ is the set of map pixels for a given Stokes parameter, $N$ is the noise covariance matrix, and $A$ is the pointing matrix.  Since $d_{(+,-)}$ are filtered timestreams, the resulting sky maps $m_{(I,Q/U)}$ are also filtered.  For this work, it is sufficient to assume that the noise is uncorrelated, making $N$ diagonal and simple to invert.  The variance associated with samples in a single half-scan is assumed to be time-independent, and is calculated from the samples lying outside the Galactic mask after polynomial subtraction.  For the pair-sum data, the pointing matrix consists of ones and zeros, indicating whether or not the telescope is pointing at a particular map pixel.  This simplifies Equation~\ref{eq:linear_fitting}, only requiring the calculation of the weighted mean of $d_{+}$ to determine the total intensity map $m_I$.  To determine the polarization maps $m_{Q/U}$ from the pair-difference data $d_-$, the pointing matrix consists of a combination of the $\alpha$'s and $\beta$'s, and Equation~\ref{eq:linear_fitting} can be written as:
 \begin{eqnarray} \label{eq:qu_diff}
   \sum_{i=1}^{S} w_{i}
   \left(\begin{array}{c}
     d_{-~i}~\alpha_{i} \\
     d_{-~i}~\beta_{i} \\
   \end{array}\right) 
   = \sum_{i=1}^{S}  w_{i}
   \left(\begin{array}{cc}
     \alpha^2_{i} & \alpha_{i}\beta_{i} \\
     \alpha_{i}\beta_{i} & \beta^2_{i}\\
   \end{array}\right)
   \left(\begin{array}{c}
     Q_{j,k} \\
     U_{j,k} \\
   \end{array}\right).
 \end{eqnarray}
 In Equation~\ref{eq:qu_diff}, $d_{-~i}$ is a single pair-difference timestream sample, $w_i$ is the inverse variance for a single sample, index $j$ corresponds to a given map pixel, $k$ corresponds to a given band (100, 150, or 220 GHz), and $S$ is the total number of samples from all the feeds per band in a given map pixel.  After summing over $i$, the matrices per pixel per band are inverted to solve for $Q$ and $U$.  For each map pixel, the seven quantities used to recover $I$, $Q$, and $U$ are the weighted pair-sum data, weighted pair-sum hits, weighted pair-difference data multiplied by $\alpha$ and $\beta$, and weighted $\alpha^2$, $\beta^2$, and $\alpha \beta$.  An eighth quantity, integration time per pixel, is also recorded in order to measure noise properties and observing efficiency.

%%%%%%%%%%%%%%%%%%%%%%%%%%%%%%%%%%%%%%%%%%%%%%%%%%%%%%%%%%%%%%%%%%%%%%%%%%%%%%%%%%%%%%%%%%%%%%%%%%%%%%%%%%%%%%%%%%%%%%%%%%%%%%%%%%%%%%%%%%%%%%%%%%%%%%%%%%%%%%%%%%%%%%%%%%%%%%%%%%%%%%%%%%%%%%%%%%%%%%%%%%%%%%%%%%%%%%%%%%%%%%%%%%%%%%%%%%%%%%%%%%%%

%%%%%%%%%%%%%%%%%%%%%%%%%%%%%%%%%%%%%%%%%%%%%%%%%%%%%%%%%%%%%%%%%%%%%%%%%%%%%%%%%%%%%%%%%%%%%%%%%%%%%%%%%%%%%%%%%%%%%%%%%%%%%%%%%%%%%%%%%%%%%%%%%%%%%%%%%%%%%%%%%%%%%%%%%%%%%%%%%%%%%%%%%%%%%%%%%%%%%%%%%%%%%%%%%%%%%%%%%%%%%%%%%%%%%%%%%%%%%%%%%%%%%%%%%%%%%%%%%%%%%%%%%%%%%%%%%%%%%%%%%%%%%%%%%%%%%%%%%%%%%%%%%

\section{Results}\label{sec:results}

This section begins with a qualitative discussion of \bicep's Galaxy maps, followed by a section describing uncertainties in the maps, a direct comparison to the maps observed by the \wmap\ satellite, and then finally a quantitative analysis of the properties in the maps.  For quantitative calculations, two analysis regions were defined (Figure~\ref{fig:Integration_time_B}).  The ``100/150 GHz analysis region'' consists of 147 one degree \healpix\ map pixels that have intensity values greater than zero, located at less than two degrees in Galactic latitude, and a polarization fraction magnitude less than 20\%.  Since there were only two 220 GHz feeds installed in \bicep, this limited the sky coverage for that band.  Therefore, a ``220 GHz analysis region'', using 53 of the 147 pixels from the 100/150 GHz analysis region, defines a subset of map pixels that can be analyzed at all three bands.

\subsection{Intensity and Polarization Maps}\label{sec:maps}
\begin{figure}  %figure 6
  \begin{center}
    \resizebox{\columnwidth}{!}{\includegraphics{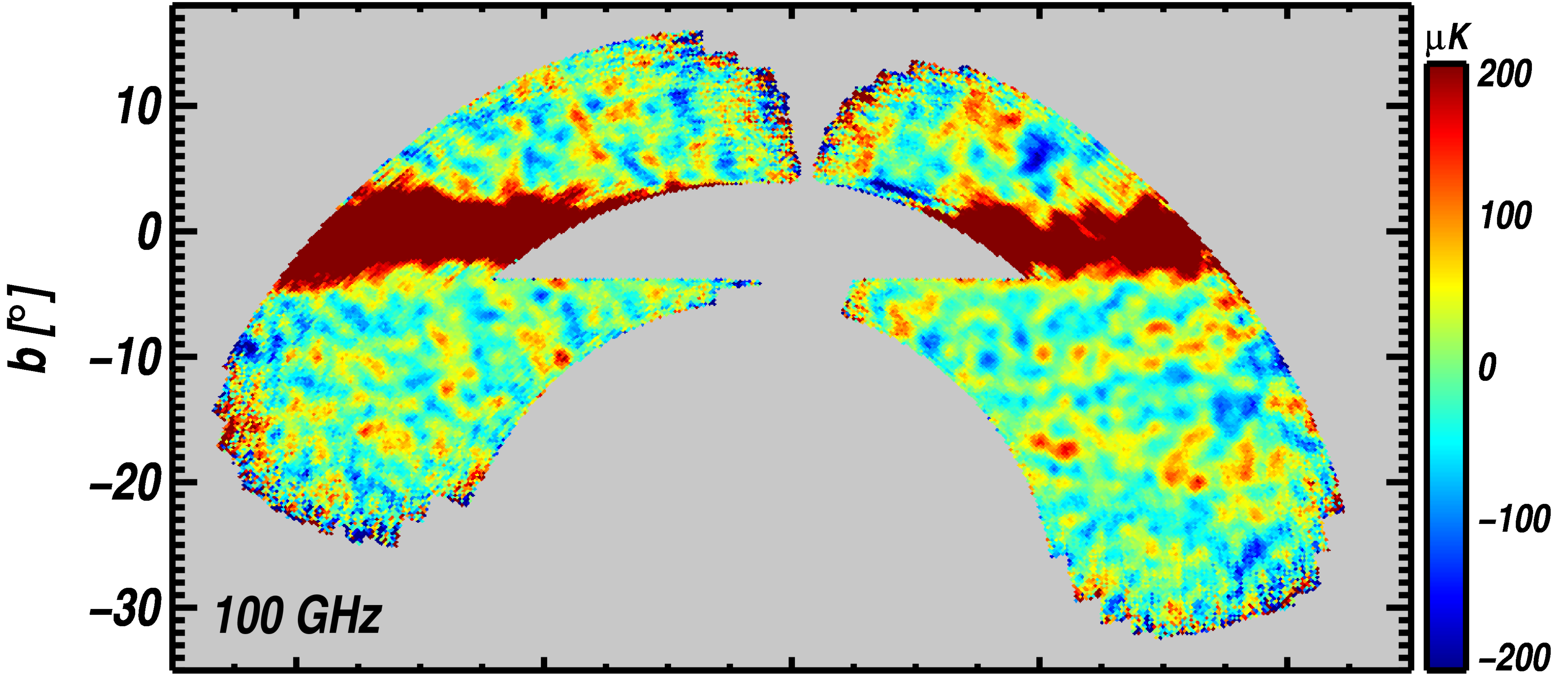}}
    \resizebox{\columnwidth}{!}{\includegraphics{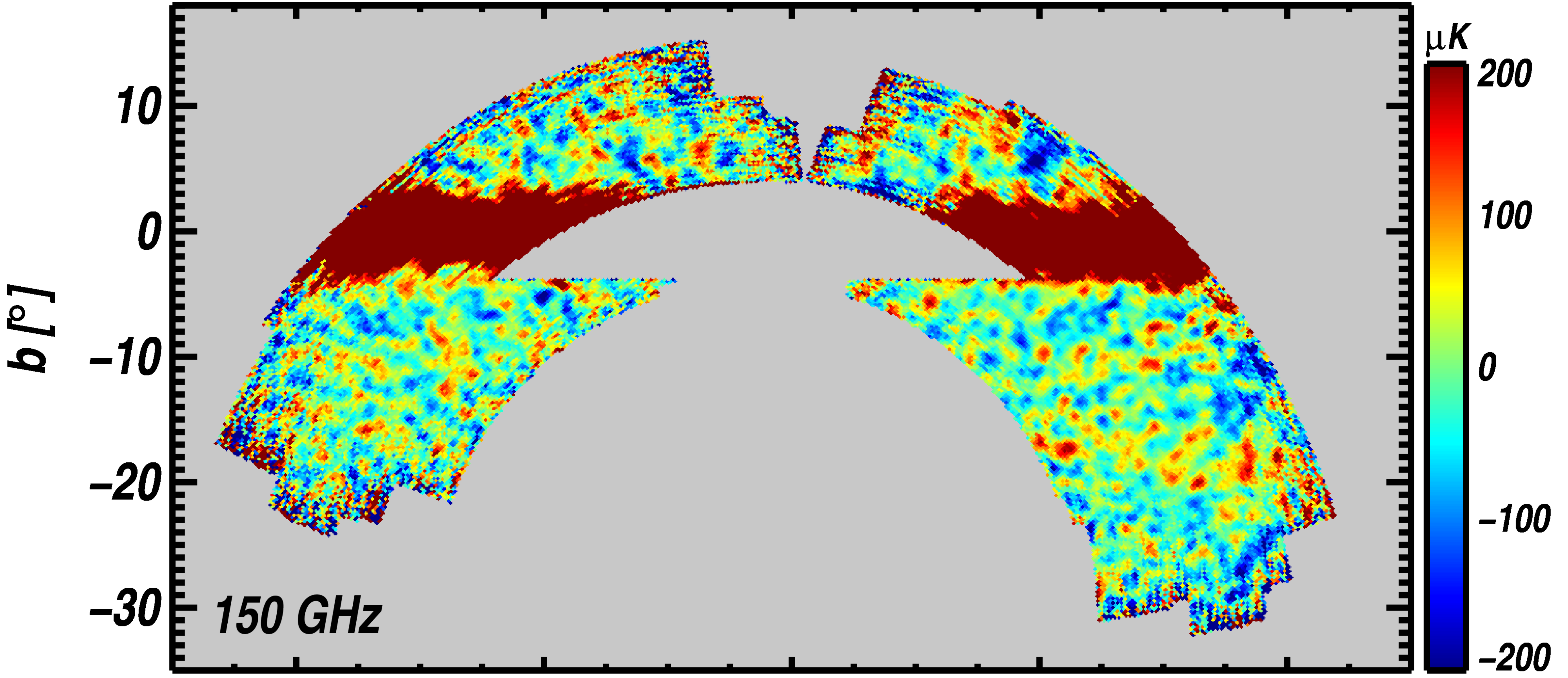}}
    \resizebox{\columnwidth}{!}{\includegraphics{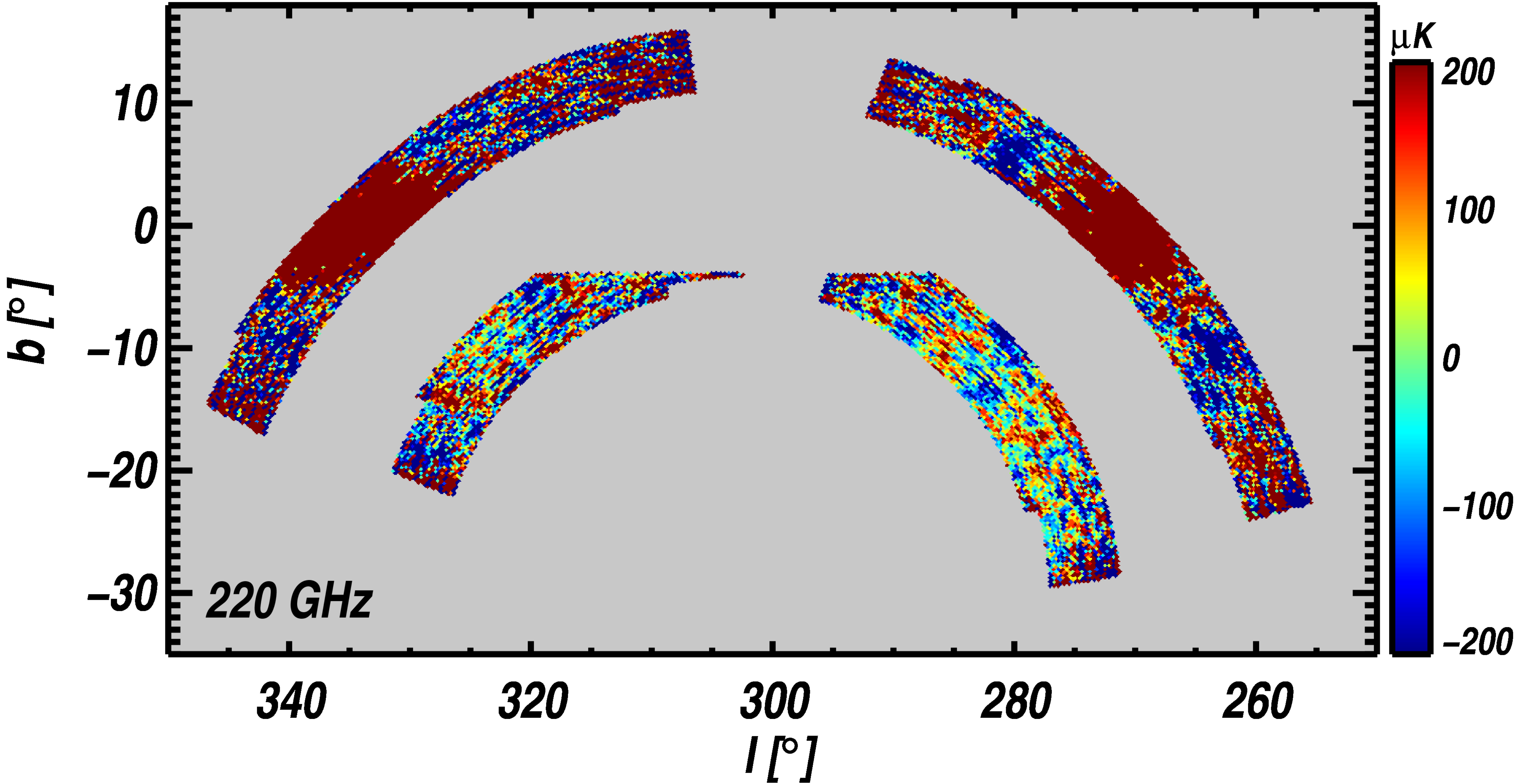}}
  \end{center}
  \caption{\bicep\ intensity maps from all three seasons co-added over all four boresight angles at 0.25$\deg$ \healpix\ resolution, in Galactic coordinates, at 100, 150, and 220 GHz from top to bottom respectively.  The color scale has been chosen to emphasize the CMB anisotropy, which is visible in all three bands.}
\label{fig:Traw}
\end{figure}

\begin{figure*}  %figure 7
  \begin{center} 
    \resizebox{\textwidth}{!} {
    $ \begin{array}{cc}
    \includegraphics{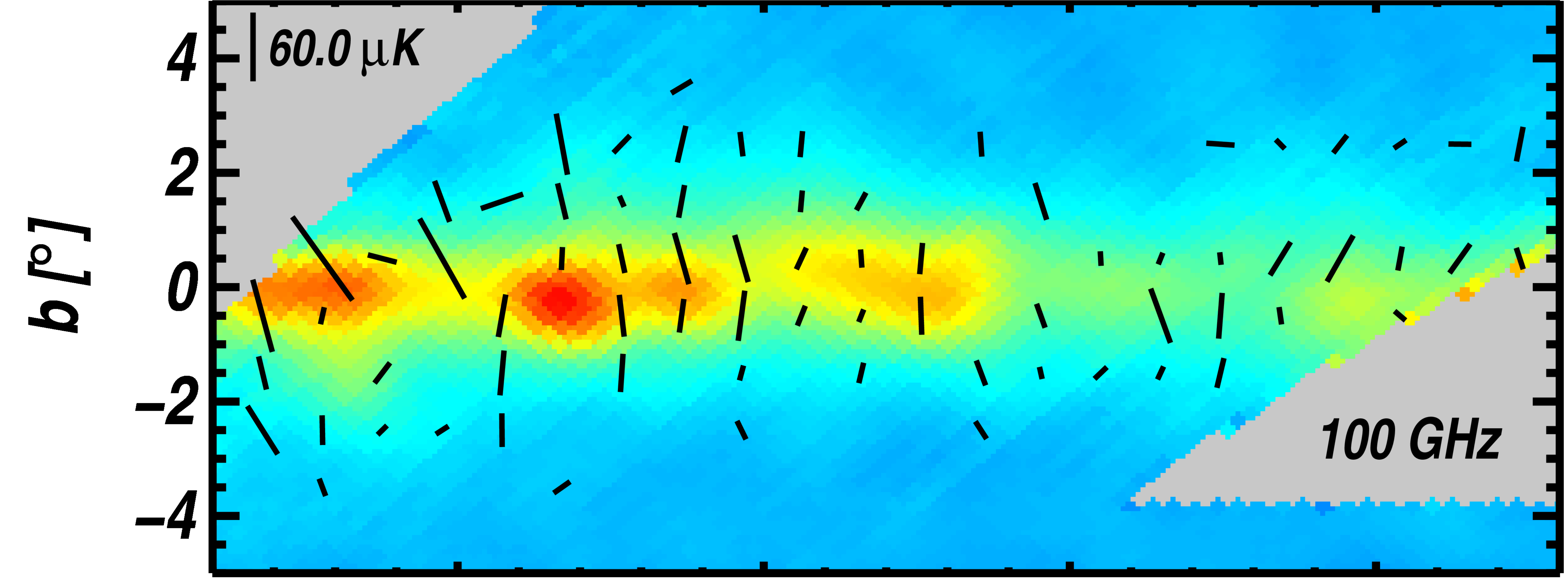} & \includegraphics{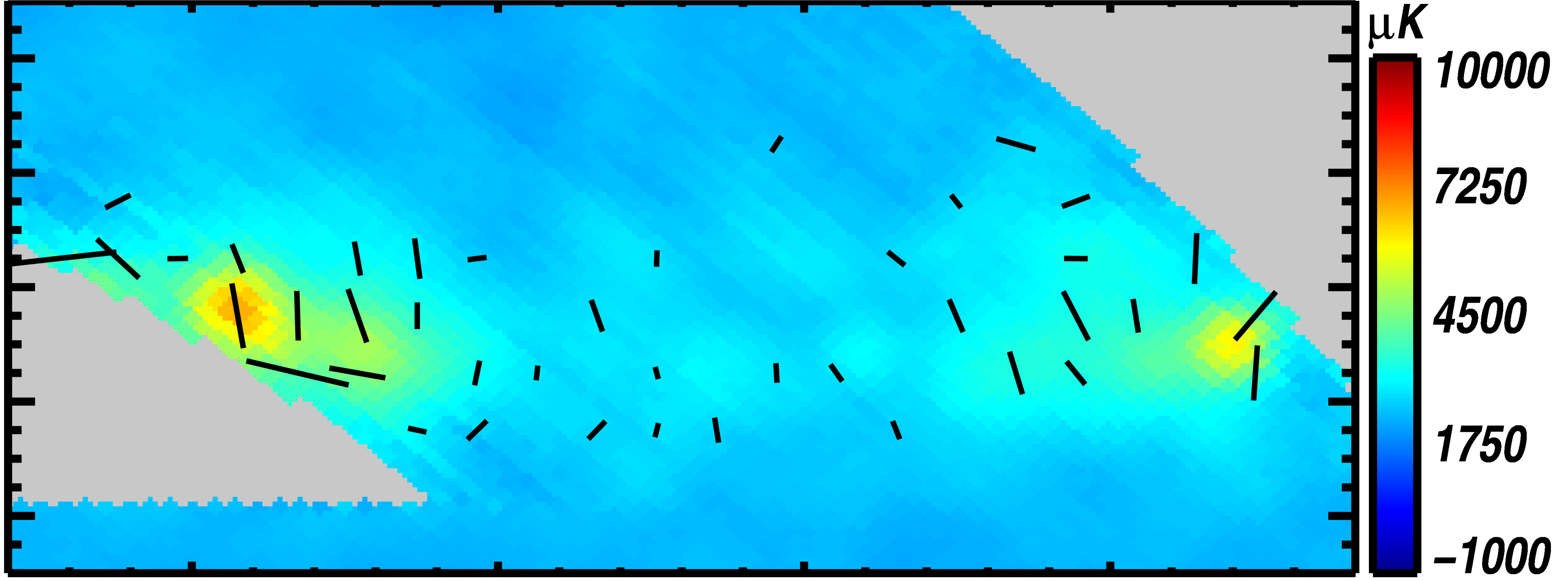} \\
    \includegraphics{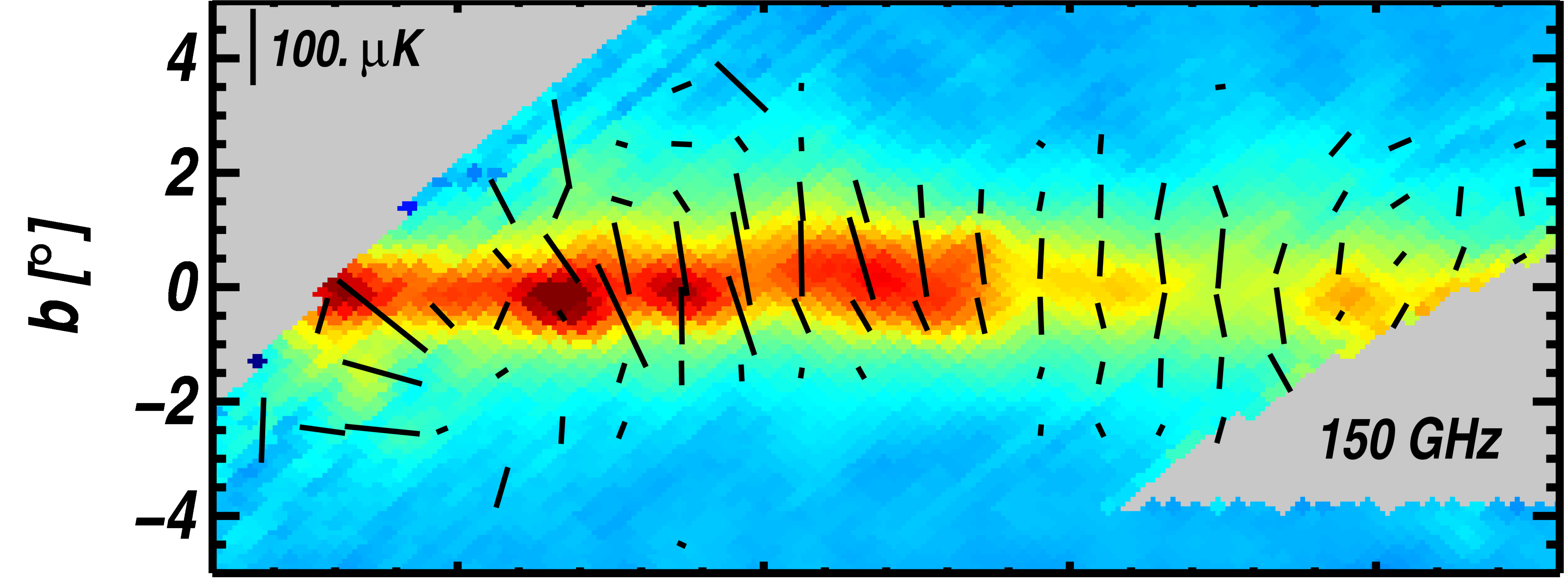} & \includegraphics{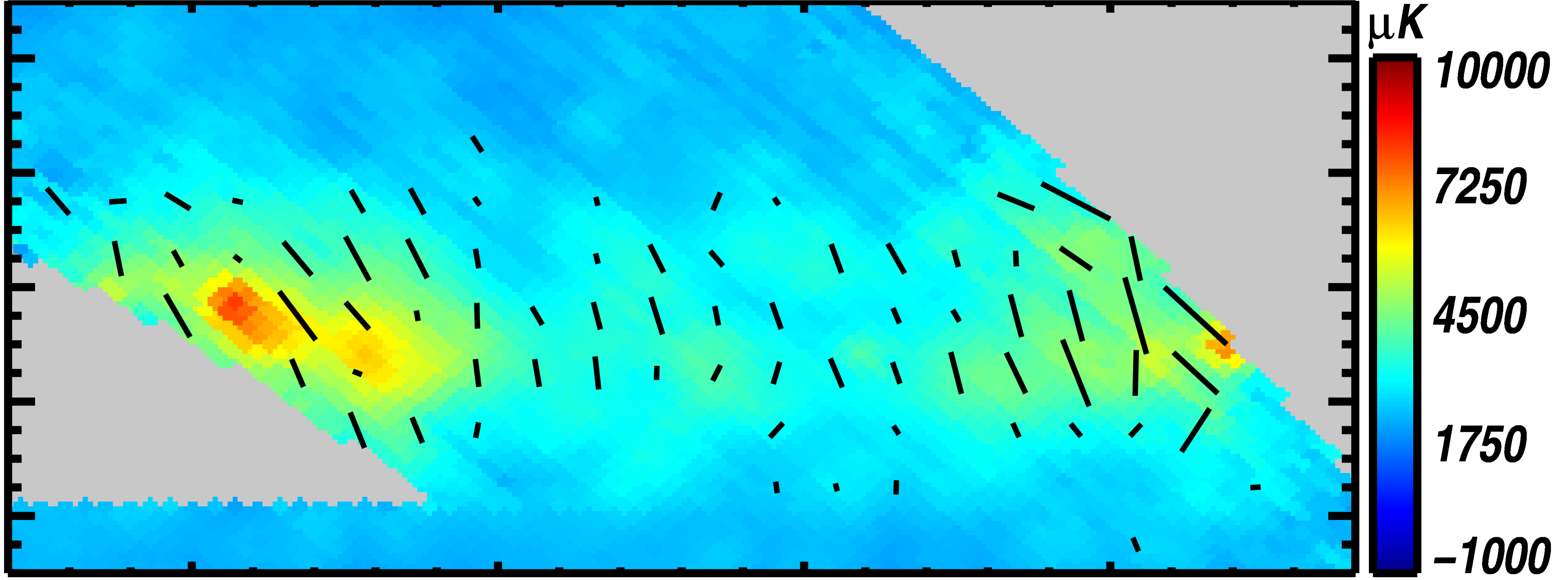} \\
    \includegraphics{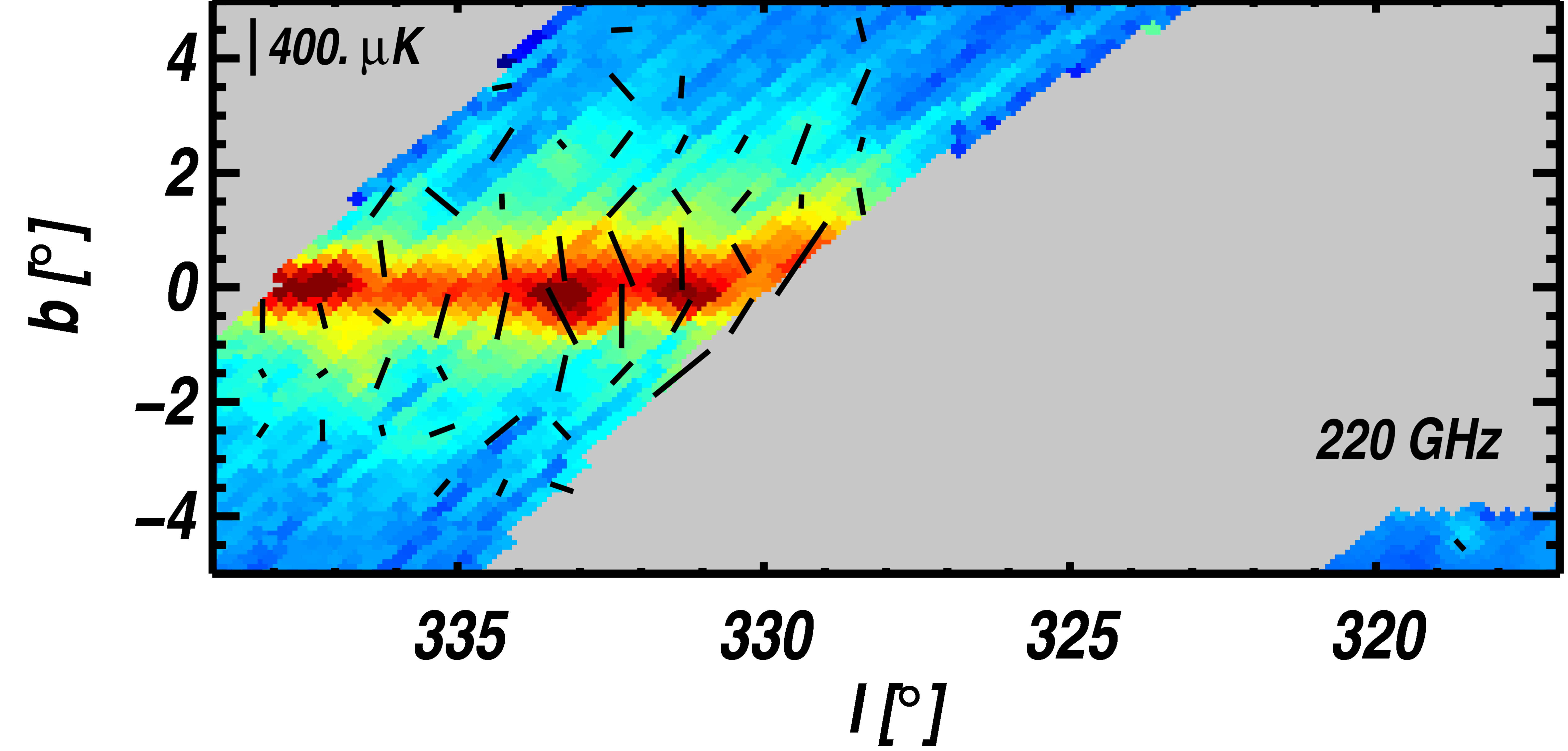} & \includegraphics{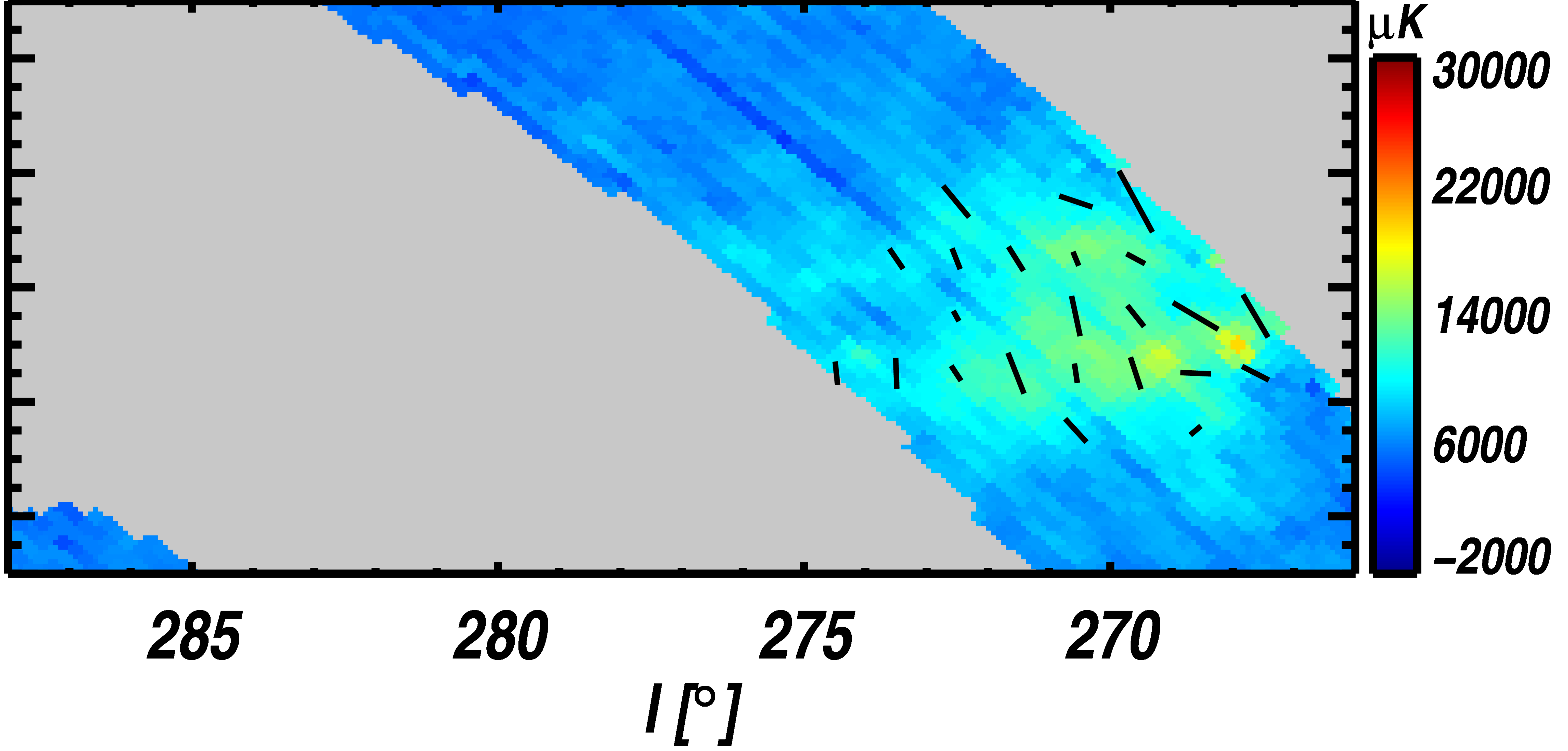} \\
    \end{array} $
      }
  \end{center}
  \caption{\bicep\ intensity maps from all three seasons co-added over all four boresight angles with polarization vectors, in Galactic coordinates, at 100, 150, and 220 GHz from top to bottom respectively. The color scale is chosen to emphasize the emission in the Galactic plane.  Polarization vectors are only displayed if the map pixel has a signal-to-noise ratio above 10 for intensity and above 4 for total polarization $P$.  Polarization vectors are predominantly perpendicular to the Galactic plane, implying that the magnetic field in the medium sampled by \bicep\ is parallel to the plane of the Galaxy.  While intensity values can't physically be negative, the observing and analysis strategy filters the maps, causing negative values in the maps as can be seen near higher Galactic latitudes.}
\label{fig:Tvec}
\end{figure*}

\begin{figure}  %figure 8
\begin{center}
    \resizebox{\columnwidth}{!} {
    $ \begin{array}{cc}
    \includegraphics{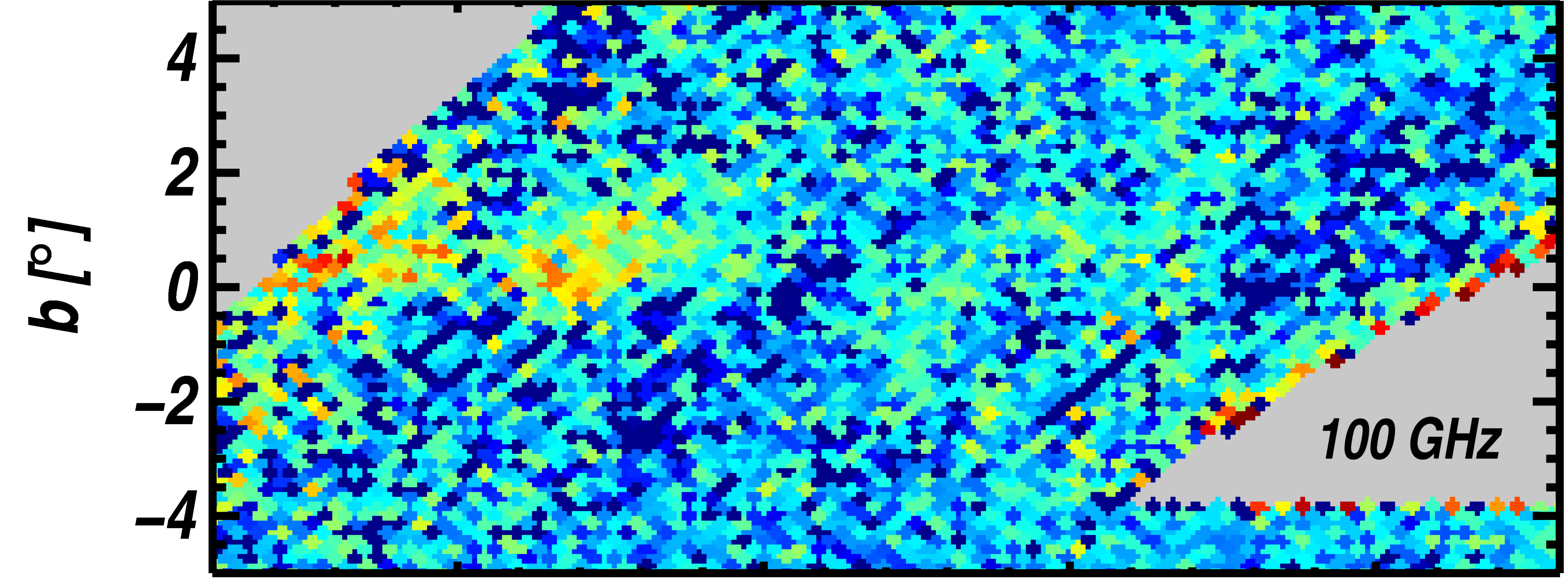} & \includegraphics{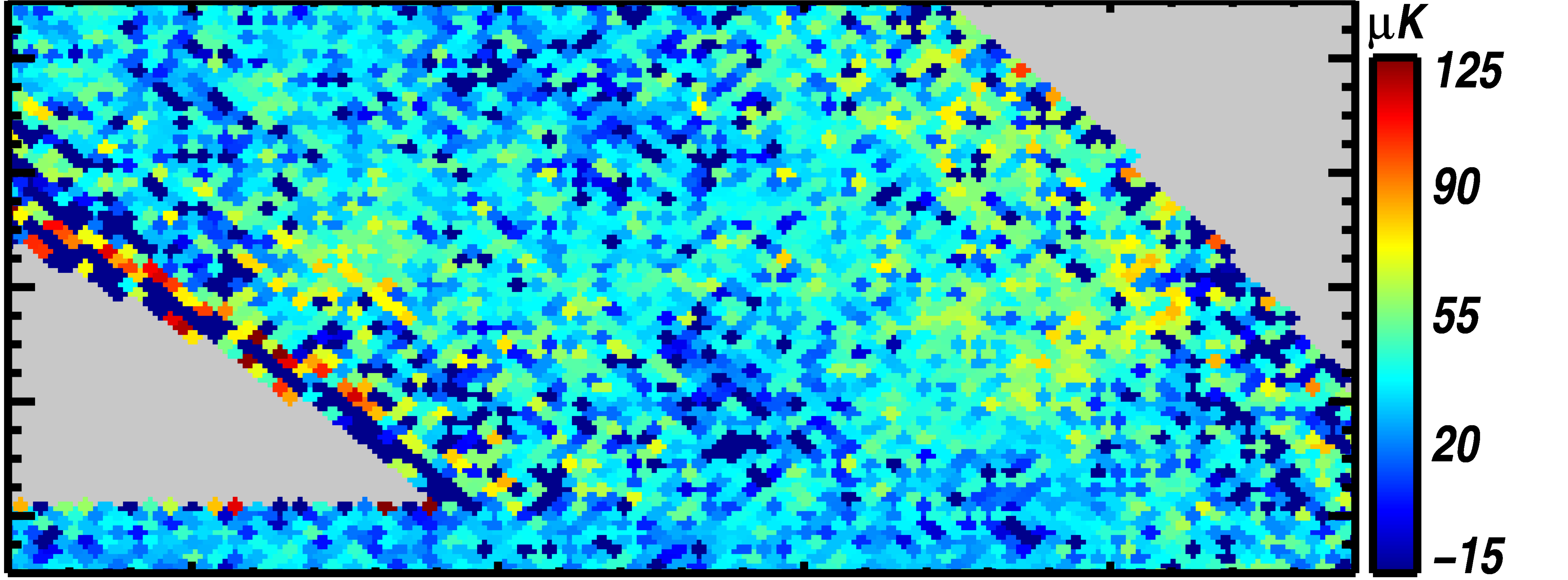} \\
    \includegraphics{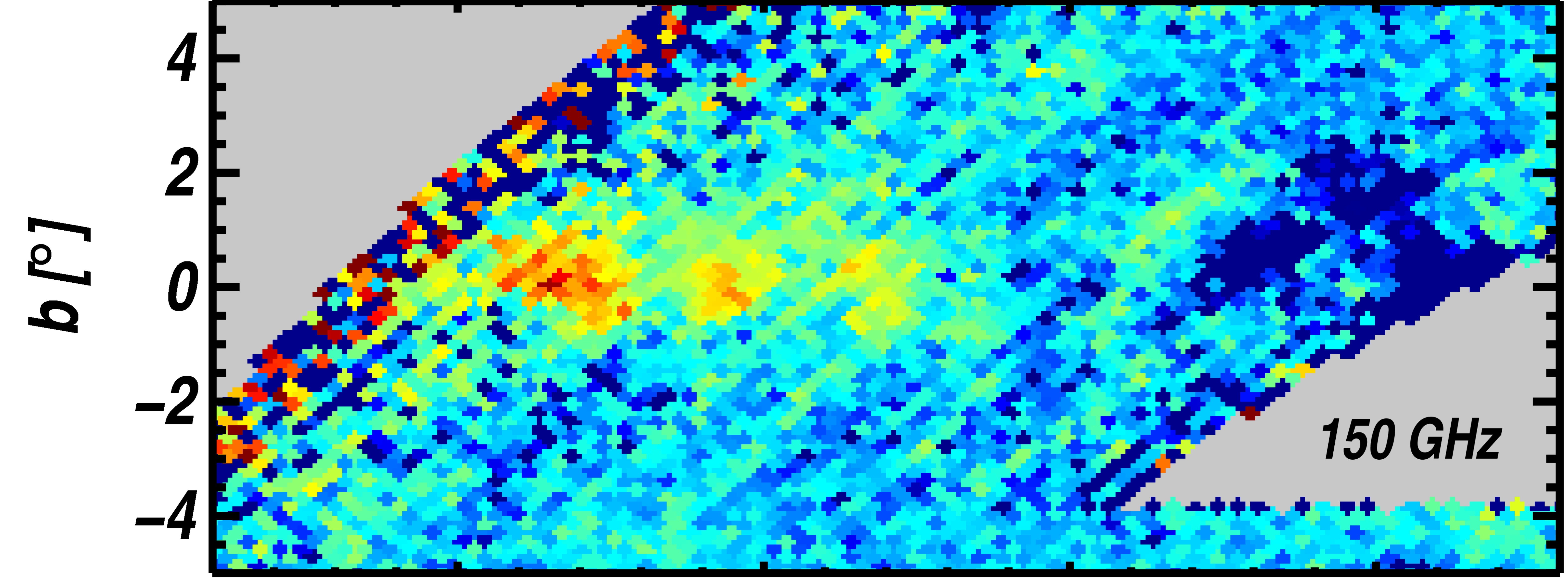} & \includegraphics{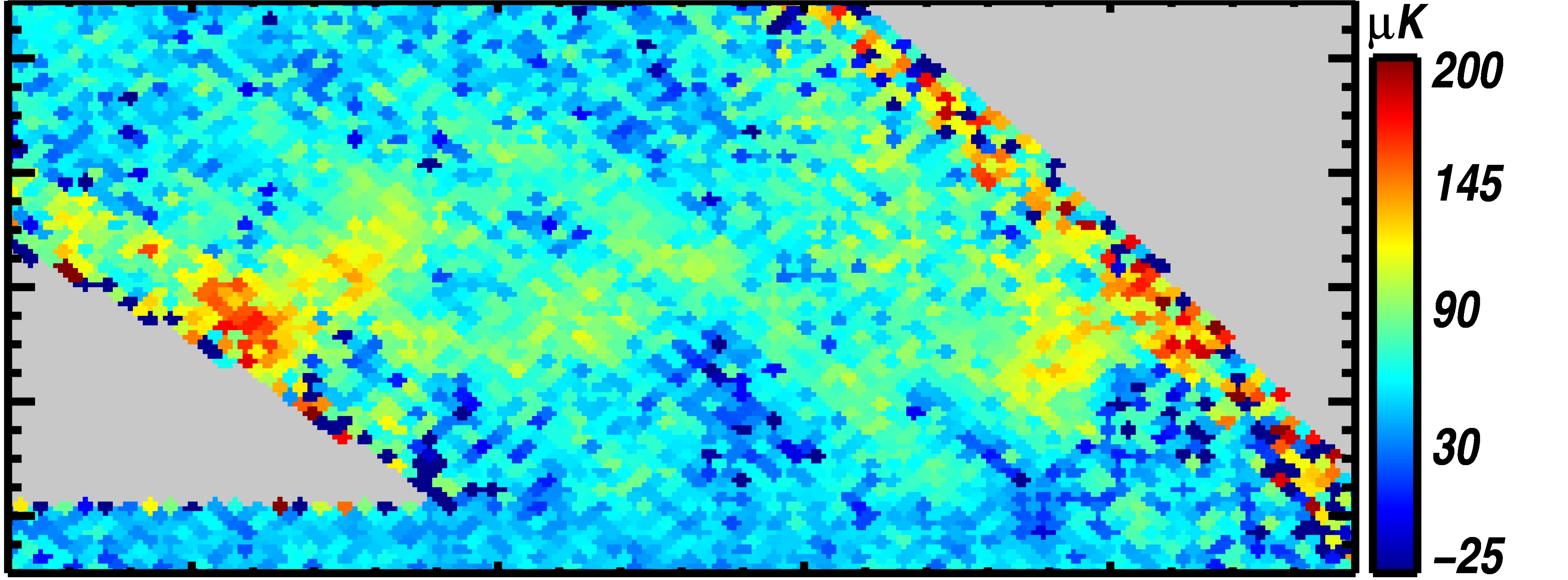} \\
    \includegraphics{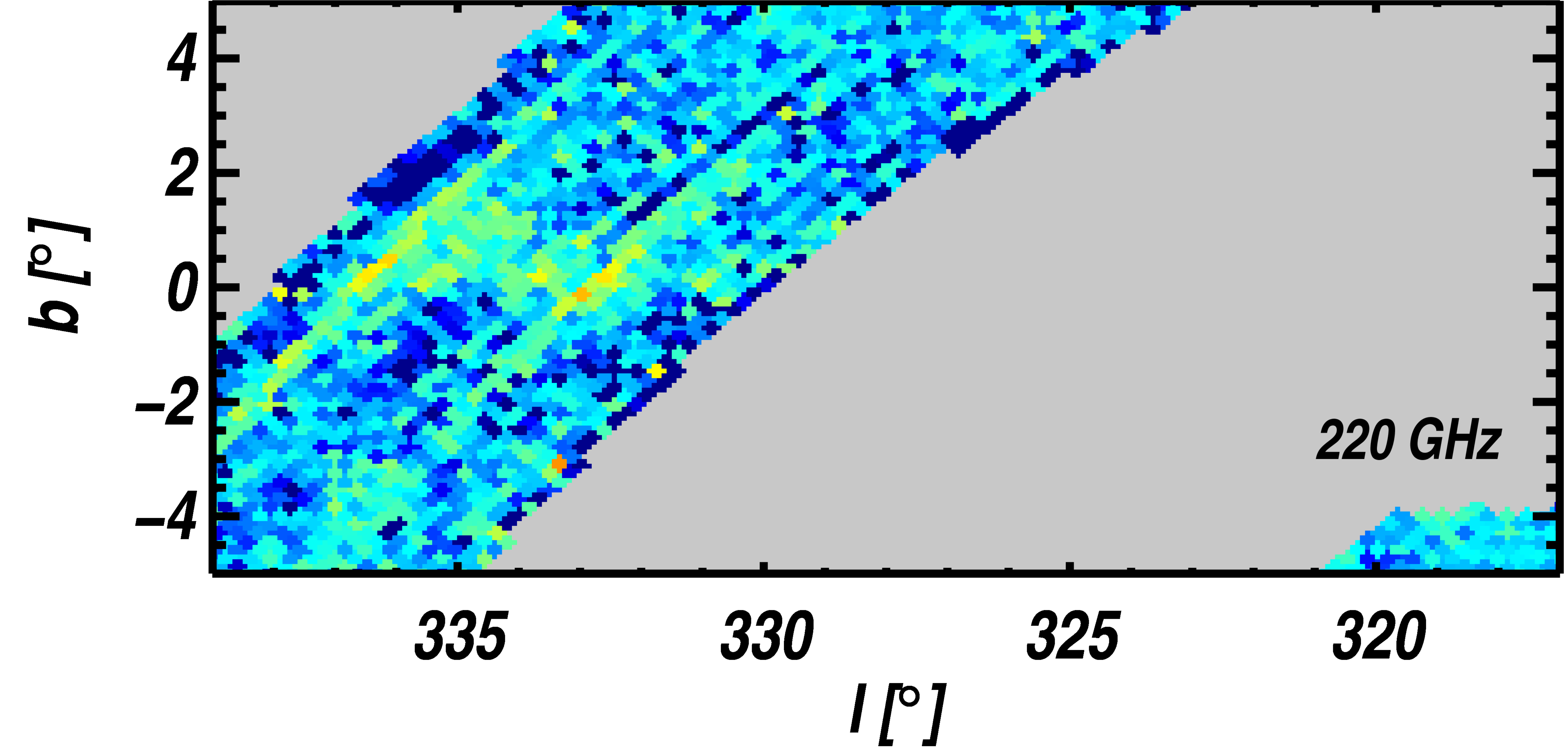} & \includegraphics{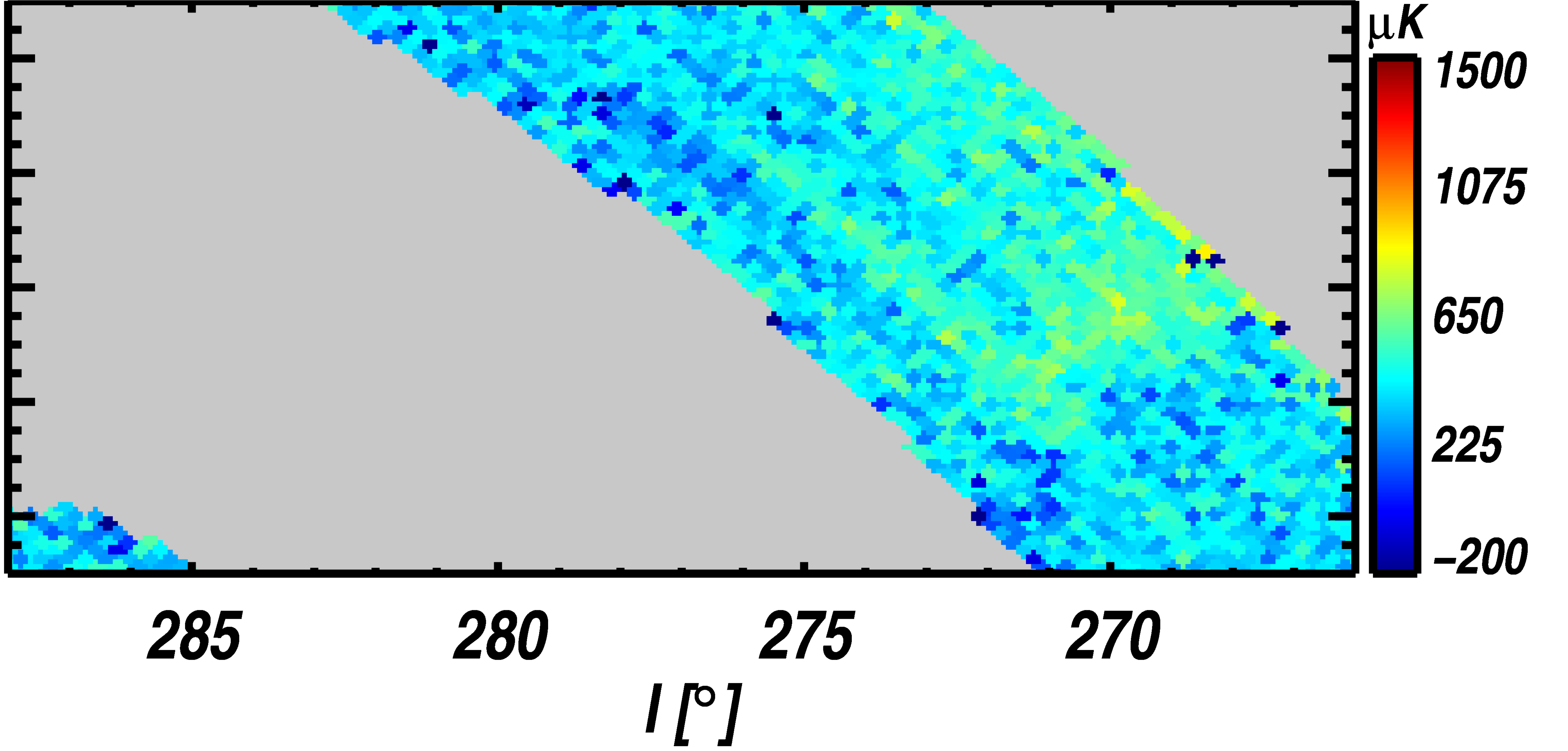} \\
    \end{array} $
      }
\end{center}
\caption{\bicep\ $U$ polarization maps from all three seasons co-added over all four boresight angles, in Galactic coordinates, at 100, 150, and 220 GHz, from top to bottom respectively.  Galactic +$U$ polarization corresponds to a polarization vector in the $+b$ and $+\ell$ direction. Power can be seen in all three bands indicating areas where the Galactic magnetic field is not exactly aligned with the Galactic plane.  The significant detection of negative $U$ power in 150 GHz map at Galactic Longitude 322$\deg$ is physical and not an artifact of filtering or other systematics.}
\label{fig:Uraw}
\end{figure}

\begin{figure}  %figure 9
  \begin{center}
    \resizebox{\columnwidth}{!} {
    $ \begin{array}{cc}
    \includegraphics{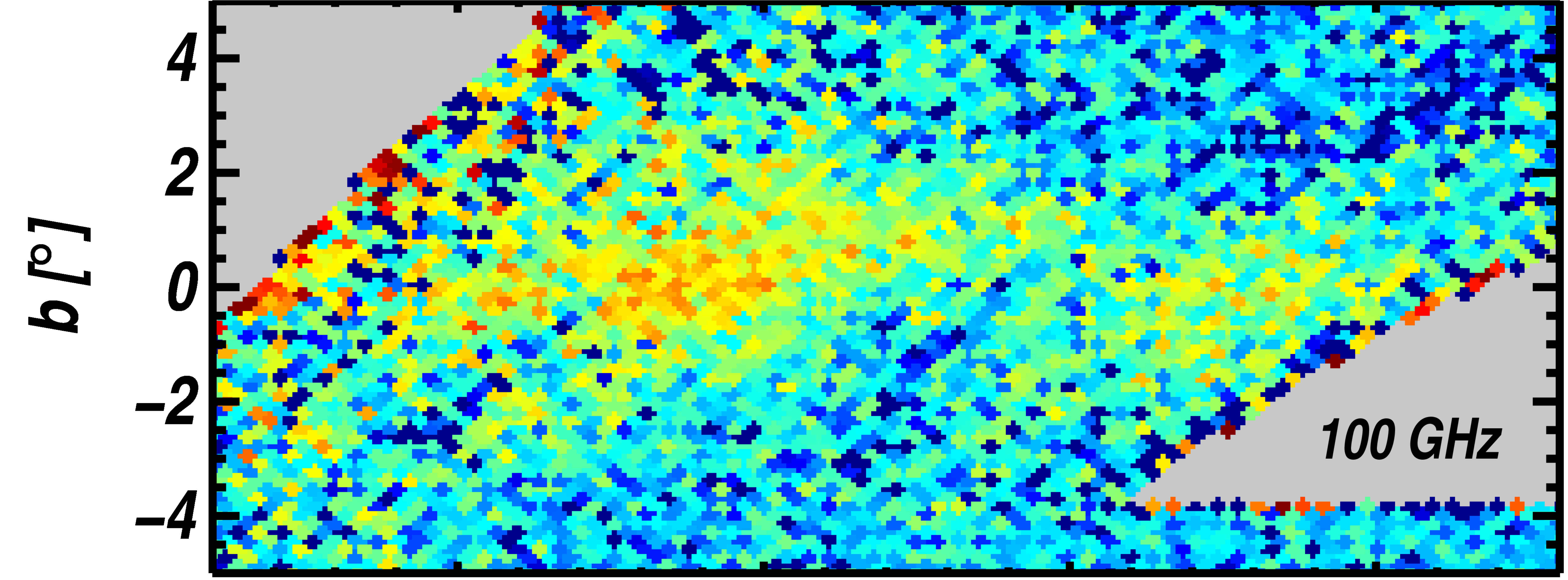} & \includegraphics{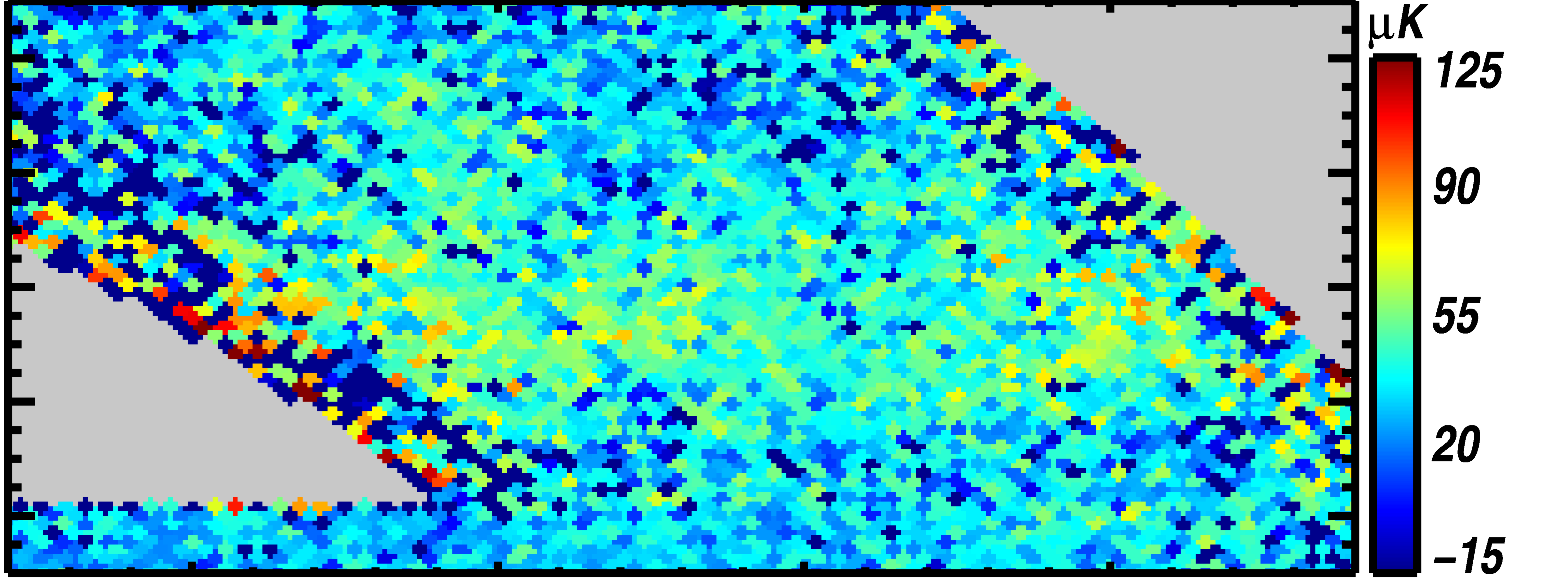} \\
    \includegraphics{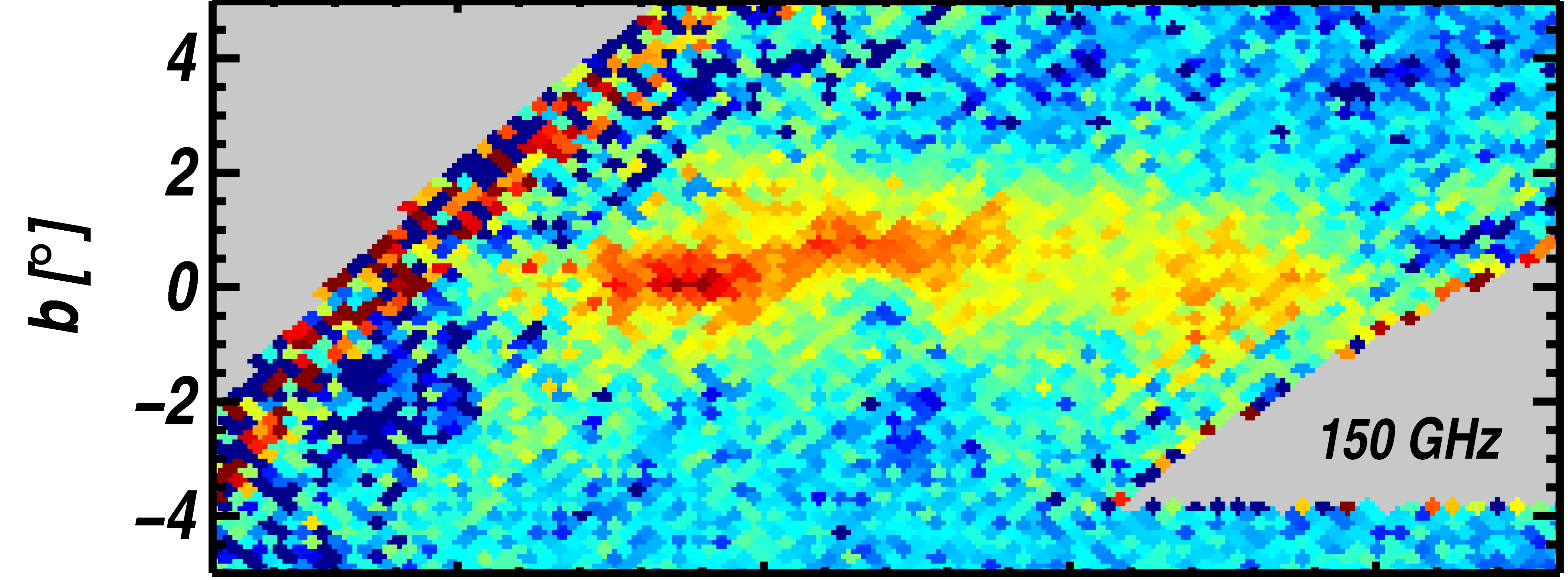} & \includegraphics{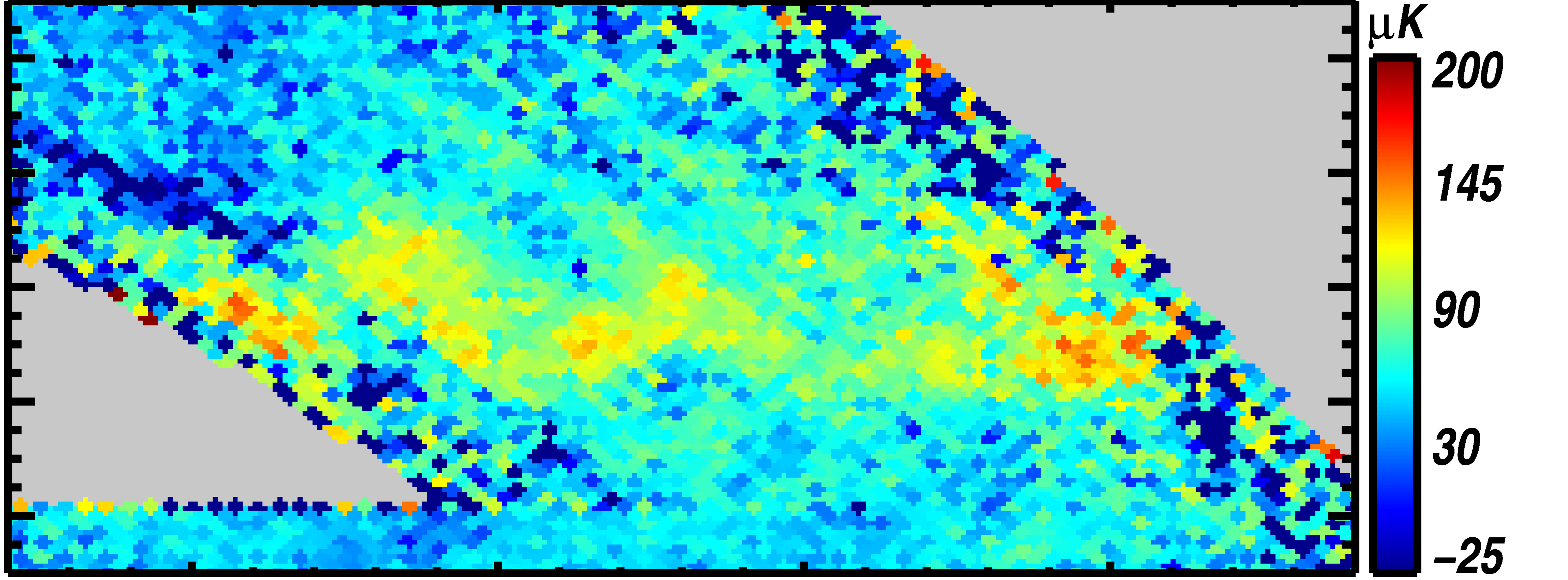} \\
    \includegraphics{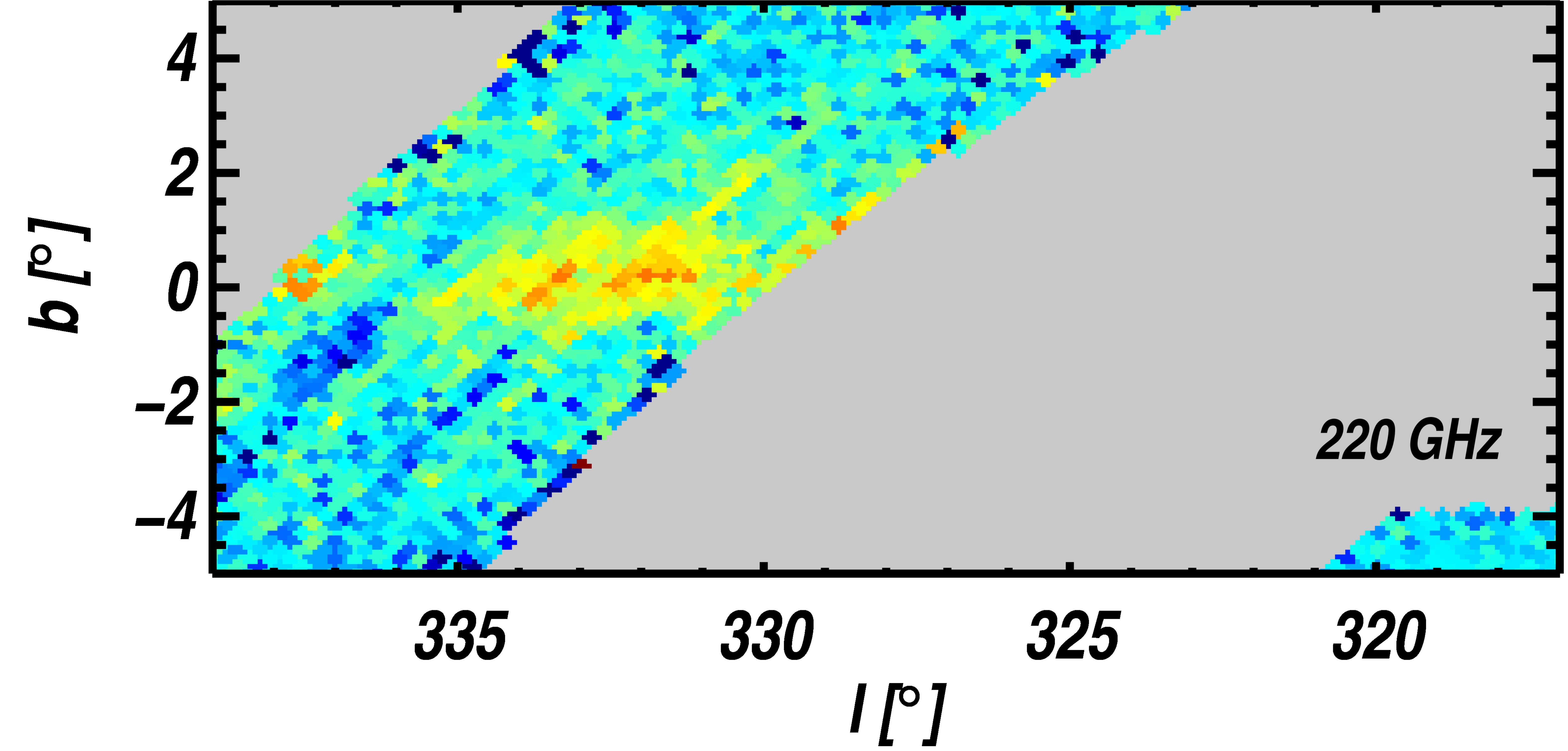} & \includegraphics{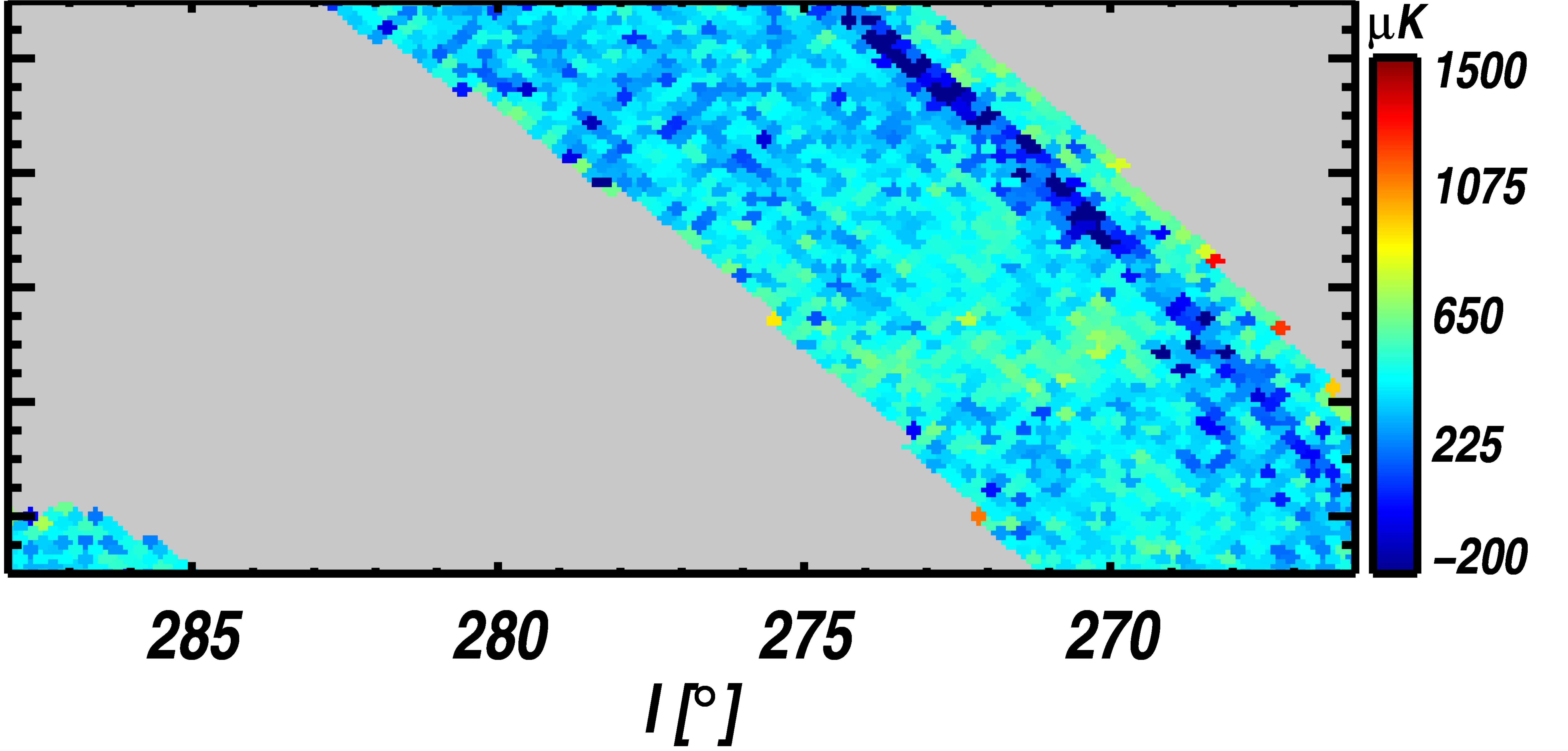} \\
    \end{array} $
      }

  \end{center}
  \caption{\bicep\ $Q$ polarization maps from all three seasons co-added over all four boresight angles, in Galactic coordinates, at 100, 150, and 220 GHz, from top to bottom respectively. Galactic +$Q$ polarization (red), corresponding to a vector that is perpendicular to the Galactic plane, dominates the maps.} 
\label{fig:Qraw}
\end{figure}

Figures~\ref{fig:Traw}-\ref{fig:Qraw} that show \bicep\ Galactic maps in three different bands, binned into 0.25$\deg$ \healpix\ pixels, using second-order polynomial filtering while masking data $\left|b\right|<4\deg$.  Polarization angles are defined counterclockwise from the meridian at the map pixel, in accordance with the IAU definition~\citep{Weiler1973,HamakerIII1996}.  \bicep\ absolute calibration casts the maps in thermodynamic temperature, which shows CMB anisotropy with the same value at all frequencies.  This unit is convenient for CMB analysis and gives a consistent reference frame for emission from sources other than the CMB.  

The total intensity maps in Figure~\ref{fig:Traw} show the large contrast between the CMB temperature anisotropy ($\left|b\right|>5\deg$) and the emission near the Galactic plane.  The noise in the intensity maps is not white; however, it is barely visible at either 100 or 150 GHz, even off the plane, except at the edges of the map.  The map pixels at the lower elevation have a lower signal-to-noise ratio due to the \healpix\ pixelization scheme that bins data into equal area pixels, while the scan strategy follows a Mercator projection.  This effect is most readily visible in the 220 GHz maps when comparing the upper and lower portions of the map.

The total intensity maps overlaid with polarization vectors in Figure~\ref{fig:Tvec} show that the brightest Galactic emission is within two degrees of the plane and consists of smooth large scale features and compact sources along the plane.  The maps show that the intensity signal dominates over the noise in the plane at all three bands and the polarization vectors are mostly perpendicular to the plane. 

Figure~\ref{fig:Uraw} shows there is $U$ signal in the plane corresponding to polarization vectors that are not perfectly perpendicular to the Galactic plane.  There are regions of positive $U$ polarization in the plane at all three bands.  However, there is a region in the 150 GHz maps near Galactic longitude $\ell=322\deg$ that shows a significant amount of negative $U$ signal that is not an artifact of the filtering or systematics.  The noise appears mostly white over the whole observing region with some residual striping along the scan direction at this map resolution.  

Figure~\ref{fig:Qraw} shows there is no significant negative $Q$ polarization in the Galactic plane, which would have produced polarization vectors generally parallel to the Galactic plane.  The noise is nearly identical in nature to the noise in the $U$ maps because both are derived from the same pair-difference data with identical filtering.

%\clearpage

%%%%%%%%%%%%%%%%%%%%%%%%%%%%%%%%%%%%%%%%%%%%%%%%%%%%%%%%%%%%%%%%%%%%%%%%%%%%%%%%%%%%%%%%%%%%%%%%%%%%%%%%%%%%%%%%%%%%%%%%%%%%%%%%%%%%%%%%%%%%
%%%%%%%%%%%%%%%%%%%%%%%%%%%%%%%%%%%%%%%%%%%%%%%%%%%%%%%%%%%%%%%%%%%%%%%%%%%%%%%%%%%%%%%%%%%%%%%%%%%%%%%%%%%%%%%%%%%%%%%%%%%%%%%%%%%%%%%%%%%%

\subsection{Map Quantities and Polarization Fraction Model}\label{sec:polynoise}

The map quantities studied are $I$, the pixel intensities (Section~\ref{sec:ivsfreq}); $q$, the polarization fraction perpendicular to the Galactic plane; $\eta$, the exponent of a simple power law describing the relationship between $q$ and $I$(Section~\ref{sec:flatvsslope}); and $\theta$, the polarization angle in Galactic coordinates (Section~\ref{sec:propvsfreq} ).  The derived variables $q$, $u$, $p$, and $\theta$ are given by:
\begin{eqnarray}  \label{eq:q_qe}
  q~&\equiv&~\frac{Q}{I},~u\equiv\frac{U}{I},~p\equiv \sqrt{q^2~+~u^2},~\theta\equiv\frac{1}{2} \tan ^ {-1} \Big( \frac{U}{Q} \Big)
\end{eqnarray}
The main polarization quantity studied is $q$ as opposed to $p$ or $u$, because $p$ suffers noise bias being a positive definite quantity, and there is relatively little signal in $u$.

It is known that the observed polarization fraction from an astronomical source will be lessened by disorder in the magnetic field.  To explore the Galaxy's magnetic structure more, \bicep\ maps are fit to a phenomenological power law as:
\begin{eqnarray}    \label{eq:qvslogt}
  q~&=&~  q_{(I_{median}) } \times \Big( \frac{I}{I_{median}} \Big) ^ {\eta}
\end{eqnarray}
where $I_{median}$ is calculated from the map pixels in the 220 GHz analysis region, $\eta$ is the slope parameter and $q_{(I_{median}) }$ is the overall polarization fraction normalization parameter.  The parameter, $\eta$, approximately represents the disorder in the magnetic field which is traced by millimeter-wave polarization.

The simplest Galactic magnetic field model to compare \bicep\ data to is one where $\eta=0$, which has no disorder in the field.  The measured polarization is a direct imprint of the intensity signal, related by a constant polarization factor $q_0$.

Other models describe the Galaxy's magnetic fields as being much more random.  For example, a simple toy model for this case is one where the Galaxy is uniformly thick and consists of a constant polarized ($q_0$) diffuse component emitting with a weak intensity, $I_0$.  Scattered throughout are dense star forming regions with random polarization angles which, when integrated along the line of sight or over the beam width, will integrate down to very low net polarization but will contribute to total intensity with $I_1$.  In this case the polarization fraction would be:
\begin{eqnarray}  \label{eq:q_simplemodel}
  q~&=&~\frac{q_0 ~ I_0}{I_0 + I_1} \propto I ^ {-1}.
\end{eqnarray}
In this case $\eta=-1$, implying increasing dust column density that contributes no additional polarized intensity.  

While both of the models here are strictly empirical, other studies have used similar methods.  These methods involve fitting starlight polarization to a similar power law model; the exponent for polarization by absorption is related to the power law exponent for polarization by emission in the millimeter-wave band by $\eta_{em}=\eta_{abs}-1$.  \cite{Fosalba2002} fit starlight polarization data to $p$ vs. $E(B-V)^{\eta_{abs}}$ finding $\eta_{abs}=0.8$, which implies $\eta_{em}=-0.2$.  \cite{Fosalba2002} then relates this fit parameter to a magnetic field model from \cite{Burn1966} using the assumption above, namely that the polarization fraction is a product of the ratio of uniform and random components of the magnetic field.  \cite{Jones1989} fits starlight polarization data to $p$ vs. $\tau_{K}^{\eta_{abs}}$, finding $\eta_{abs}=0.75$, which implies $\eta_{em}=-0.25$.  \cite{Jones1989} takes the fitting process a step further and runs Monte Carlo simulations for observing Galactic magnetic field arrangements with varying level of randomness and claims a more accurate fit than using an analytic equation.  Lastly, these other works study polarization fraction $p$, while in this paper $q$ is studied, which does not necessarily follow the same power law trend as $p$.  For example, a Galaxy with completely random magnetic field directions would predict a polarization fraction given by a power law exponent of $\eta=-0.5$; however in this scenario, $q$ would oscillate about zero and give an average $\eta=0$.

%%%%%%%%%%%%%%%%%%%%%%%%%%%%%%%%%%%%%%%%%%%%%%%%%%%%%%%%%%%%%%%%%%%%%%%%%%%%%%%%%%%%%%%%%%%%%%%%%%%%%%%%%%%%%%%%%%%%%%%%%%%%%%%%%%%%%%%%%%%%
%%%%%%%%%%%%%%%%%%%%%%%%%%%%%%%%%%%%%%%%%%%%%%%%%%%%%%%%%%%%%%%%%%%%%%%%%%%%%%%%%%%%%%%%%%%%%%%%%%%%%%%%%%%%%%%%%%%%%%%%%%%%%%%%%%%%%%%%%%%%

\begin{figure}  %figure 10
  ~~~~~~~~~~~~~Before Spectral Correction~~~~~~~~~~~~~After Spectral Correction
  \begin{center}
    \resizebox{\columnwidth}{!} {
    $ \begin{array}{cc}
    \includegraphics{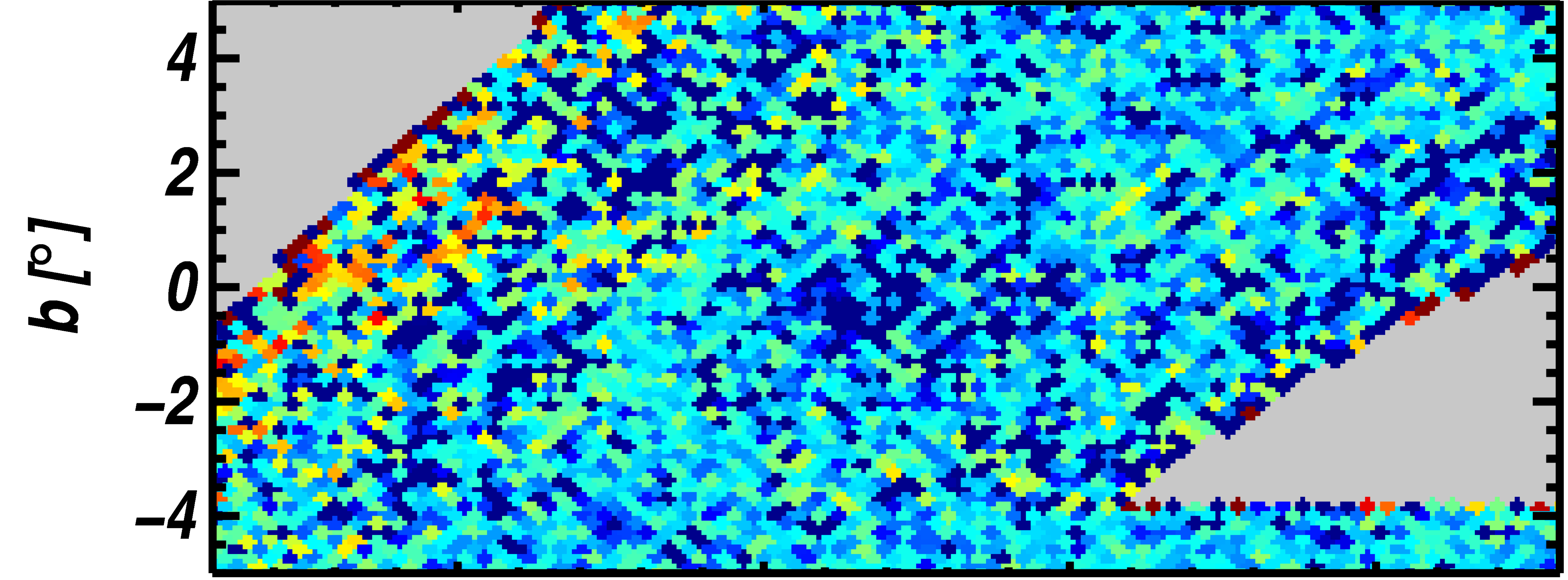} & \includegraphics{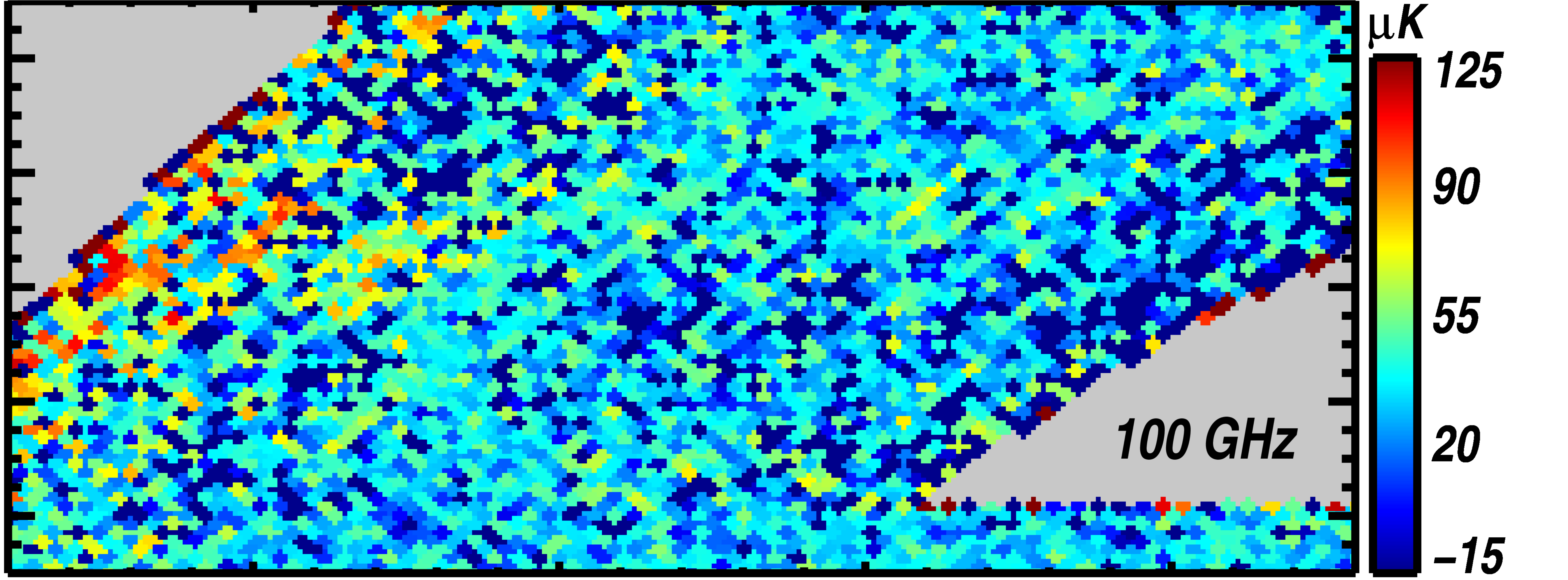} \\
    \includegraphics{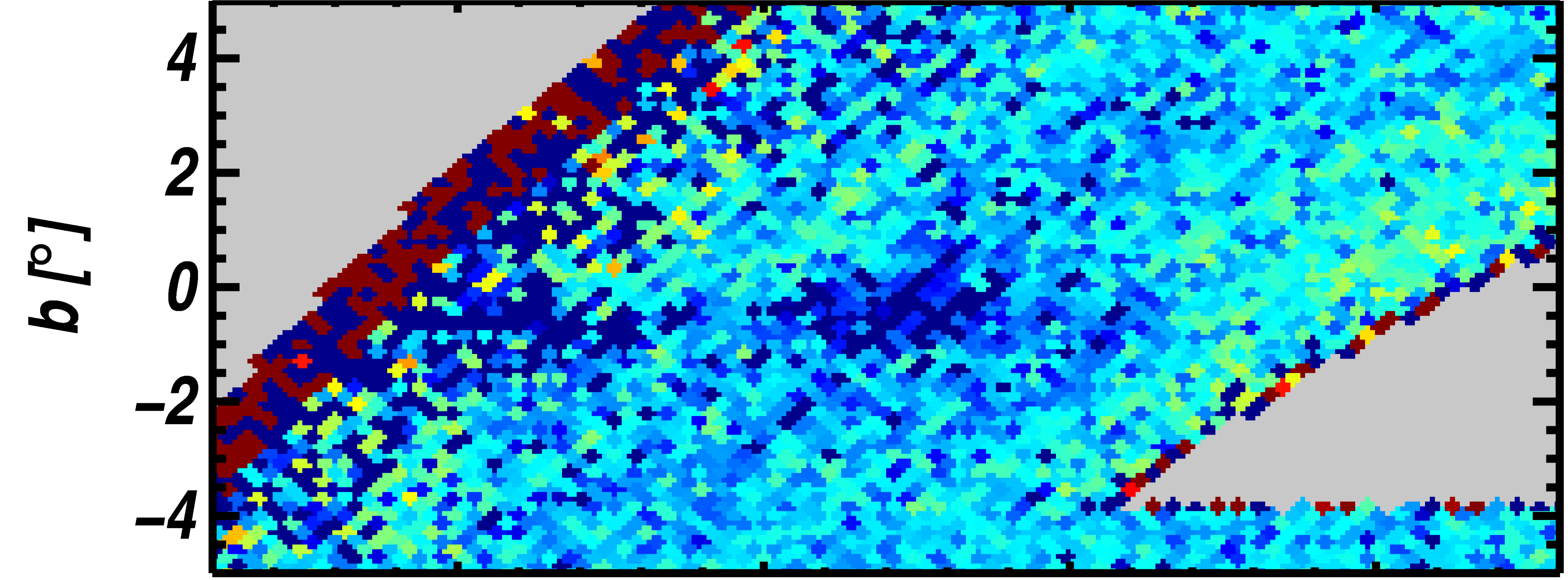} & \includegraphics{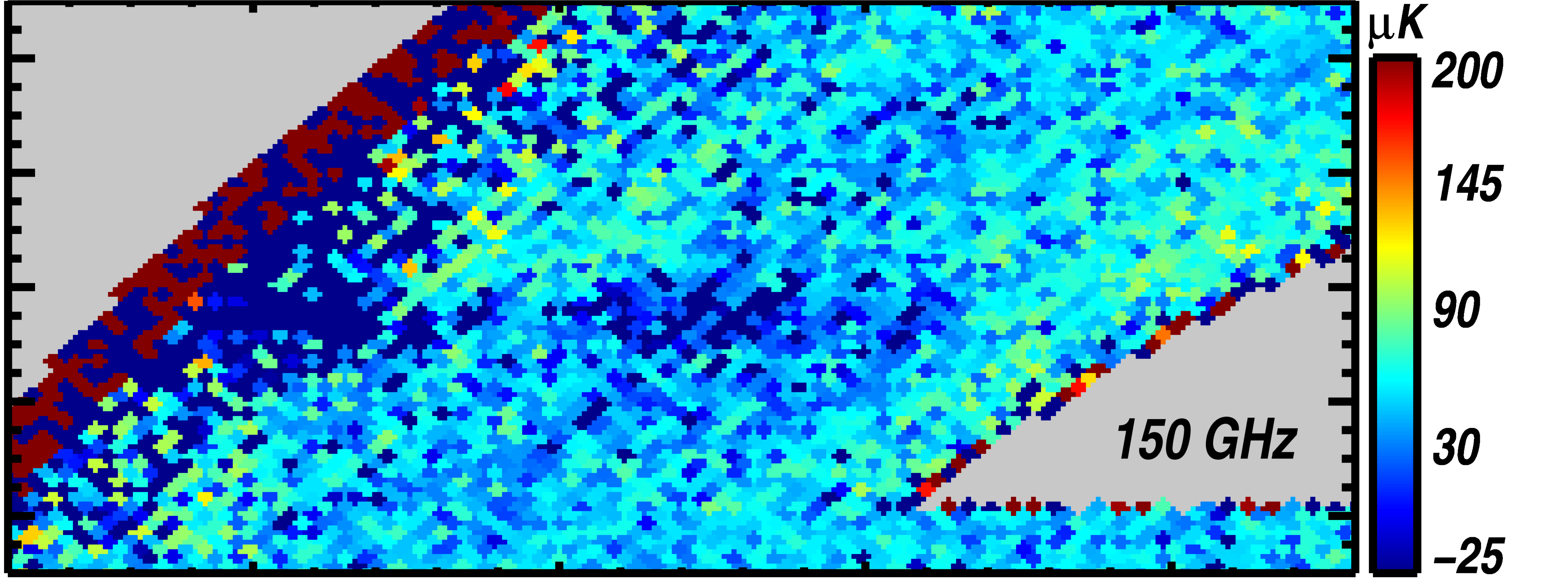} \\
    \includegraphics{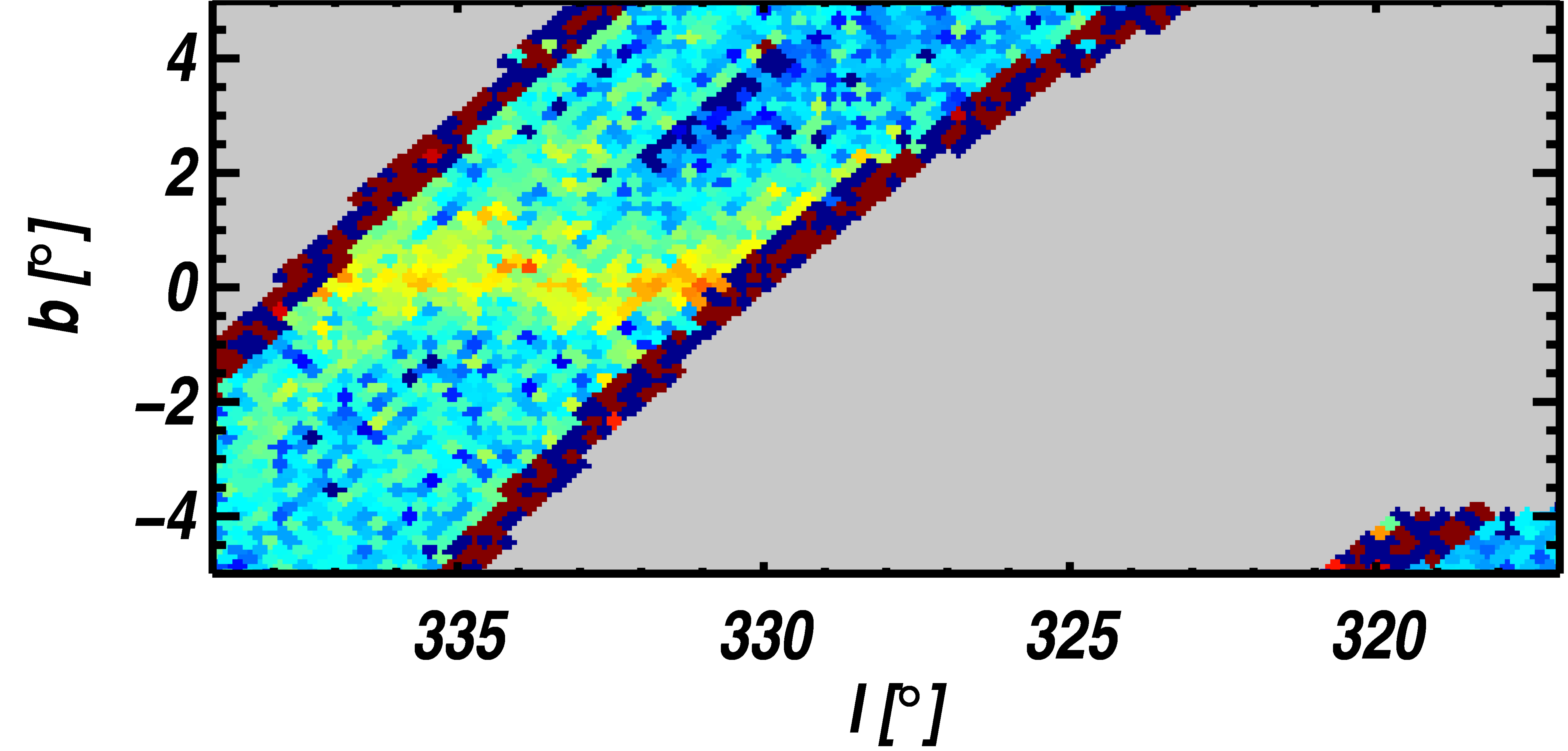} & \includegraphics{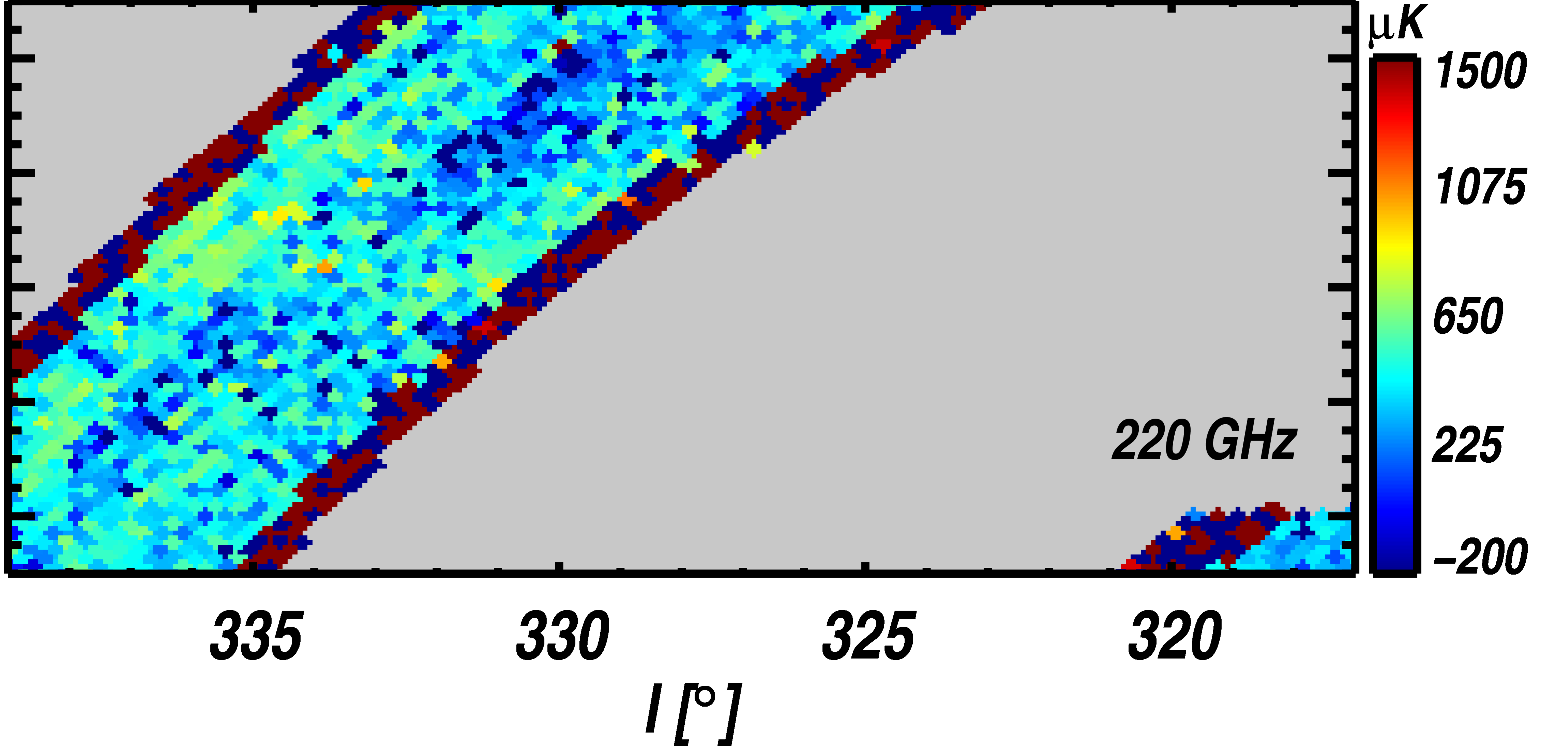} \\
    \end{array} $
      }
  \end{center}
  \caption{Difference between $Q$ Gal-bright maps made at boresight rotation angles \{0,315\} and \{135, 180\}, divided by two, in Galactic coordinates for 100, 150, and 200 GHz from top to bottom respectively.  The left hand plots are the difference maps uncorrected for spectral gain mismatch while the right hand plots have been corrected.  Some of the features in these raw jackknife maps are due to differences in integration time and polarization coverage such as near the edges of the observing area.  Other features, such as the faint blue excess on the Galactic plane at 150 GHz map arise from scan-fixed or telescope systematic contamination.  Most of the excess at 220 GHz is corrected by accounting for spectral gain mismatch while there is marginal change in the 100 and 150 GHz maps.  There is some faint striping nearly orthogonal to the Galactic plane in the 150 and 220 GHz maps from residual $1/f$ noise leaking through the polynomial mask.}

\label{fig:150qjack}
\end{figure}

%%%%%%%%%%%%%%%%%%%%%%%%%%%%%%%%%%%%%%%%%%%%%%%%%%%%%%%%%%%%%%%%%%%%%%%%%%%%%%%%%%%%%%%%%%%%%%%%%%%%%%%%%%%%%%%%%%%%%%%%%%%%%%%%%%%%%%%%%%%%
%%%%%%%%%%%%%%%%%%%%%%%%%%%%%%%%%%%%%%%%%%%%%%%%%%%%%%%%%%%%%%%%%%%%%%%%%%%%%%%%%%%%%%%%%%%%%%%%%%%%%%%%%%%%%%%%%%%%%%%%%%%%%%%%%%%%%%%%%%%%

\subsection{Noise and Systematic Error Evaluation}\label{sec:maperrors}
To confirm the integrity of the maps presented here, they were cross-checked with an independently written pipeline.  The two pipelines produced nearly identical \healpix\ maps, reducing the possibility of coding errors or other non-physical errors.  For example, taking the difference between the two pipeline's 150 GHz $Q$ maps gave a pixel $rms$ level five times smaller than the noise level, with no obvious signal features left.

The data have $1/f$ noise from atmospheric fluctuations, electronic readout drifts, thermal instabilities, and scan-fixed contamination.  The polynomial subtraction removes most of this contamination but the polynomial masking allows some of this noise to leak back into the Galaxy maps.  The polynomial fitting is only applied to data outside of the Galactic plane mask, causing $\chi^2$ not to be minimized with respect to the noise within the mask.  However, since the noise is correlated, the polynomial fit well-approximates the noise close to the mask edges, but decreases in effectiveness the further the pixels are from the edge of the mask.  Therefore, the larger the mask used, the worse the polynomial fit inside the mask approximates the noise, because there is a larger gap over which the polynomial must be extrapolated.  The smaller the mask used, the better the fit approximates the noise; however, this also increases the filtering of the signal in the plane.

Optical imperfections, such as beam mismatch, cross-polarization, depolarization, and sidelobe response can cause systematic changes in the maps.  These ``telescope systematics'', studied and characterized by \cite{Yuki2010} for the CMB B-mode analysis, are controlled very well partly because of \bicep's simple, compact design.  The overall magnitude of the telescope systematics and $1/f$ leakage in the Galactic maps are estimated by splitting the data into two halves, according to the telescope's orientation angle about the boresight.  Angles 0$\deg$ and 315$\deg$ are called ``boresight map A'', while 135$\deg$ and 180$\deg$ ``boresight map B''.  Since there are only two 220 GHz feeds, 220 GHz boresight map A is determined from a single feed, while 220 GHz boresight map B is determined from the other feed.  This split is the most probative for the polarimeter because the Galactic sky coverage in each half comes mostly from a different set of feeds, taken at different times, under different weather conditions, and with the telescope oriented differently with respect to gravity.  For this study the two \bicep\ boresight maps per frequency band are studied in this paper, which can be used to gauge the general level of residual systematic contamination in the maps.  

\subsubsection{Boresight Difference Maps}\label{sec:boresightdifmaps}
To illustrate the efficacy of the boresight difference maps, Figure~\ref{fig:150qjack} shows the difference between the two groups for the $Q$ maps for all three bands, both before and after spectral gain mismatch correction.  This is a qualitative jackknife test, as it does not test quantitatively against the expected noise or signal leakage.  Figure~\ref{fig:150qjack} shows some power in the $Q$ boresight map jackknife at all three bands, which is representative of all the Galactic jackknife maps.  The residual power in the 220 GHz raw maps was used to validate the spectral gain mismatch model and the post-corrected maps show the level of correction achieved.  While there is still some residual power in the 100 and 150 GHz channels, it does not affect the results claimed in this paper.  The post-corrected 220 GHz maps give consistent polarization results between the two detectors; however, there is still uncertainty in the spectral gain mismatch parameter due mostly to changes in atmospheric conditions.  

\subsubsection{Uncertainty due to Spectral Gain Mismatch}\label{sec:spectralgainmismatcherror}
Simulations were run to show the level of uncertainty induced on the quantitative parameters analyzed in this paper due to uncertainty in the spectral gain mismatch calculation.  Two sets of simulations were run; one based on the uncertainty in the atmospheric conditions and one based on the uncertainty in the measured spectrum.  

The uncertainty in atmospheric conditions was derived from computing the spectral gain mismatch using Austral winter and summer conditions, using a low precipitable water vapor (pwv) value, a median value, and a high value for different observing angles.  The mean and standard deviation of these conditions were computed, where the mean values are the nominal spectral gain mismatch and the standard deviation is used as the uncertainty.  The pwv conditions used were quite conservative and actual observing conditions most likely had a much smaller range of weather conditions.  The uncertainty from spectral measurements was calculated from the eight separate spectral measurements and computing the mean and error in the mean for each spectrum for each bolometer at each frequency.  

A single simulation consisted of taking the nominal spectral gain mismatch parameters for each feed and generating a modified list of gains using a random number generator and calculated uncertainty.  Then using the modified list, the mapmaking steps of Section~\ref{sec:filtering}~and~\ref{sec:mapmaking} are repeated, and $q_{(I_{median}) }$ and $\eta$ are found.  This procedure was repeated ten times for each type of uncertainty, which then allowed a standard deviation from the set of simulations to be computed.  

The percent error computed from the simulations based on different atmospheric conditions is $\frac {\delta q_{(I_{median}) }}{q_{(I_{median}) }} = [1.0\%, 1.3\%, 3.6\%]$ and $\frac{\delta\eta}{\eta} = [8.9\%, 15.\%, >100\%]$ for 100, 150, and 220 GHz.  The uncertainty from spectral measurement errors $\frac {\delta q_{(I_{median}) }}{q_{(I_{median}) }} = [0.2\%, 0.2\%, 0.3\%]$ and $\frac{\delta\eta}{\eta}= [2.4\%, 1.8\%, 41.0\%]$ for 100, 150, and 220 GHz.  The uncertainty on the 220 GHz parameters are larger than the other two bands because the two 220 GHz feeds have the largest spectral gain mismatch and the smallest measured $\eta$ value.  The spectral measurement uncertainties are five to ten times smaller than the uncertainties from the various weather conditions.  The combined spectral gain mismatch uncertainties are relatively minor for the determination of $q_{(I_{median}) }$; however, these uncertainties provide the largest single source of error on $\eta$ at all three bands.

\subsubsection{Uncertainty due to Polynomial Filtering}\label{sec:polyfiterror}
Polynomial subtraction removes signal as well as noise, causing systematic filtering effects.  To estimate the filtering effects, simulations are carried out to compare maps before and after various filtering strategies as well as to compare to the real data.  The simulations used \wmap\ 94 GHz DA 1 (Section~\ref{sec:wmap}) and FDS~\citep{Fink1999} as the basis for map intensity.  For the \wmap\ simulation, the map was first deconvolved with the beam provided by \wmap\ and convolved with the \bicep\ Gaussian beam at a given band.  Then, the maps were downsampled from 0.125$\deg$ \healpix\ pixels to 0.25$\deg$ pixels.  An additional simulation was carried out using the higher resolution pixels, but this caused less than a 5\% change in the resulting simulation values.  For the FDS simulation, tools are provided to predict the power over the entire sky at an arbitrary frequency\footnote{http://astro.berkeley.edu/dust/index.html}.  A full sky prediction is made at 0.016$\deg$ \healpix\ pixel resolution for each \bicep\ band for each measured spectral bandpass data point (approximately 200 separate frequency points per band), which can then be summed together to make three full sky simulated \bicep\ band dust maps.  The maps are convolved with the \bicep\ beam at each band and downsampled to 0.25$\deg$ \healpix\ resolution.

The polarization data are simulated based on these intensity maps by choosing parameters for Equation~\ref{eq:qvslogt} based on \bicep\ maps.  For the simulations the polarization fraction magnitude was always chosen to be ~2\% at the median intensity and the power law exponent $\eta$ was varied between 0.75, 0.50, 0.25.  The polarization angle is set to 7$\deg$, the approximate average polarization angle found in the 220 GHz analysis region, which sets the value of $U$ in the simulation.  Using this model, the polarization fraction has the possibility of taking unphysical values, therefore, the maximum polarization fraction allowed is set to 10\%.  A value of 25\% was also implemented but this did not change the results appreciably.

\bicep\ pointing was used to extract $I$, $Q$, $U$ values from the simulated maps to generate the simulated timestreams.  The simulated timestreams were processed with the same filtering as described in Section~\ref{sec:filtering}, leading to identical coverage as the real data.  In addition to the nominal 4$\deg$ polynomial mask, separate simulations were performed using 2$\deg$ to 9 $\deg$ polynomial masks for the \wmap\ simulations.  Mask values above 5$\deg$ gave marginal improvements in recovery of studied parameters but these higher mask values also decreased the amount of mapped area and also created more map pixels near the edges of the observation area with abnormal properties.

\begin{figure} %figure 11
   \hspace{-0.1in}\includegraphics[width=3.5in,height=7.0in]{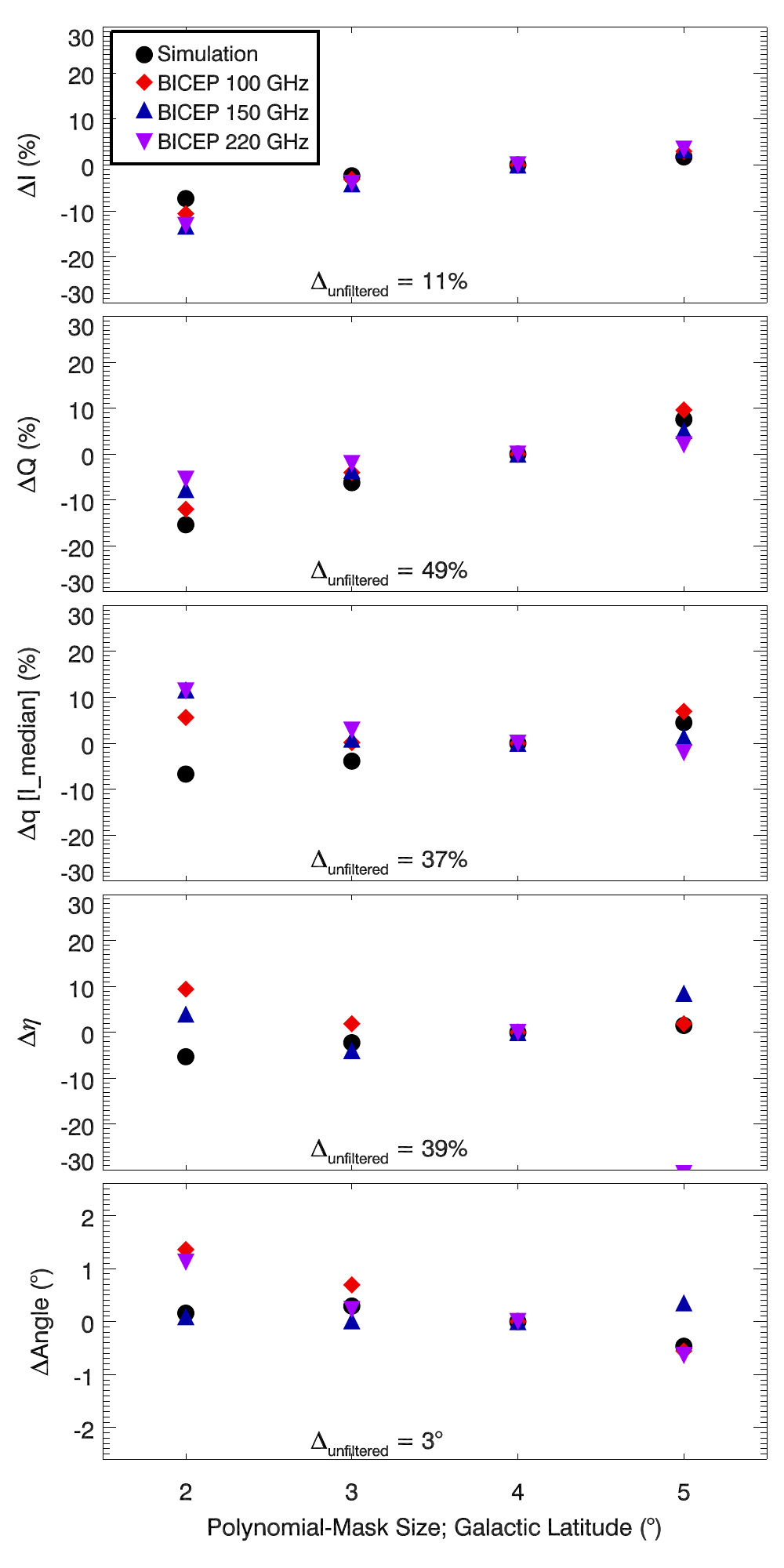}
   \caption{Fractional change in the average map parameters vs. polynomial-mask size for 100 GHz (red diamonds), 150 GHz (blue triangles), 220 GHz (purple upside-down triangles), and the 150 GHz \wmap\ simulation (black circles).  The parameters are intensity, $Q$ polarization, $Q$ polarization fraction at the median intensity, $\eta$, and polarization angle.  The units for the top four plots are relative percent difference between a given average map parameter, $X$, and the 4$\deg$ polynomial-mask average map parameter, $X_{4\deg}$; $\Delta X = (X/X_{4\deg}-1)$.  At the bottom of each plot is the fractional difference, $\Delta_{unfiltered}$, in the parameter value derived from the average values of unfiltered simulation maps compared to that from filtered simulation maps with a 4$\deg$ polynomial-mask.  The bottom plot is the difference in the average polarization angle compared to that derived from the map filtered with a 4$\deg$ polynomial-mask.   The simulation shows the size of filtering effects on the maps while the different polynomial-mask sizes show the general level of uncertainty in the filtering effects.  The changes in derived parameters due to polynomial-mask size are generally smaller than the changes in the parameters between filtered and unfiltered maps.}
\label{fig:qpsivsmask}
 \end{figure}

Polynomial masks ranging from 2$\deg$ to 5$\deg$ were also applied to the real data.  Then, for each map, the same pixels were fit to Equation~\ref{eq:qvslogt} and the average angle and standard error in the angle were computed.  The conclusion from the simulations and comparison to the real data, Figure~\ref{fig:qpsivsmask}, is that while the \bicep\ maps have been systematically filtered, the magnitude of this effect is small enough not to affect the general results from the analysis of the maps.  However, these effects do prevent additional quantitative analysis of the maps beyond what is performed here.  The systematic filtering of the intensity is much better understood than the filtering in the polarization data.  The uncertainty in the filtering correction for the intensity data is $\pm 5\%$, which comes mostly from the \wmap\ and FDS inputs rather than any variation across the bands.  The uncertainty in the polarization filtering correction is only known to $\pm 25\%$, which comes equally from variations of simulation polarization parameters across each band and from the difference between \wmap\ and FDS simulations.  The average correction factors for $I$, $q_{(I_{median}) }$, $\eta$ calculated from these simulations for the 220 GHz analysis region are 11\%, 37\%, 39\%.  The average correction factor for computing the average $q$ is 23\%.  These values are used to correct results for further analysis in this paper.

\subsubsection{General Map Noise}\label{sec:generalmapnoise}
\begin{figure}  %figure 12
\resizebox{\columnwidth}{!}{\includegraphics{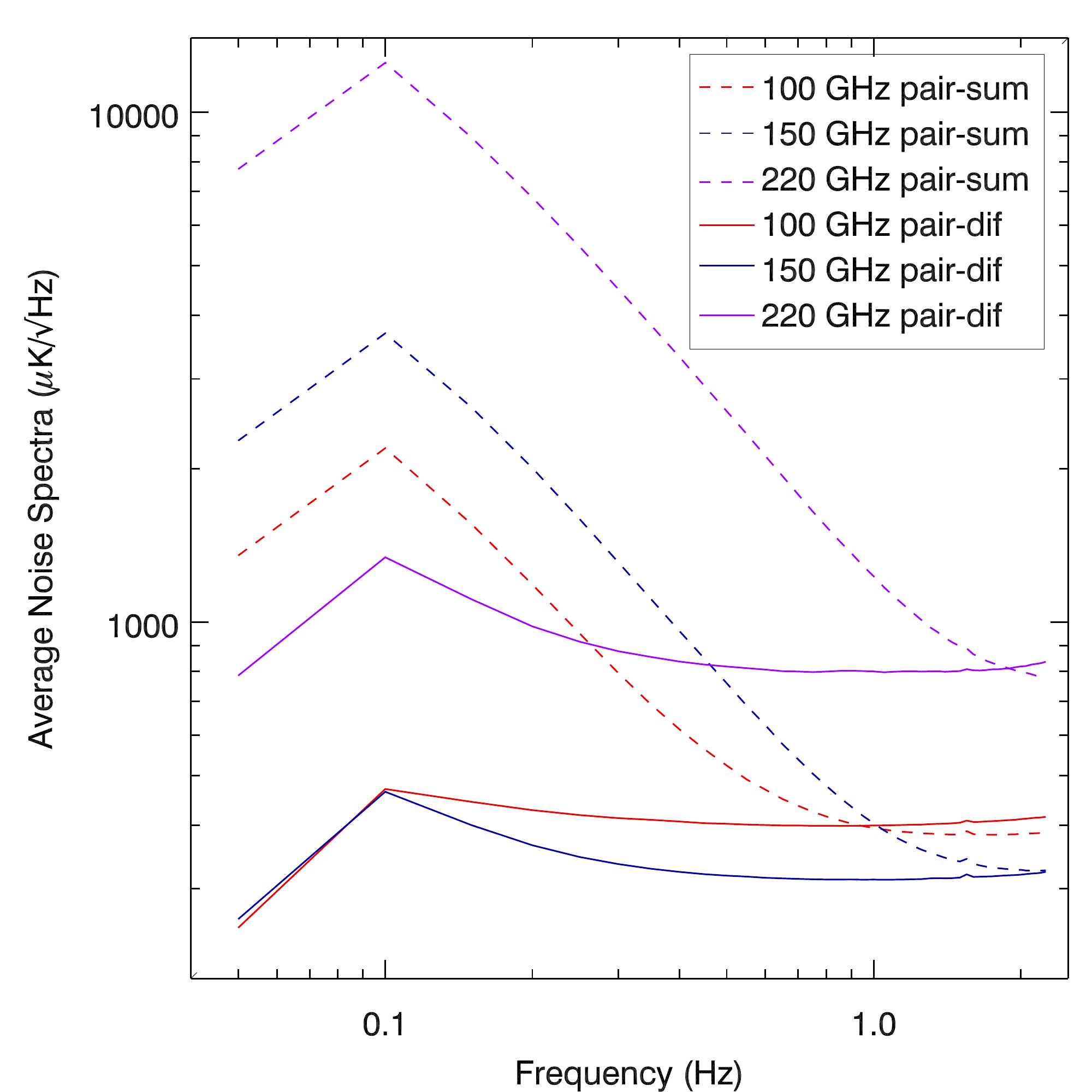}}
\caption{Average periodogram for the pair-sum and pair-difference timestreams from all of the Galactic scans, after second-order polynomial removal.  The pair-sum data suffers from increasing levels of $1/f$ atmospheric contamination from 100 GHz to 220 GHz.  Most of the $1/f$ atmospheric noise is removed from the polarization data by pair differencing within a feed, resulting in nearly white noise at all three bands above 0.1 Hz.  The second-order polynomial filtering is apparent at 0.05 Hz (the lowest frequency bin), where the power is lower than the white noise level.  The rise at high frequency is due to the deconvolution of the detector and system time constant.}
\label{fig:220_psd}
\end{figure}

Background photon and detector noise dominates the white noise floor in the maps.  The average polarization map pixel sensitivity can be calculated by taking the calibrated timestreams and computing the periodogram for each half scan.  The average periodogram per feed for each frequency band is then calculated (Figure~\ref{fig:220_psd}).  Averaging the pair-difference periodogram from 0.1 Hz to 1.0 Hz (corresponding to an angular size of $\approx 0.5\deg - 5\deg$) gives Noise Equivalent Temperature (NET) per detector values of 520, 450, and 1040 \ukrts  for 100, 150, and 220 GHz respectively.  After accounting for polarization efficiencies, these correspond to NEQ per feed values of 410, 340, and 880 \ukrts  for 100, 150, and 220 GHz respectively.  There is very good agreement between the noise estimate from the periodograms in~\cite{Yuki2010}, which used two years of third-order polynomial filtered data from the CMB region, and this calculation using three years of data from the Gal-bright and Gal-weak regions that has second-order polynomial filtering.

The noise in a given map pixel is calculated using the NEQ per feed values, the integration time per 1$\deg$ pixel, and assuming the integration time is split evenly between the $Q$ and $U$ maps.  The average sensitivity for $Q$ or $U$ map pixels in the 220 GHz analysis region is 2.8, 2.8, 16.0 $\muK$-rms for 100, 150, and 220 GHz respectively. 

Another method to gauge the noise in the maps is to split the data in half and take the difference, which cancels off the signal and leaves the residual noise.  Maps were made from right going and left going scans separately, and then differenced.  This split is used to test for  detector time constant mismatch or general telescope thermal effects; however, it tends to be one of the least probative for \bicep\ because the two halves are taken at nearly the same time, under the same weather conditions, and with the same set of feeds.  Pixels from the $Q$ and $U$ differenced maps, with a significant amount of integration time, were multiplied by the square root of integration time per pixel and the standard deviation of all the pixels was computed.  This was done for all the pixels and the 220 GHz analysis region pixels.  Using all the pixels, the average noise values are 420, 320, and 930\ukrts for 100, 150, and 220 GHz respectively, which are comparable to the NEQ per feed noise estimate from the periodogram method indicating good agreement between the two methods.  The noise for the 220 GHz analysis region pixels as compared to the entire map is 20\%, 40\%, and 45\% higher for 100, 150, and 220 GHz respectively, due to excess $1/f$ noise leaking inside the mask.

\subsection{\wmap\ 94 GHz Band Comparison}\label{sec:wmap}

\wmap\ is a millimeter-wave satellite that has ten differencing assemblies (DAs), all producing $I$, $Q$, and $U$ maps over the whole sky using one channel at 23 GHz, one at 33 GHz, two at 41 GHz, two at 61 GHz, and four at 94 GHz.  The combination of \bicep\ data with \wmap\ data provides frequency coverage in the 220 GHz analysis region from 23 to 220 GHz.  As an important cross check on the validity of the \bicep\ observations, a direct comparison is made between the \wmap\ 94 GHz band and \bicep\ 100 GHz intensity maps.

There are two systematic differences that prevent direct comparison.  First, the \bicep\ beam is a different size than the \wmap\ beam and the maps are filtered as described in Section~\ref{sec:polynoise}.  Therefore, to make a direct comparison, each \wmap\ DA map is smoothed to \bicep's beam resolution, sampled into \bicep-like timestreams, and filtered.  For consistency, sampling of the \wmap\ maps was performed using both 0.0125$\deg$ maps, and the nominal 0.025$\deg$ maps.  Since both sets of maps are filtered, there is no need for systematic filtering correction in this case.

Secondly, there is a difference in spectral bandpass between the two experiments.  \bicep\ is calibrated to \wmap\ by comparing the CMB fluctuations in the low astronomical foreground region.  However, Galactic emission has a different spectrum than the CMB, and combined with the different bandpass response, this could cause a systematic difference between the two experiments.  In a very similar manner to the spectral gain mismatch calculation in Section~\ref{sec:specchar}, the expected miscalibration is calculated using the average \bicep\ 100 GHz spectrum, the average \wmap\ 94 GHz band spectrum, the CMB anisotropy spectrum as seen from space and from the South Pole, and the typical Galactic source spectrum as seen from space and the South Pole.  

Different atmospheric observing conditions were used as in Section~\ref{sec:specchar}.  The mean and standard deviation of spectral gain mismatch from the different atmospheric conditions between \bicep\ 100 GHz and \wmap\ 94 GHz band is $0.322\% \pm 0.001\%$ (the positive sign indicates an increase in \bicep\ power relative to \wmap).  The small difference results from the CMB and the typical Galactic source spectrum being very similar at this band, so even though \bicep\ and \wmap\ have relatively different bandpasses, there is no spectral gain mismatch between the experiments.  The extremely small standard deviation on this quantity results from the insensitivity of the \bicep\ 100 GHz band to different atmospheric conditions, especially the emission lines outside the band.  Instead of using the average 100 GHz spectra, this calculation was repeated for each individual detector's spectrum giving an average and standard deviation of anomalous gain factors of $0.315\% \pm 0.055\%$, consistent with the first method.

The data from the two experiments were compared in a similar manner to the absolute calibration routine from ~\cite{Chiang2010}, except the comparison was done on maps instead of on angular power spectrum.  The anomalous gain factor comparing the \bicep\ 100 GHz and \wmap\ 94 GHz band intensity maps is calculated as
\begin{eqnarray}
  g_{anom} = \frac{ \langle MAP^{\rm WMAP-1} MAP^{\rm BICEP} \rangle}{ \langle MAP^{\rm WMAP-1} MAP^{\rm WMAP-2} \rangle} - 1
\label{eq:wmap_bicep_gain}
\end{eqnarray}
where the angle brackets represent a weighted average of the 220 GHz analysis region pixels weighted by the \bicep\ integration time and the \wmap\ and \bicep\ terms represent the two \bicep\ boresight maps and the four \wmap\ DAs at 94 GHz.  As a consistency check on this method, the gain was computed for the CMB observing region as well and compared to the absolute calibration numbers from \cite{Chiang2010}, which found the gains using the angular power spectrum, giving consistent results.  The two boresight maps give values 5.0\% and 4.7\% gain increases using the $0.25\deg$ maps and 4.2\% and 4.0\% gain increases using the $0.125\deg$ maps.  Each has a 0.5\% statistical error derived from using the different combinations of \wmap\ 94 GHz band DA maps.

To check the uncertainty based on pixel selection or sky variance, a simulation was done using all pixel values $\left|b\right|<3\deg$ in the \bicep\ observing region (335 one degree pixels) for the $0.125\deg$ maps.  For each of 1000 different trials, 53 pixels were chosen at random and the anomalous gain was computed.  The mean and standard deviation of the distribution gave 3.1 $\pm$ 2.1\% and  2.5 $\pm$ 1.9\% for the two boresight maps.  For comparison, the same procedure was repeated for the 253 one degree pixels between $-3\deg$ and $-6\deg$ in Galactic latitude giving anomalous gain values of 1.0 $\pm$ 8.5\% and -5.0 $\pm$ 6.5\%

\cite{Chiang2010} quotes the absolute gain uncertainty to be 2\% for the \bicep\ maps, which decreases the relative significance of the anomalous gain factor found here.  The difference in anomalous gain factor due to the bandpass differences was calculated to be a very small; however, this calculation used a simple two-component continuum Galactic spectrum.  The real Galactic spectrum is undoubtedly more complicated, which could lead to a larger spectral factor difference.  Another likely cause of anomalous gain is from systematic uncertainties in the processing of raw to \bicep-filtered \wmap\ maps (Section~\ref{sec:polynoise}), either from the beam correction or flat interpolation procedure.  Irregardless of the underlying cause, the ~4\% gain difference does not represent a significant detection of a deviation between the two experiments in the Galaxy.  A larger difference was found for QUaD~\citep{Culverhouse2010}, despite the similarity between the instruments and analysis approach.  However, QUaD's high-frequency cutoff is higher and low-frequency cutoff lower than either \bicep\ or \wmap, making it sensitive to certain emission lines to which \bicep\ and \wmap\ are insensitive.

\subsection{Intensity vs. Frequency} \label{sec:ivsfreq}
By studying the spectrum of the unpolarized and polarized emission for a given point on the sky, it is possible to understand the composition of the ISM across the sky.  The spectral response plots of intensity vs.~frequency can show the fraction of dust and synchrotron in the Galactic plane.  \wmap\ provides a vast amount of information on this topic; however, their highest frequency band still has an appreciable amount of synchrotron emission.  Above 150 GHz, most Galactic emission comes from dust, which is where \bicep's 150 and 220 GHz channels add unique information.  

\begin{figure} % Figure 13
   \begin{center}
     \resizebox{\columnwidth}{!}{\includegraphics{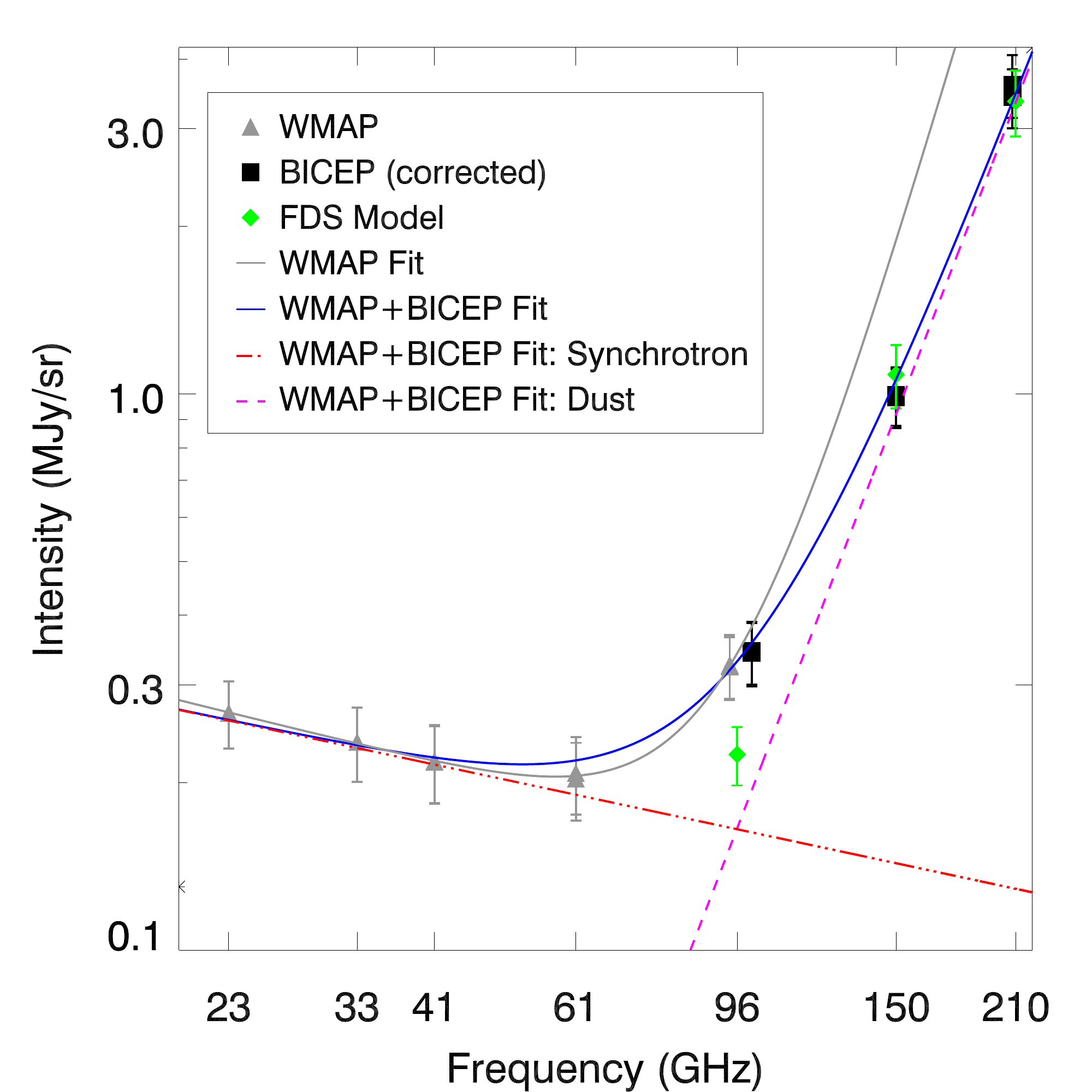}}
   \end{center}
   \caption{The mean intensity as a function of frequency for the 220 GHz analysis region for \wmap's ten DAs (gray triangles) and \bicep's two boresight maps for each of the three bands (black squares).  \bicep\ maps have been filtered by the mapmaking process, systematically lowering the intensity values.  To correct for this, each \bicep\ point  and error bar has been increased by 11\% (Section~\ref{sec:polyfiterror}).  \bicep's 100 GHz points are shifted to the right 5 GHz from the calculated band centers for clarity.  \wmap\ points come from raw, unfiltered maps, downsampled to 1$\deg$ resolution.  The noiseless FDS model 8~\citep{Fink1999} predicted dust maps (green diamonds) have been beam smoothed, spectral bandpass filtered, and downsampled to 1$\deg$ to match \bicep's bands but not polynomial filtered.  The error bars for the FDS points and other bands come from sky variance, not measurement errors.
The fit to Equation~\ref{eq:Imodel} only using \wmap\ points (gray) gives a dust spectral index $\zeta_d=5.06$, larger than when \bicep\ data (blue) are included, which gives $\zeta_d=3.70$.  The pink and red dashed lines are the individual positive and negative spectral index components from fitting to both \wmap\ and \bicep\ points. \bicep\ has two data points per band that are computed from the two boresight maps.  The fact that two data points are nearly identical indicates that the systematic uncertainties are not the dominant source of uncertainty for the average intensity of the maps, even for the 220 GHz channels, consistent with the discussion in Section~\ref{sec:maperrors}.  \wmap\ points are from each DA separately, which can indicate the approximate level of residual systematic uncertainty in a given band.}
\label{fig:medtvsfreq}
 \end{figure}

A common unit for distributed astrophysical millimeter-wave emission is differential intensity as opposed to thermodynamic temperature, where the conversion factor from thermodynamic units can be calculated as:
\begin{eqnarray}
  dI&=&\frac{dB}{dT} \Big| _{\mathrm{2.725~K}} \times dT, \nonumber
\end{eqnarray}
and $B$ is the Planck blackbody function.  The derivative with respect to temperature is:
\begin{eqnarray}
  \frac{dB}{dT}&=&\frac{2 k_B \nu ^ 2}{c ^ 2} \frac{z ^ 2 e ^ z}{ (e ^ z - 1) ^ 2}; ~ ~ \mathrm{where} ~ ~ z = \frac{h \nu}{k T}. \nonumber
\end{eqnarray}
Using the values of the physical constants and the \bicep\ band centers, the conversion factors are:
\begin{eqnarray}
  \frac{dB}{dT}&=&~ \mathrm{0.22, 0.40, 0.48}~ \frac {\mathrm{MJy/sr}}{\mathrm{mK}_{\mathrm{CMB}}},  \nonumber \\
  \label{eq:MJtemp}
\end{eqnarray}
for 100, 150, and 220 GHz respectively.  Figure~\ref{fig:medtvsfreq} and the subsequent spectral analysis in this section uses \bicep, \wmap, and FDS data in these differential intensity units.

Figure~\ref{fig:medtvsfreq} shows a plot of the mean and standard error of the mean intensity of the analysis area pixels from both \bicep\ and \wmap, calculated from each of the ten \wmap\ DAs and two boresight maps for \bicep's three bands in differential intensity units.  The points in Figure~\ref{fig:medtvsfreq} are fit to
\begin{eqnarray}  \label{eq:Imodel}
  I(\nu)~=~A_s~\nu^{\zeta_s}~+~A_d~\nu^{\zeta_d},
\end{eqnarray}
a simple two-component model.  \bicep's points and error bars have been increased by 11\% using the systematic filter correction found in Section~\ref{sec:polyfiterror}.  The best fit spectral index parameters and errors from using only the \wmap\ points in Equation~\ref{eq:Imodel} are $[\zeta_s, \zeta_d]=[-0.36,5.06] \pm [0.02, 0.33]$, while if \bicep\ data is included then $[\zeta_s, \zeta_d]=[-0.39,3.70] \pm [0.07, 0.11]$.  For reference, FDS model 8 predicts $\zeta_{d} = 3.5$ in the frequency range 23 - 220 GHz. From signal simulations, the uncertainty in systematic filter correction could be as large as 5\%; however, this factor has a relatively minor effect on the fit.  For example, a 5\% uncertainty in this correction factor leads to an uncertainty in $\zeta_d$ of $\sigma_{\zeta_d}=1\%$.  This reduced sensitivity to input uncertainty is due partially to the non-linear nature of the fitting model but also due to the \wmap\ points not suffering from \bicep's filtering bias. Repeating the fit procedure with only one set of boresight maps, causes a change in $\zeta_d$ of $\sigma_{\zeta_d}=2\%$.  Splitting the map into the Gal-bright and Gal-weak regions gave $\zeta_d=[3.81, 3.27]$ with $\sigma_{\zeta_d}=[0.14, 0.15]$ indicating a difference in dust spectral index between the two regions at 2.7 sigma significance.  

Therefore, in this case, the uncertainty in this fit itself is larger than the uncertainty due to systematic contamination from weather or telescope systematics which is larger than the uncertainty due to systematic filtering bias uncertainty.  Analyzing the \wmap\ DAs separately is a way to include systematics from that experiment.  However, the separation between points in any \wmap\ band (See Figure~\ref{fig:medtvsfreq}) is smaller than the uncertainty in the average value from that band, making systematics minimally important.

The simple two-component model fits the data well; however, additional components are not excluded.  The plot shows the importance of measurements at higher frequency bands to fully understand the composition of the ISM.  At 95 GHz, using the two-component fit parameters, dust makes up 51\% of the total emission, while at 150 GHz it makes up 87\% and at 220 GHz it makes up 97\%.  \cite{Fink2003} gives a template for the full sky free-free emission based on $H_\alpha$ measurements.  Taking their input map at 0.0625$\deg$ resolution, downsampled to 1$\deg$ resolution, converted to MJy/sr using the factors given in their table 1 at 100 GHz, the average map pixel value in the 220 GHz analysis region is 0.004 MJy/sr.  Therefore, free-free is approximately 40 times smaller than either the dust or synchrotron component in \bicep\ maps.  Further measurements at intermediate frequencies with finer spacing would also provide useful information about the transition from the synchrotron to dust dominated regime and give more information about other possible emission sources.

% \clearpage

\subsection{Polarization Fraction vs. Intensity} \label{sec:flatvsslope}

The Galactic magnetic field cannot be measured directly, therefore measurements of the field's effect on the intervening interstellar matter are crucial.  The Galactic magnetic field produces polarization on large and small scales across the electromagnetic spectrum.  Millimeter-wave polarization measurements probe emission that spans the full extent of the Galactic plane making it possible to exclude or motivate models for the Galactic magnetic field structure.

\subsubsection{Polarization Fraction Data} \label{sec:qfracdat}
\begin{figure} % Figure 14
  \begin{center}
    \resizebox{\columnwidth}{!}{\includegraphics{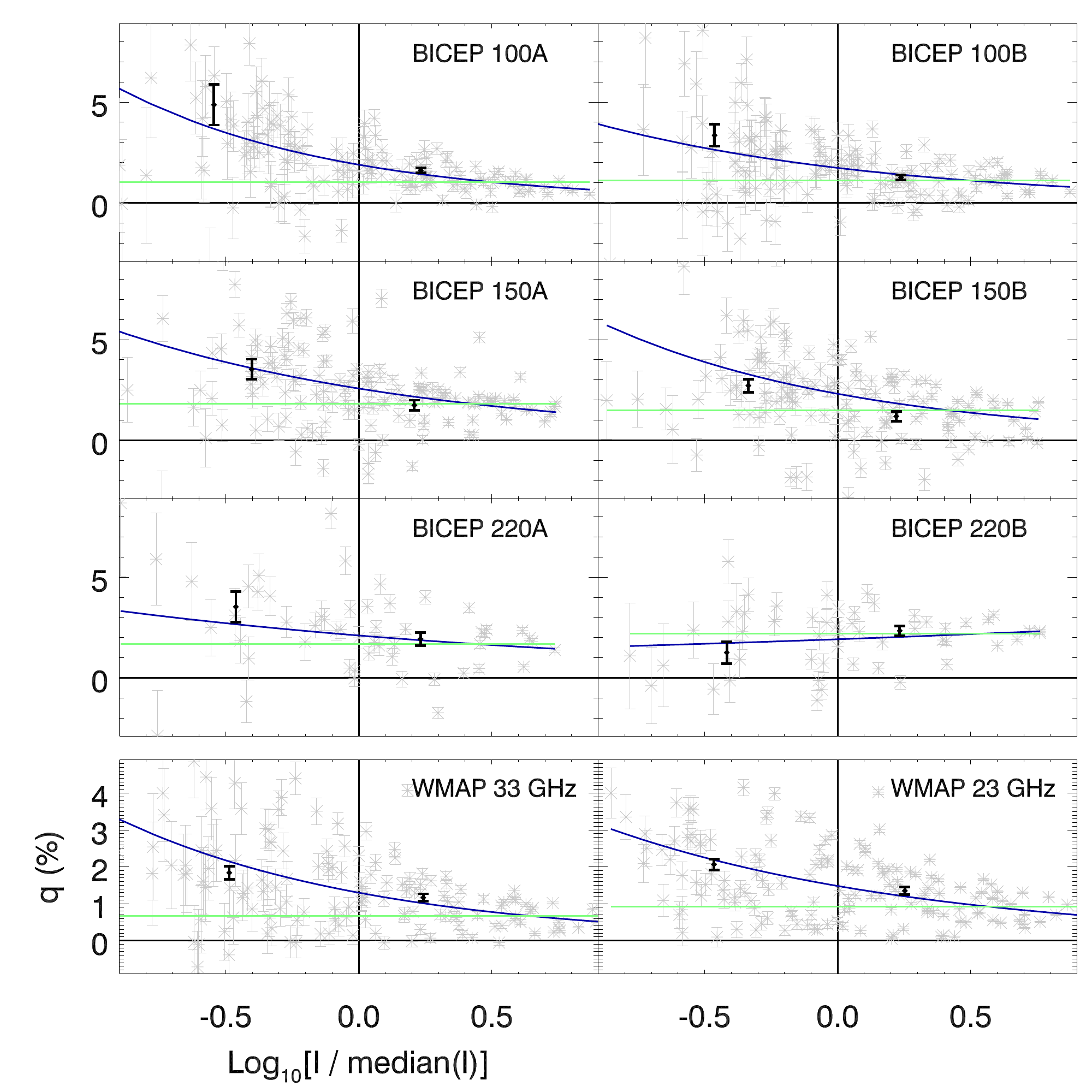}}
  \end{center}
   \caption{Polarized fraction, $q$, as a function of total intensity, $I$, using 147 one degree \healpix\ pixels (gray) from \bicep's three bands and two boresight maps along with \wmap\ 23 and 33 GHz bands.  Out of the 147 pixels, 53 were used for the \bicep\ 220 GHz detectors because of the reduced observed region available at that band. The error bars are those due to the white noise in the maps and do not account for systematic errors due to the atmosphere, instrumental systematics, or filtering.  The best fit line (blue) and constant model (green) are shown.  The weighted average and uncertainty (black) of $q$ for pixels less than the median value and greater than the median value clearly show the trend of decreasing polarization fraction as intensity increases.  All \bicep\ maps detect this trend with greater than 98\% confidence; except \bicep\ 220 GHz, which has ten times fewer feeds than \bicep's 100 or 150 channels. Two \wmap\ channels are also shown for comparison.}
\label{fig:qtscatter}
 \end{figure}

One prediction of most Galactic magnetic field models is a trend of observed polarization fraction as a function of unpolarized intensity.  To determine if there is a trend and what the nature of the trend may be,  Figure~\ref{fig:qtscatter} shows scatter plots of $q$ vs. $I$ for the various maps.  The plotted map pixels for both \bicep\ and \wmap\ are from the analysis area and the error bars are calculated using the noise per band or DA and the integration time per pixel.  Over-plotted on the scatter plots is the weighted linear least-squares fit following Equation~\ref{eq:qvslogt} and the weighted mean $q$.  Since the white noise floor value is used, all of the fits had reduced-$\chi^2$ values much greater than one, indicating, to some extent, the total absolute uncertainty, including genuine variation about this trend across the sky.  To better approximate the total absolute uncertainty in the resulting fit parameters, the uncertainty for each fit parameter is multiplied by the $\sqrt{\chi_{reduced}^{2}}$, which are the error bars used in Figure~\ref{fig:mvsfreq}.

Figure~\ref{fig:qtscatter} may not sufficiently support the hypothesis that there is a significant trend of decreasing polarization fraction with increasing intensity as opposed to a constant model for $q$.  The difficulty is that the error in polarization fraction increases as the intensity value decreases because the uncertainty in $q$ is proportional to $I^{-1}$, weighting the $q$ values measured more strongly at higher $I$.  The fact that there is a slope detected for all bands from 23 GHz to 150 GHz favors a sloped model over a flat one (See Table~\ref{tab:qifitpar}).

\subsubsection{Model Fit Parameters} \label{sec:qfracmodelfit}
\begin{figure} % Figure 15
   \begin{center}
     \resizebox{\columnwidth}{!}{\includegraphics{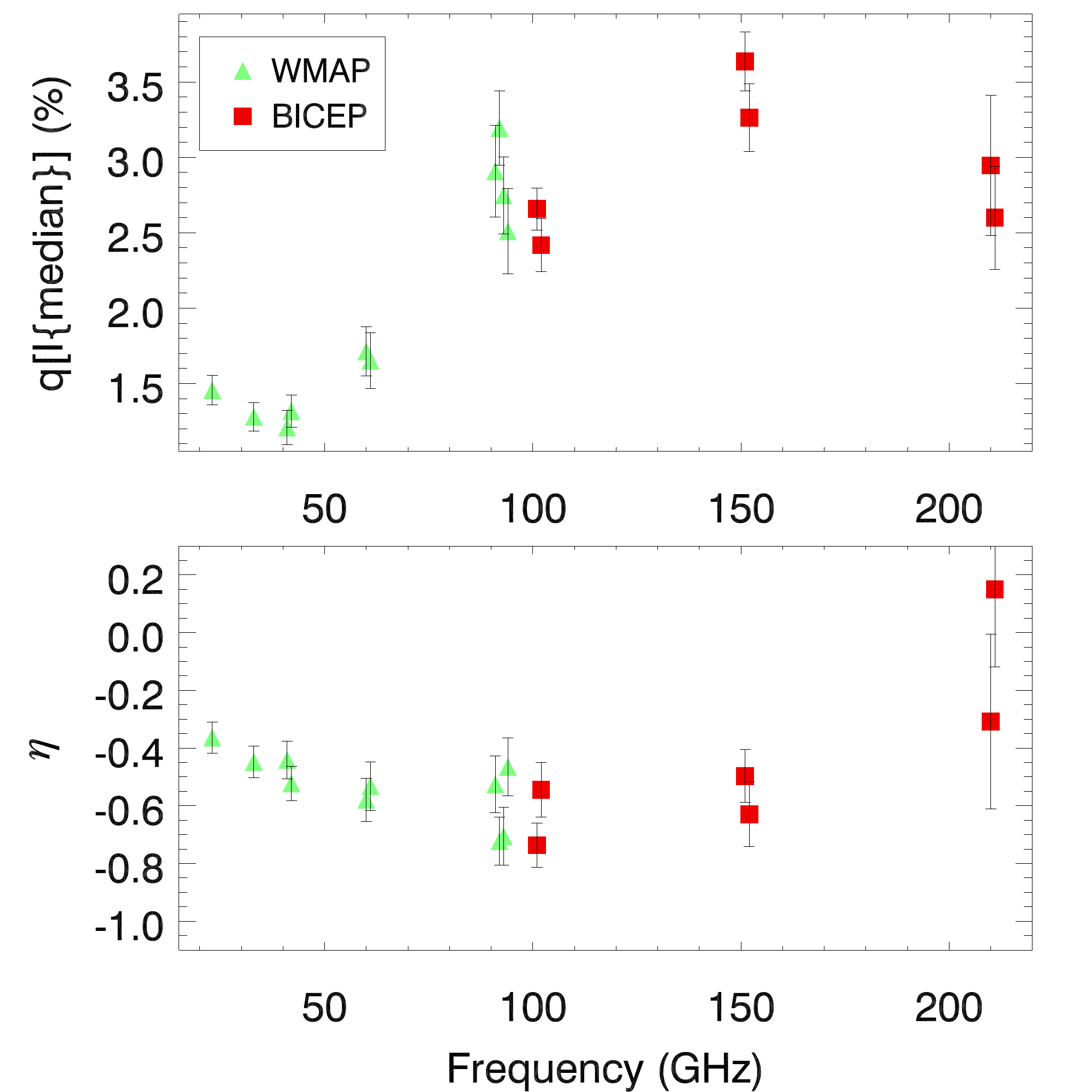}}
   \end{center}
 \caption{$q_{(I_{median}) }$ and $\eta$ as a function of frequency using \bicep\ (red) and \wmap\ (green) data.  All frequencies use the 100/150 GHz analysis region except for 220 GHz, which uses pixels from the smaller 220 GHz analysis region.  The error bars are the quadrature sum of fitting errors and the uncertainty from atmospheric conditions.  \bicep's points have been corrected for filtering effects, increased by 37\% and 39\% for $q$ and $\eta$ respectively, based on signal simulations.  The trend of increasing $q$ with increasing frequency is apparent.  \bicep\ has two data points per band that are computed from the two boresight maps.  For $q_{(I_{median}) }$ the two data points are close to each other relative to the error bars, while for $\eta$, the separation is close to the same size as the error bars.  This indicates the general level of systematic uncertainties is not the dominant uncertainty for $q_{(I_{median}) }$, consistent with the discussion in Section~\ref{sec:maperrors}, but is approaching an unacceptable level for $\eta$.  \wmap\ points are from each DA separately, which can give some idea of possible systematic uncertainty in a given band.}
\label{fig:mvsfreq}
 \end{figure}

Figure~\ref{fig:mvsfreq} shows a plot of the fit parameters from Equation~\ref{eq:qvslogt}, and Table~\ref{tab:qifitpar} lists these average parameters with uncertainties from three sources and band properties.  One type of uncertainty is from the nonlinear regression covariance matrix, and these uncertainties have been increased by $\sqrt{\chi_{reduced}^{2}}$ to better approximate the true parameter uncertainty and include other noise sources such as genuine spatial variation across the sky.  This type of error can be decreased by including more map pixels (possible at 100 and 150 GHz, but not at 220 GHz) or by increasing integration time or detector sensitivity.

Second, there is uncertainty due to the dispersion of different maps within a band, the two boresight maps for \bicep\ and the various DAs for \wmap.  This uncertainty, especially for \bicep, indicates the approximate level of systematic contamination from the atmosphere, instrument thermal properties, and telescope systematics other than spectral gain mismatch.  This type of uncertainty is prevalent at all three bands and hinders further quantitative analysis of $\eta$.  

Third, there is uncertainty from spectral gain mismatch that is quantified from simulations in Section~\ref{sec:specchar} using different atmospheric conditions.  This type of uncertainty compared to the other two is only important at 220 GHz.  Overall, there is a strong detection of $q_{(I_{median}) }$ at all three bands and a significant detection of $\eta$ at 100 and 150 GHz.  There is not a detection of $\eta$ at 220 GHz due to the combination of all three types of uncertainties exacerbated by the fact that \bicep\ only has two 220 GHz feeds for only two of the three observing seasons (See Appendix~\ref{sec:s220} for more information).  Fitting for $\eta$ is somewhat biased by the uncertainty in each pixel being proportional to $I^{-1}$, as the higher intensity pixels tend to dominate the fit.  For example, it is difficult to tell whether a third parameter is needed to fit the data properly or whether there are separate effects happening for lower intensity pixels that some models would predict.

The top plot of Figure~\ref{fig:mvsfreq} shows a trend of increasing $q_{(I_{median}) }$ vs. frequency with a sharp increase between 60 and 100 GHz.  At higher frequencies where dust emission dominates the polarization fraction is above 2.5\%, rather than below 1.5\% at lower frequency where synchrotron emission dominates.  Comparing the polarization fraction levels for the \wmap\ maps found here to Figure 5 from \cite{Kogut2007}, which shows a histogram of polarization fraction from pixels within the Galactic plane, the pixels chosen here are at the very low end of the distribution of polarization fractions within the Galactic plane.  This is consistent with the choice to only analyze regions very close to zero Galactic latitude that are relatively close to the Galactic center.  It's important to note that unpolarized emission mechanisms such as free-free emission contribute to the measured intensity but not the polarization.  This can lower the measured polarization fraction at bands where free-free contributes significantly, although that is not the case for \bicep\ bands and sky coverage region.

The bottom plot in Figure~\ref{fig:mvsfreq} shows $\eta$ as a function of frequency; however, no model of this function is proposed in this paper.  One important point is that $\eta$ changes the relative positions of the points in $q$ vs. frequency.  For example, since the average $\eta$ value at 220 GHz is close to zero, the value of $q$ at 220 GHz is relatively constant as a function of total intensity.  Therefore, if $q$ were evaluated at a very large value of total intensity and the plot of $q$ vs. frequency remade, then it would appear that the polarization fraction increases relative to 150 GHz, even though the actual polarization fraction value at 220 GHz was nearly constant.

The weighted mean and standard error of $\eta$ across all bands, assuming the \bicep\ data points are corrected for filtering, is $\eta=-0.47 \pm 0.05$.  A simple, flat polarization model,$\eta=0$,  is ruled out as is the simple toy model of unpolarized embedded sources in a uniformly polarized background in the Galaxy, from Section~\ref{sec:polynoise}, $\eta=-1$.  A  more complicated Galactic magnetic field model is needed to explain the measured value.  A comparison of $q$ and $\eta$ between the Gal-bright and Gal-weak regions was performed, but there was no detectable difference between the two regions.

\begin{figure} % Figure 16
   \begin{center}
     \resizebox{\columnwidth}{!}{\includegraphics{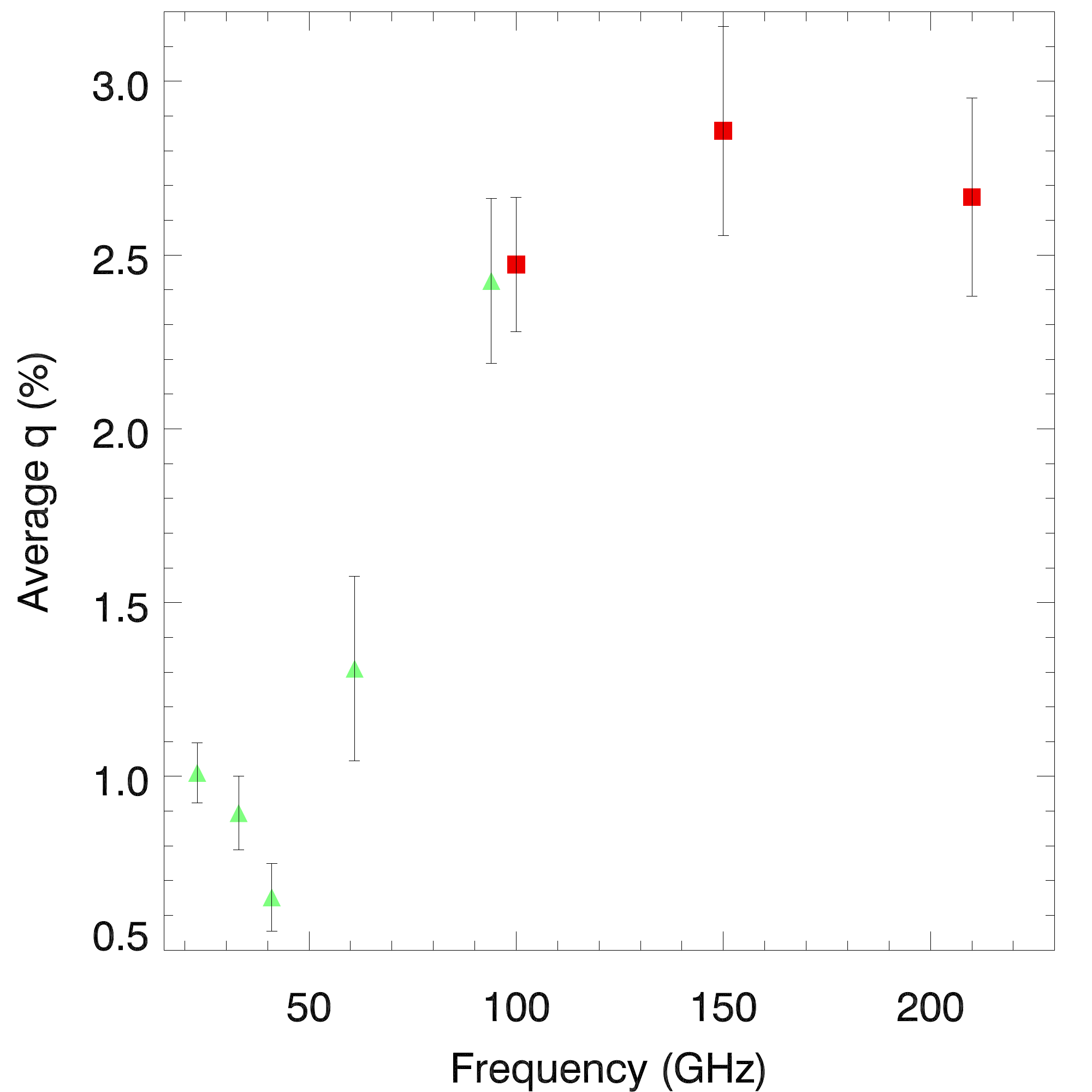}}
   \end{center}
 \caption{The average and error of $q$ from the 220 GHz analysis region pixels combining all the data at each \bicep\ (red) and \wmap\ band.  This shows a different method of computing $q$ vs. frequency as compared to Figure~\ref{fig:mvsfreq}, which uses a model to find the value of $q$, splits the data by boresight angle or DA, and uses a different numbero f pixels at 220 GHz.  Once again a general trend of increasing polarization fraction vs. frequency is found with the minimum $q$ occurring near 40 GHz.  There is a steep increase in polarization fraction between 60 GHz and 100 GHz with the minimum fraction occurring near 40 GHz.}
\label{fig:qmeanvsfreq}
 \end{figure}

An inherent problem in analyzing the $q$ vs. frequency plot is that it depends at what intensity value $q$ is evaluated.  While evaluating $q$ at the median intensity gives a $q$ value at a moderate level of intensity, other methods, such as computing the average of $q$, effectively results in finding $q$ at the weighted mean intensity, as in Figure~\ref{fig:qmeanvsfreq}.  The average has two advantages: a model including $\eta$ is not needed and no value of $I$ needs to be chosen at which to evaluate $q$.  For this plot, the average from all the data is shown, neglecting different boresight angles and DAs.  As opposed to Figure~\ref{fig:mvsfreq}, this figure uses only map pixels from the 220 GHz region, simplifying the comparison of 220 GHz vs. the other bands.  While the minimum $q$ occurs near 40 GHz in this plot, depending on what value of intensity $q$ is evaluated, this frequency can change.  The polarization fraction approximately quadruples from 41 GHz to 95 GHz showing a relatively large dependence on frequency.  No trend in polarization fraction above 100 GHz is visible, although the error bars do not exclude this possibility.  In the end, Figure~\ref{fig:mvsfreq} demonstrates the systematics levels are small compared to the general trend of $q$ vs. frequency as is evident from either Figure~\ref{fig:mvsfreq} or Figure~\ref{fig:qmeanvsfreq} (even if the error bars are larger for Figure~\ref{fig:qmeanvsfreq} because fewer map pixels were used in the average).

\begin{table*}

  \caption{Raw fit parameters for $q$ vs. $I$}
   {Average fit parameters and errors for Equation~\ref{eq:qvslogt} using map pixels from \wmap\ and \bicep\ in the 100/150 GHz analysis region (except for 220 GHz which uses pixels from the 220 GHz analysis region).  The \bicep\ fit parameters have been corrected for filtering effects (Section~\ref{sec:polyfiterror}).  The systematic errors ``A'' for the \bicep\ $\eta$ parameters come from atmospheric uncertainty on spectral gain mismatch (Section~\ref{sec:specchar}).  The statistical errors for both \bicep\ and \wmap\ come from the non-linear fits, while the systematic errors, ``B'' for \bicep\ come from the two boresight maps and for \wmap, the different DAs.  The median intensities are set by taking the median value within the analysis region (53 pixels) at a given band and rounded to two significant figures.}
  \begin{center}
   \begin{tabular*}{0.8\textwidth}{lcccccccc}
     \hline
     \hline
     Experiment Channel  & Frequency (GHz) & $I_{median}$ ($\muK$)& $q_{(I_{median}) }$ (\%)    & $\Delta q$ (\%)& $\eta$   & $\Delta \eta_{stat}$  & $\Delta \eta_{sysA}$  & $\Delta \eta_{sysB}$   \\
     \hline
     \wmap\ K            & 23              & 8500                & 1.5                  & 0.10             & -0.36    & 0.05                 & N/A                   & N/A                   \\
     \wmap\ Ka           & 33              & 3400                & 1.3                  & 0.10             & -0.45    & 0.06                 & N/A                   & N/A                   \\
     \wmap\ Q            & 41              & 1900                & 1.3                  & 0.11             & -0.48    & 0.06                 & 0.04                  & N/A                   \\
     \wmap\ V            & 61              & 880                 & 1.7                   & 0.17             & -0.56    & 0.08                 & 0.02                  & N/A                   \\
     \wmap\ W            & 94              & 800                 & 2.8                   & 0.27             & -0.60    & 0.10                 & 0.06                  & N/A                   \\
     \bicep\ 100         & 95.5            & 800                 & 2.5                   & 0.16             & -0.64    & 0.09                 & 0.10                  & 0.041                 \\
     \bicep\ 150         & 149.8           & 1400                & 3.4                   & 0.20             & -0.56    & 0.10                 & 0.07                  & 0.063                 \\
     \bicep\ 220         & 208.2           & 4400                & 2.8                   & 0.40             & -0.08    & 0.29                 & 0.23                  & 0.24                  \\
     \hline
   \end{tabular*} 
  \end{center}
\label{tab:qifitpar}
 \end{table*}

\subsection{Polarization Angles} \label{sec:propvsfreq} 

\begin{figure} %Figure 17
   \begin{center}
     \resizebox{\columnwidth}{!}{\includegraphics{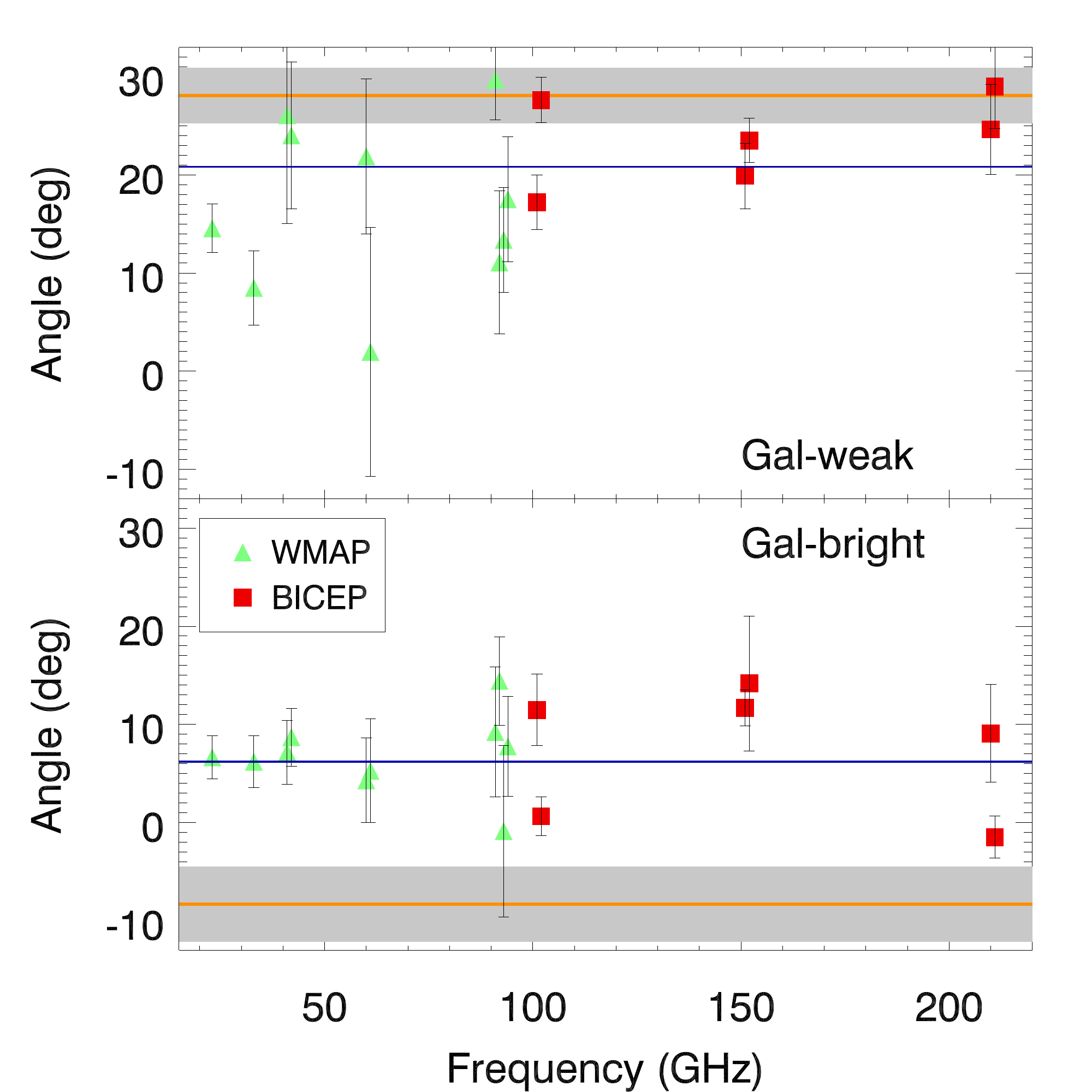}}
   \end{center}
 \caption{Polarization angle as a function of frequency for \bicep\ (red squares) and \wmap (green triangles) for both observing regions with the average angle for each region over-plotted (blue).  Also shown are the average starlight polarization angles (orange) and uncertainties in the average values (gray shaded region) in the Gal-weak region and Gal-bright regions.  The polarization angles do not have any trend as a function of frequency in either region showing consistency between the two experiments and different emission mechanisms.  However, the angles for the Gal-bright region are systematically lower than for the Gal-weak region.  The starlight angles show a similar change in polarization direction between the two regions in agreement with \bicep\ and \wmap.    This rotation of polarization angle between the two regions indicates large scale structure in the Galactic magnetic field.  There is good agreement between optical and millimeter-wave polarization angles in the Gal-weak region but poor agreement in the Gal-bright region.  The \bicep\ angles have not been corrected for filtering effects.  The two \bicep\ boresight maps per band show the general level of systematic uncertainty due to excess noise and telescope systematics.  \wmap\ points are from each DA separately, which can be indicative of possible systematic uncertainty in a given band.}
\label{fig:angvsfreqvsreg}
 \end{figure}

The Galactic magnetic field generally causes polarization angles to be perpendicular to the Galactic plane in the millimeter regime.  However, \bicep\ and \wmap\ measurements show the magnetic field is not exactly parallel to the plane and changes direction between the two observing regions (Figure~\ref{fig:angvsfreqvsreg}).  The analysis area used throughout this analysis are further split up into Gal-bright (27 pixels) and Gal-weak (26 pixels) and the weighted average polarization angle for each \bicep\ boresight maps and \wmap\ DA is found.  The average polarization angle in the Gal-bright region from both \bicep\ and \wmap\ is $6.2\deg \pm 1.7\deg$ whereas the average polarization angle in the Gal-weak region is $20.8\deg \pm 2.2\deg$ for both \bicep\ and \wmap.  This represents a 5 sigma detection of a difference in polarization angle between the two regions.  

\bicep\ and \wmap\ polarization angles can also be compared to starlight polarization data.  Millimeter-wave dust polarization is due to emission from particles that are preferentially aligned perpendicular to the local magnetic field~\citep{Lazarian2007}.  A complimentary process takes place in the optical band due to absorption, leading to polarization angle that should be rotated 90$\deg$ relative to the millimeter-wave polarization. ~\cite{Heiles2000} compiles a table of stellar polarization measurements and characteristics that can be used to compare polarimetry from the optical and millimeter-wave bands.  Out of the 9286 stars in \cite{Heiles1999}, only those with angle errors less than 10$\deg$ that are within the \bicep\ observing region were considered.  For direct comparison, the starlight polarization angles have the expected 90$\deg$ difference subtracted off.  There were 36 stars with starlight polarization measurements in the Gal-bright region giving a weighted average polarization angle of $-8.3\deg \pm 3.8\deg$ while there were 24 stars in the Gal-weak region giving a weighted average polarization angle of $28.1\deg \pm 2.8\deg$.  One caveat is that the starlight polarization measurements do not exactly track \bicep's and \wmap's measurements in space, as there are multiple stars in some map pixels and none in others.

Since the Gal-bright and Gal-weak regions are separated by 60$\deg$ in Galactic longitude, the change in polarization direction indicates the Galactic magnetic field has structure on very large scales.  Further small-scale analysis is possible, but results are inconclusive due to the uncertainty in \bicep's current polarization angle measurements.  As opposed to $q$, there is no difference in the average polarization angle between pixels whose intensity is less than the median intensity and greater than the median intensity.  No trend in polarization angle as a function of frequency is detected, showing consistent, independent polarization angle measurements from different lower frequencies where synchrotron dominates to high frequency where dust emission dominates.

\subsection{Visual Optical Depth}\label{sec:visoptdep}

In order to compare underlying astronomical objects from the optical to millimeter-wave band, the intensity of the emission can be converted to a common unit used in studies of the interstellar medium, visual optical depth, $A_V$.  The conversion can be calculated as:
\begin{eqnarray}
  \frac{dA_V}{dT}&=&\frac{dA_V}{d\tau} \times \frac{d\tau}{dI} \times \frac{dB}{dT} \nonumber
\end{eqnarray}
where the relationship between the emitted intensity and optical depth, assuming an optically thin medium, is given by:
\begin{eqnarray}
  I&=&B (1-\exp^{-\tau})\approx B \tau.\nonumber
\end{eqnarray}
 A relationship to millimeter-wave optical depth, $\tau(\lambda)$, derived by ~\cite{hildebrand1983} and~\cite{dickman1978} is:
\begin{eqnarray}
  A_V&\approx&~\mathrm{1900} \ \tau(\lambda)\ \Big( \frac{\lambda}{\mathrm{250~um}} \Big)^2 \nonumber
\end{eqnarray}
Assuming a dust temperature of 20 K, the conversion factor is calculated as:
\begin{eqnarray}
  \frac{dA_V}{dT}&\approx&~\mathrm{1900}~ \Big( \frac{\lambda}{\mathrm{250~um}} \Big)^2 \times \frac{1}{B_{(\mathrm{20~K})}} \times \frac{2 k_B \nu ^ 2}{c ^ 2} \frac{z ^ 2 e ^ z}{ (e ^ z - 1) ^ 2}.   \nonumber
\end{eqnarray}
Substituting in the values of the constants, wavelengths, and \bicep\ band centers this gives conversion factors of:
\begin{eqnarray}
  \frac{dA_V}{dT}&\approx&~{14, 4.3, 1.5}~ \frac{\mathrm{A}_{\mathrm{V}}}{\mathrm{mK}_{\mathrm{CMB}}},
\label{eq:avtemp}
\end{eqnarray}
for 100, 150, and 220 GHz respectively.  One caveat to this calculation is that the factors are only valid for dust emission.  Section~\ref{sec:ivsfreq} showed that at 95 GHz,  51\% of the emission comes from other sources such as synchrotron radiation.  Using the factors of [0.51, 0.87, 0.97] from Section~\ref{sec:ivsfreq}, intensity maps from \bicep\ can be converted to $A_V$ for the dust component in the map as:
\begin{eqnarray}
  \frac{dA_V}{dT_{BICEP}}_{Dust}&\approx&~{7.14, 3.74, 1.45}~ \frac{\mathrm{A}_{\mathrm{V}}}{\mathrm{mK}_{\mathrm{CMB}}},
\label{eq:avtempdust}
\end{eqnarray}
For 100, 150 and 220 GHz.  

\bicep\ detections span a thermodynamic temperature brightness range from [0.1, 0.3, 0.7] to [6, 8, 25] $\mathrm{mK_{CMB}}$  for 100, 150, and 220 GHz respectively.  This corresponds to integrated visual optical depths $A_V$, using Equation~\ref{eq:avtempdust}, from [0.7, 1.1, 1.0] to [42, 30, 36].  On average, \bicep\ probes a component of the ISM which is more diffuse than star-forming regions ($A_V \gtrsim 10$), but is more dense than the medium sampled by optical polarimetry ($A_V \lesssim 2$).  Due to the relative beam sizes and dust spectrum, \bicep\ measurements probe approximately the same density medium at local intensity maxima at all three bands.  In principle, experiments such as QUaD and Archeops have the potential to probe even denser cloud complexes because of their smaller beam sizes; however, Archeops had higher noise levels than \bicep\ and both experiments have so far presented their polarization data with angular resolution $\approx$ 0.5$\deg$, similar to \bicep.

%%%%%%%%%%%%%%%%%%%%%%%%%%%%%%%%%%%%%%%%%%%%%%%%%%%%%%%%%%%%%%%%%%%%%%%%%%%%%%%%%%%%%%%%%%%%%%%%%%%%%%%%%%%%%%%%%%%%%%%%%%%%%%%%%%%%%%%%%%%%%%%%%%%%%%%%%%%%%%%%%%%%%%%%%%%%%%%%%%%%%%%%%%%%%%%%%%%%%%%%%%%%%%%%%%%%%%%%%%%%%%%%%%%%%%%%%%%%%%%%%%%%

\section{Discussion}\label{sec:discussion}

Continuum polarimetry results can be interpreted with the aid of a model in which the magnetic field for a given patch of sky is nearly constant in magnitude but has an angular structure which is a superposition of a uniform and a random component~\citep{Jones1989,Miville2007}.  The very simplest models, such as a completely uniform magnetic field or a completely random one, are ruled out by \bicep\ and \wmap\, which show a statistically significant trend of decreasing $Q$ polarization fraction vs. increasing intensity (Section~\ref{sec:flatvsslope}). 

The detection of dust polarization with \wmap, QUaD, Archeops and \bicep\ constrains the degree of order in the Galactic magnetic field on large scales.  An ordered magnetic field nearly parallel to the plane is detected at all millimeter-wave frequencies.  The mean polarization angles are nearly constant as a function of frequency, but systematically change direction between \bicep's two Galactic regions.  This change in direction is also found in starlight polarization measurements.  This change shows structure in the Galactic magnetic field on scales greater than 60$\deg$ in Galactic longitude.

The average degree of polarization observed in the integrated emission from star-forming cores ($A_V \gtrsim 30$) $p\lesssim 1.5\%$\cite{Stephens2010}, is similar to the degree of polarization observed with \bicep\ (Section~\ref{sec:flatvsslope}).  However, this coincidence does not imply that the observed \bicep\ polarization arises from a superposition of unresolved star-forming cores, with no significant polarized component emitted by the diffuse medium.  While there is some evidence for coherence in the magnetic field across star-forming molecular cloud complexes up to 100 pc in size \citep{Li2009}, as a whole, those complexes and cores have a nearly random distribution of magnetic field directions and there is no evidence for coherence in the dense medium on larger scales~\citep{Glenn1999,Stephens2010}.  The contribution from these star-forming complexes to \bicep\ maps would consist of polarization with a high angular disorder, averaging to very low polarization when the beam encompasses multiple complexes.  For example, a \bicep\ 150 GHz 0.6$\deg$ beam corresponds to 50 pc for a typical Galactic source at a distance of 5 kpc in the Gal-bright field, and hence poorly resolves molecular cloud complexes.  Consequently, because \bicep\ observes a substantial degree of polarization over the whole Galactic plane at $\sim$1$\deg$ resolution on average, \bicep\ must sample a medium outside star-forming cores, one with an embedded magnetic field that retains a significant component ordered on the Galactic scale. 

On the other hand, \bicep\ rules out a model where no additional polarization intensity comes from the higher density ISM.  The polarization fraction drops somewhat as surface brightness increases, $q \propto I^{-0.47}$, but the decline is more shallow than $q \propto I^{-1}$ as predicted by the toy model of Section~\ref{sec:polynoise} having bright unpolarized sources embedded in a polarized medium.  Therefore, \bicep\ measurements probe aligned grains and magnetic fields over the full range of observed column densities approaching star-forming cores.  \cite{Jones1989} and \cite{Fosalba2002} find an equivalent power law exponent from starlight data closer to $\eta_{em}=-0.25$, implying a more ordered magnetic field than what is implied from millimeter-wave measurements.  A caveat to this comparison is those previous works analyzed total polarization fraction $p$, which does not necessarily have the same functional relationship as $q$.  No attempt to directly model $q$ using magnetic fields is included in this paper.

The detection of statistically significant polarization across the Galactic plane at all intensity levels shows that there are aligned dust grains across the entire map.   The $\eta$ parameter indicates the nature of the Galactic magnetic field and the disorder within the beam and along the line of sight.  The polarization fraction $q$ increases as a function of frequency, which was somewhat unexpected; other work has indicated that the opposite can be true at higher latitudes due to the much larger polarization fraction intrinsic to the synchrotron emission compared to dust polarization~\citep{Kogut2007}.  In contrast, \bicep\ has shown that dust emission can have a higher polarization fraction in the Galactic plane.  This could indicate unpolarized emission mechanisms such as free-free play a more important role in the map regions analyzed here or it could be an indication that material emitting synchrotron radiation exists in more randomized Galactic magnetic field regions on average, as opposed to dust grains which may lie in more ordered field regions.

The observations discussed here are consistent with a Galactic magnetic field whose structure and order vary inversely with density.  The increasing disorder in denser components of the ISM may be due to gravitational or dynamical accumulation of gas and/or feedback from star formation.  The high angular resolution and sensitivity of the Planck~\citep{Tauber2010} 350 GHz polarization channels should be able to map this structure with greater precision to discriminate among more detailed models for the neutral ISM.

\section{Conclusion}\label{sec:conclusion}
\bicep's Galactic observations have added new information and insight into Galactic physics, while also confirming previous measurements.  \bicep\ samples an intermediate optical depth of the ISM and polarization is detected everywhere within a two degrees of the Galactic plane with values ranging up to a few percent at 100, 150, and 220 GHz.  \bicep\ detects a significant trend of decreasing polarization fraction as intensity increases in the maps and increasing polarization fraction as a function of frequency.  Polarization angles were found to be consistent from 23 GHz to 220 GHz and in general agreement with polarization angles measured by stars in out observing regions.  Adding 220 GHz capability to \bicep\ helped to refine the understanding of in-plane foreground emission.  Improvements in \bicep's scan strategy and data analysis pipeline could result in increased mapping efficiency.  \bicep\ data have shown astronomical foregrounds to be complex, and simply modeling polarization as a percentage of unpolarized intensity is insufficient.  While polarized foreground models in the Galactic plane are becoming more precise, less is known at higher latitudes and it is not obvious that one can extrapolate models from the plane to these latitudes.  Future CMB polarimeters are poised to build upon \bicep's results.

%%%%%%%%%%%%%%%%%%%%%%%%%%%%%%%%%%%%%%%%%%%%%%%%%%%%%%%%%%%%%%%%%%%%%%%%%%%%%%%%%%%%%%%%%%%%%%%%%%%%%%%%%%%%%%%%%%%%%%%%%%%%%%%%%%%%%%%%%%%%%%%%%%%%%%%%%%%%%%%%%%%%%%%%%%%%%%%%%%%%%%%%%%%%%%%%%%%%%%%%%%%%%%%%%%%%%%%%%%%%%%%%%%%%%%%%%%%%%%%%%%%%

\acknowledgments

  We would like to especially recognize Andrew Lange whose passing has deeply affected everyone on the team and who never saw the culmination of this project. Without his guidance and friendship this paper would not have been possible. 

\bicep\ is supported by NSF Grant OPP-0230438, Caltech Discovery Fund, Caltech President's Fund PF-471, JPL Research and Technology Fund, and the late J. Robinson. We thank our colleagues in the ACBAR, Boomerang, QUaD, Bolocam, POLARBEAR, and SPT collaborations for advice and helpful discussions, Kathy Deniston for logistical and administrative support, and the South Pole Station staff for their support. We gratefully acknowledge support by the NASA Graduate Fellowship program (E.M.B. and H.C.C.), the John B. and Nelly Kilroy Foundation (J.M.K.), the NSF PECASE Award AST- 0548262 (B.G.K.), the U.S. DOE contract to SLAC No. DE-AC02-76SF00515 (C.L.K. and J.E.T.), KICP (C.P. and C.S.), and the NASA Science Mission Directorate via the US Planck Project (G.R.). \\

%%%%%%%%%%%%%%%%%%%%%%%%%%%%%%%%%%%%%%%%%%%%%%%%%%%%%%%%%%%%%%%%%%%%%%%%%%%%%%%%%%%%%%%%%%%%%%%%%%%%%%%%%%%%%%%%%%%%%%%%%%%%%%%%%%%%%%%%%%%%%%%%%%%%%%%%%%%%%%%%%%%%%%%%%%%%%%%%%%%%%%%%%%%%%%%%%%%%%%%%%%%%%%%%%%%%%%%%%%%%%%%%%%%%%%%%%%%%%%%%%%%%
%%%%% References %%%%%

\bibliographystyle{apj}
\bibliography{ms}{}
%\bibliography{galpap}{}

%%%%%%%%%%%%%%%%%%%%%%%%%%%%%%%%%%%%%%%%%%%%%%%%%%%%%%%%%%%%%%%%%%%%%%%%%%%%%%%%%%%%%%%%%%%%%%%%%%%%%%%%%%%%%%%%%%%%%%%%%%%%%%%%%%%%%%%%%%%%%%%%%%%%%%%%%%%%%%%%%%%%%%%%%%%%%%%%%%%%%%%%%%%%%%%%%%%%%%%%%%%%%%%%%%%%%%%%%%%%%%%%%%%%%%%%%%%%%%%%%%%%%%%%%%%%%%%%%%%%%%%%%%%%%%%%%%%%%%%%%%%%%%%%%%%%%%%%%%%%%%%%%%%%%%%%%%%%%%%%%%%%

\begin{appendix}

\section{\bicep\ 220 GHz channels} \label{sec:s220}
Prior to the second observing season, two feeds designed to observe through the 220 GHz atmospheric transmission window were installed into \bicep.  These feeds represent the first attempt to observe CMB and Galactic polarization in this frequency band using PSBs.  Due to the relative sensitivity of a 220 GHz channel to thermal dust emission and its polarization over lower frequency channels, the 220 GHz feeds serve a three-fold purpose.  First, to constrain the polarization and intensity of the three main millimeter-wave sources (synchrotron, thermal dust, and the CMB) solely from \bicep\ observations, without an ad hoc foreground model, requires three frequency bands.  Second, thermal dust emission and its polarization increase understanding of the physics of the Galaxy.  Finally, it was unclear whether the CMB B-modes, \bicep's primary science target, would be contaminated by foreground emission, and would need to be modeled and removed.  The 220 GHz feeds can carry more weight, per feed, than the other two bands for measurement of dust polarization due to the steep spectral index of millimeter-wave dust emission.  More 220 GHz feeds were planned to be added in 2008 but logistical constraints at the South Pole in 2008 ultimately prevented this, limiting the experiment to using two feeds for the two final observing seasons.

Emission from the Galaxy has the potential to contaminate the faint B-mode signal in all observing fields.  Since dust emission follows a thermal spectrum, by monitoring polarized interstellar dust at higher frequency than the primary B-mode bands (100 and 150 GHz), it is possible to monitor dust contamination with higher signal-to-noise per feed.  While \bicep's CMB region, located at high Galactic latitude, certainly exhibits minimal dust column-density and therefore low-intensity millimeter-wave emission, this doesn't imply it will exhibit low-polarized emission due to the details of magnetic grain alignment~\citep{Lazarian2007}.  Little is known about the polarization properties of thermal dust emission in the high-frequency bands above 100 GHz used for CMB observations.  \bicep's 220 GHz feeds are a unique link between the low-frequency CMB polarimetry of \wmap\ and the high-frequency survey of the Archeops experiment; from centimeter wavelengths to the sub-millimeter. The combination of \wmap, QUaD, \bicep, and Archeops polarization observations of the Galactic plane cover more than a decade of frequency.

\begin{figure}  %figure 17
  \begin{center}
  \resizebox{0.5\columnwidth}{!}{\includegraphics{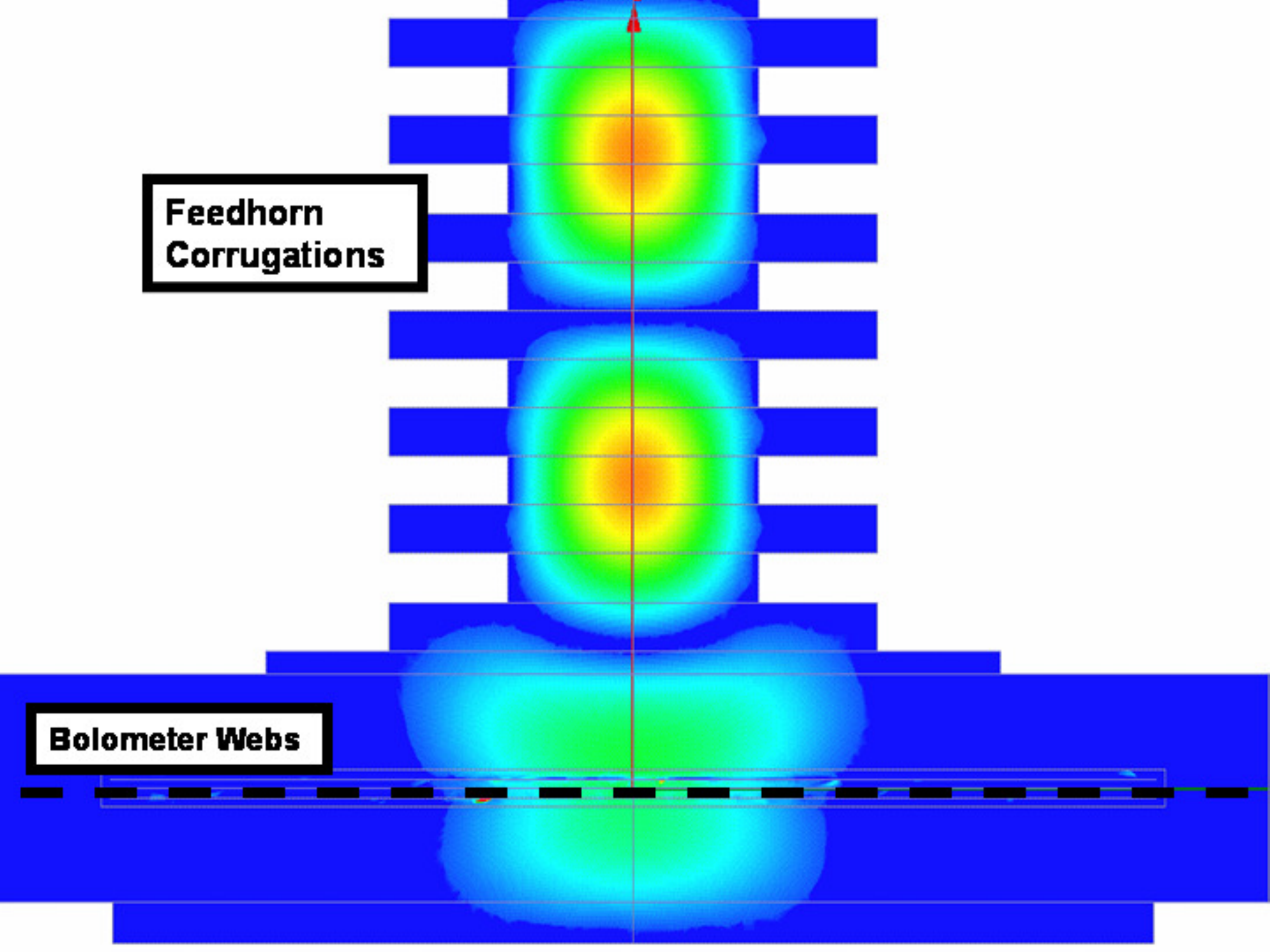}}
  \end{center}
  \caption{HFSS simulation of the 220 GHz coupling feedhorn to 150 GHz bolometer.  The color scale represents the electric field intensity.  The phase was chosen to maximize the field intensity at the bolometer webs.  The shape of the coupling feedhorn and distance between the last corrugation in the feedhorn to the position of the bolometer webs is optimized to increase the coupling efficiency and minimize the cross-polarization response. }
\label{fig:hfss220}
\end{figure}

\bicep\ PSBs were only fabricated for 100 and 150 GHz operation.  The 220 GHz pixels therefore were modified from 150 GHz PSBs using in the first season of observing.  The 220 GHz feed installation consisted of replacing two 150 GHz feedhorn stacks with a set of 220 GHz feedhorn stacks.  Each feedhorn stack consists of three separate feedhorn pieces, all of which used the same outer-forms as the 150 GHz feedhorns.  Two of the feedhorn inner-forms were made with a similar profile to the 150 GHz feedhorns.   The most difficult part of the design process was to optimize the feedhorn that couples the radiation to the bolometer.  In this case, the 220 GHz coupling feedhorn and corrugation profile had to be matched to the existing 150 GHz bolometer and housing.  Using HFSS\footnote{Ansoft's HFSS: http://www.ansoft.com/products/hf/hfss/}, the feedhorn-to-bolometer housing coupling, co-polar beam shape, total throughput, and cross-polarization beam response were optimized.  Specifically, a straight 220 GHz feedhorn profile inserted into the 150 GHz housing resulted in the best overall simulated performance (Figure~\ref{fig:hfss220}).    The lower frequency cutoff of the 220 GHz band is defined by the feedhorn waveguide cutoff while the higher cutoff is defined by a set of metal mesh filters, borrowed from the ACBAR experiment~\citep{Runyan2003}.

\section{Telescope Characterization}\label{sec:signalcharacterization}
Summarized here are the characterizations of parameters in Table~\ref{tab:bicepsummary} not mentioned previously.  For more information, see ~\cite{Yuki2010}.  The detector polarization angle, $\psi$, is measured with a 0.7$\deg$ uncertainty across all three bands, while the relative angle uncertainty in the pair is measured with a $0.1\deg$ accuracy ~\citep{Yuki2010}.  Nominally this is the mechanical orientation of the bolometer web in the focal plane, however in practice, this is a parameter characterized after deployment for more precise analysis. The ability to measure this parameter with such low statistical and systematic error is one of \bicep's design advantages.

The average cross-polarization response, $\epsilon$, measures the relative magnitude of the residual signal in the orthogonal or cross polarization direction given a purely polarized input in the co-polar direction.  The HFSS simulations predicted the 220 GHz feeds should have an $\epsilon$ less than 0.07, similar to the measured value.  Depolarization is larger for the 220 GHz channel due to the coupling effects between the 220 GHz coupling feedhorn and the bolometer and housing optimized for 150 GHz observations.  The resulting polarization efficiency, $\gamma$, is the loss in polarization signal compared to the unpolarized signal used for calibration.  $\gamma$ directly affects the observed polarization fraction, however, it has been measured to 2\% uncertainty, which is much smaller than other errors in the measurement.

The optical efficiencies (OEs) were derived by taking load curves while observing blackbody radiation at different temperatures and give the total end-to-end sensitivity on the sky.  While this parameter is not used in the mapmaking pipeline, it is implicitly included in the telescope noise estimates.  The OEs for the last observing season were 20.8\%, 19.8\%, and 15.8\%  for 100, 150, and 220 GHz respectively.  HFSS simulations for the 220 GHz feeds predicted a coupling efficiency over 98\% (similar to the 100 and 150 GHz feeds) with 2\% reflected power.  The OE of the feeds that were converted from 150 GHz to 220 GHz declined by an average of 15\%, while the other feeds did not change between seasons.  The simulated transmission and reflectivity did not include the lenses, which were optimized for 125 GHz operation (midway between 100 and 150 GHz), not 220 GHz.  

The beam functions, $P(\Omega)$, were measured using a circularly polarized broadband noise source for 100 and 150 GHz feeds and Jupiter for 220 GHz feeds.  The beam response functions are fit with an elliptical Gaussian giving an average beam Full Width Half Max (FWHM) of 0.93$\deg$, 0.60$\deg$, and 0.42$\deg$ for 100, 150, and 220 GHz respectively.  The differential beam size and ellipticities are very small for \bicep\ and do not affect the results of this paper.  Both polarized and unpolarized sidelobe response are characterized by observations of a broadband source and are determined to be negligible for this analysis.

The absolute telescope pointing is calculated from observations of multiple stars taken by an optical camera mounted to the telescope.  The absolute pointing is the same as used in ~\cite{Chiang2010} and is independent of a given feeds transmission band.  Relative radio pointing for each feed is calibrated using individual maps of CMB temperature anisotropy.  The differential pointing mismatch between PSB pairs has similar values across all three bands, with a mean of 0.0041$\deg$, 0.0046$\deg$, and 0.0047$\deg$ (1.0\%, 1.8\%, 2.6\% of the beam size) for 100, 150, and 220 GHz respectively.  The 220 GHz intensity maps show some small evidence for differential pointing between the two PSBs within a feed and between the two feeds.

Reflections in the optical system created a ``ghost image'' opposite to the primary image with respect to the boresight for each feed.  The average ghost image power, as determined by observations of the Moon, was 0.41\%, 0.50\%, and 1.3\%  for 100, 150, and 220 GHz respectively relative to the primary image. The magnitudes of the ghost beam images relative to the primary image are nearly identical for a given pair of PSBs in a feed, giving power differences of 0.02\%, 0.04\%, 0.04\%  for 100, 150, and 220 GHz bands respectively.  Since the Moon's high intensity lowers the responsivity of the detectors, the ghost image systematics given here are upper limits.  These differences are below the noise level in the Galactic polarization maps for all three bands.  While this effect has been measured well using a bright source such as the Moon, there is no evidence for ghost effects in the Galactic maps.  

A miscalibration of the time-domain impulse response, $K_t$, would act like a beam size mismatch leaking intensity to polarization.  However, the time-domain response for each of the three bands is measured to a very high precision and deconvolved from the raw 50-Hz-sampled timestream.  There is no evidence for time constant mismatch from either the Galaxy data or CMB data.   
\end{appendix}

%%%%%%%%%%%%%%%%%%%%%%%%%%%%%%%%%%%%%%%%%%%%%%%%%%%%%%%%%%%%%%%%%%%%%%%%%%%%%%%%%%%%%%%%%%%%%%%%%%%%%%%%%%%%%%%%%%%%%%%%%%%%%%%%%%%%%%%%%%%%%%%%%%%%%%%%%%%%%%%%%%%%%

\end{document}